\documentclass[journal,final,onecolumn,twoside]{IEEEtran}
\usepackage{cite}

\ifCLASSINFOpdf
   \usepackage[pdftex]{graphicx}
   \graphicspath{{./pdf/}{./jpeg/}}
   \DeclareGraphicsExtensions{.pdf,.jpeg,.png}
\else
   \usepackage[dvips]{graphicx}
   \graphicspath{{./eps/}}
   \DeclareGraphicsExtensions{.eps}
\fi
\usepackage[cmex10]{amsmath}
\usepackage{amssymb}

\usepackage[caption=false,font=footnotesize]{subfig}
\usepackage{stfloats}
\begin{document}
%
\title{Improving Low Bit-Rate Video Coding using Spatio-Temporal Down-Scaling}
%
%
%

\author{~\\Yehuda~Dar and Alfred~M.~Bruckstein
        \\~\\Technion -- Israel Institute of Technology\\Haifa 32000, Israel
		\\E-mail: ydar@cs.technion.ac.il, freddy@cs.technion.ac.il
}



\maketitle

\begin{abstract}
Good quality video coding for low bit-rate applications is important for transmission over narrow-bandwidth channels and for storage with limited memory capacity.
In this work, we develop a previous analysis for image compression at low bit-rates to adapt it to video signals. 
Improving compression using down-scaling in the spatial and temporal dimensions is examined. We show, both theoretically and experimentally, that at low bit-rates, we benefit from applying spatio-temporal scaling. The proposed method includes down-scaling before the compression and a corresponding up-scaling afterwards, while the codec itself is left unmodified. 
We propose analytic models for low bit-rate compression and spatio-temporal scaling operations. Specifically, we use theoretic models of motion-compensated prediction of available and absent frames as in coding and frame-rate up-conversion (FRUC) applications, respectively. The proposed models are designed for multi-resolution analysis.
In addition, we formulate a bit-allocation procedure and propose a method for estimating good down-scaling factors of a given video based on its second-order statistics and the given bit-budget. 
We validate our model with experimental results of H.264 compression.
\end{abstract}


%
\IEEEpeerreviewmaketitle

\section{Introduction}
%
%
%
%


The digital video is a 3D spatio-temporal signal, a sequence of 2D images (frames) captured over time. Therefore, a sampled video represents a large amount of information compared to other signals such as image or audio. As a result, video transmission and storage systems demand efficient coding from a rate-distortion perspective. Specifically, video coding for low bit-rate applications is important due to narrow-bandwidth transmission channels and low-memory devices.

\begin{figure*}[!t]
\centering
\subfloat[]{\includegraphics[width=6in]{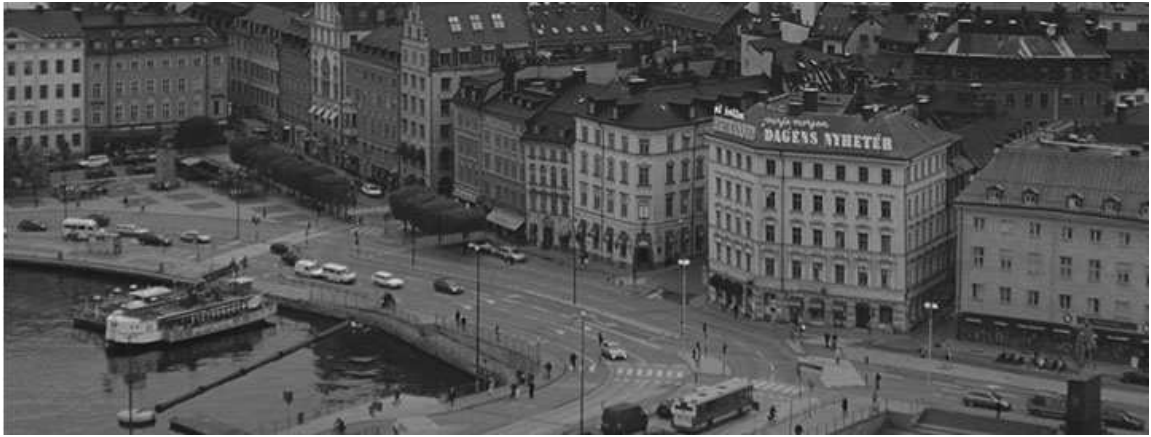}
\label{Fig:oldtowncross_original_corresponding_to_direct_compression_at_180kbps}}
\\
\subfloat[]{\includegraphics[width=6in]{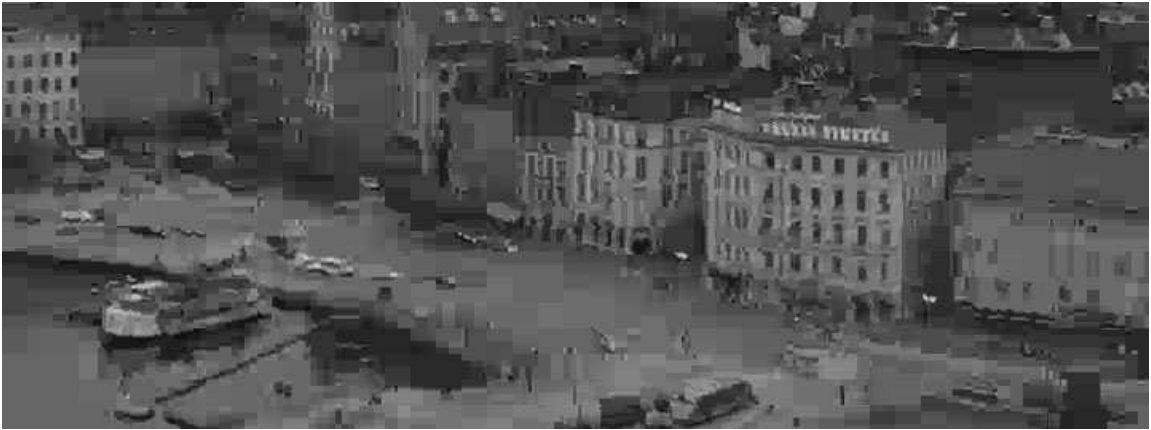}
\label{Fig:oldtowncross_direct_compression_at_180kbps}}
\caption{Demonstration of video compression at low bit-rates. Part of a frame from 'Old town cross' (720p, 50fps). (a) original (b) compressed at 180kbps.}
\label{Fig:compression_at_low_bit_rates}
\end{figure*}

Modern hybrid video compression systems encode the video by splitting its frames into 2D blocks encoded by predictive and transform coding techniques. 
The H.264 standard has led the video compression field over the last decade.
H.264 employs a variety of spatial and temporal prediction methods, and performs considerably better than previous standards. However, as in prior standards, coding at low bit-rates results in reconstructed video with severe artifacts such as blockiness (Fig. \ref{Fig:compression_at_low_bit_rates}). This poor quality is due to the reduced bit-budget that may be allocated to each block.

It has been experimentally known that image compression at low bit-rates can be improved by down-scaling the image before compression and up-scaling it to its original size after reconstruction. 
For a block-based compression method with a fixed block size, a smaller image contains fewer blocks. Therefore, the per-block bit-budget grows as the image gets smaller, and the compression distortion decreases. However, image down-sampling implies removal of high-frequency information; hence, it also reduces the quality. This trade-off between compression and down-sampling errors makes the down-sampling profitable at low bit-rates. 
Bruckstein et al. \cite{RefWorks:2} proposed an analytic explanation for these observations by modeling the JPEG compression standard as an example of a block-based codec that utilizes transform-coding. In addition, they presented an estimation procedure for the optimal down-scaling factor of a given image using its second-order statistics and the given bit-budget.

While video is a three-dimensional signal, modern hybrid compression methods perform the transform-coding on two-dimensional spatial blocks within each frame (usually, after subtracting a corresponding prediction). Hence, the spatial and temporal dimensions of the video affect the number of blocks to encode. Consequently, reducing the video dimensions will result in higher bit-budget per each block and therefore smaller compression error. 
Whereas this is similar to the static image case, video compression includes a more complex relation between a block's bit-budget and its compression error. In image compression, the bit-budget affects only the quantization. However, applying compression on video is a significantly more complex procedure than for an image; therefore, the block's bit-budget in video compression has wider effect than just adjusting the quantizer parameters. First to be affected is the chosen coding-mode, i.e., the prediction type (e.g., spatial or temporal). Next to be influenced is the prediction result, since it depends on previously decoded data. Then, the prediction error is transform-coded and quantized according to the bit-budget.

Extensions of the scaling-compression approach for video were proposed in \cite{RefWorks:65, RefWorks:74, RefWorks:76, RefWorks:73, RefWorks:64}, also referred as \textit{down-sampling based video coding}. However, these studies suggested only spatial scaling, whereas the temporal dimension was left untouched.
In \cite{RefWorks:74}, spatial decimation was performed adaptively on DCT coefficient blocks before quantization. While this method improves results by allowing variable down-sampling factors within a frame, the decisions per block are done technically by comparing errors of all the scaling possibilities. Moreover, while \cite{RefWorks:74} includes down-scaling of motion-compensated prediction residual (i.e., its application for inter-coded blocks), a theoretic justification is not given for it. Dong and Ye \cite{RefWorks:64} proposed a practical method for finding the optimal spatial down-scaling factor for video coding. They decomposed the overall error to down-scaling and coding errors and empirically estimated them using periodgram and empirical rate-distortion curve fitting, respectively.

Further studies (see e.g. \cite{RefWorks:76, RefWorks:73}) proposed to spatially down-sample video frames and to interpolate them back to their original size using super-resolution techniques. 
Shen et al. \cite{RefWorks:73} proposed to down-scale only the inter-coded frames before the motion estimation and compensation. A theoretic justification is given by expanding the analysis of \cite{RefWorks:2} to transform-coding of motion-compensated prediction residuals. However, while they simplified their analysis by assuming that frame differences come from a translational motion only, actual differences raise also from additional factors that are of great importance, e.g., reconstruction error of the reference frame, and non-translational motion. Furthermore, the motion-estimation error is modeled independently from signal characteristics such as frame rate and the complexity of the contained motion.
In addition, they substitute the second-order statistics of the motion-compensated residual into the formula from \cite{RefWorks:2} that estimates the distortion of spatial down-sampling and compression for a given signal second-order statistics and down-scaling factors; however, while in their proposal the down-scaling is applied on the original frames, the transform coding is done on the prediction-residual, and therefore their direct usage of the scaling-compression formula from \cite{RefWorks:2} does not reflect the real coding process in the system.

Temporal resolution reduction for compression at low bit-rates is mainly addressed in
studies on frame skipping mechanisms \cite{RefWorks:37, RefWorks:35, RefWorks:38, RefWorks:36, RefWorks:71}. 
While suggestions \cite{RefWorks:37, RefWorks:35, RefWorks:38} are motivated by technical considerations, \cite{RefWorks:36, RefWorks:71} explain frame-skipping by general rate-distortion analysis.
Vetro et al. \cite{RefWorks:71} studied spatio-temporal aspects of frame-skipping on the rate-distortion performance of the compression. However, the coding was represented using a general rate-distortion expression without considering properties of the coded signal and their variations due to frame-skipping. Moreover, they generally expressed the temporal interpolation error for a frame-repetition method.
Liu and Kuo \cite{RefWorks:36} proposed and explained a method for practical spatio-temporal bit-allocation consisting of adaptive frame-skipping and quantization. Due to the practical nature of their work, the analysis was based on a general rate-distortion framework that disregards specific signal properties. 

Song and Kuo \cite{RefWorks:63} proposed a rate-control algorithm that balances between spatial and temporal quality, using adaptive frame-rate selection and frame-level bit-allocation. They aimed at reducing temporal artifacts such as motion-jerkiness by allowing P-frames (that include inter-frame prediction) with constant or slowly-varying quality. Their method is based on a relation among motion-activity, bit-rate and frame-rate. The method involves an integration in the rate-control stage in a real encoder as a practical algorithm; and no theoretic analysis was done for this method. 

Many studies limit their scope to a rate-distortion analysis without considering the special statistical properties of the video signal (e.g., \cite{RefWorks:64, RefWorks:71}), or use a quantization-distortion framework where the starting point is at the transform-coefficients stage. Usually relations between the pixel domain and the signal to be transform-coded (e.g., prediction residuals) are separately studied (e.g., \cite{RefWorks:68, RefWorks:10, RefWorks:8}). 
These choices for considering a partial scope of the problem are surely due to
the difficulty in providing an accurate mathematical modeling of the video signal and the very complex video compression systems, as discussed in \cite{RefWorks:70}.
Here we aim at theoretically model the compression at low bit-rates in a wider scope than usual. Specifically, we provide an elaborate compression model that includes analysis of the coding-mode usage, the motion-compensated prediction and the transform coding. Furthermore, we express the compression distortion as function of a bit-budget and spatio-temporal properties of the input video in the pixel domain.

\begin{figure*}
\centering
\includegraphics[width=7in]{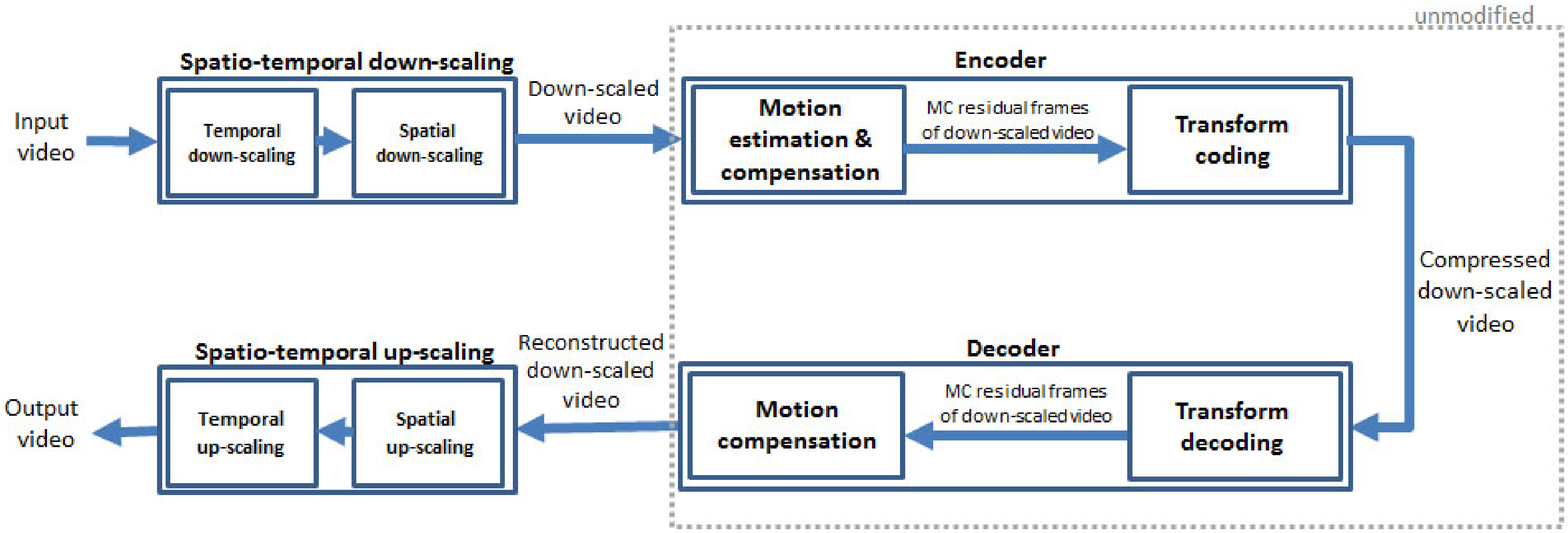}
\label{Fig:compression_scaling_system_structure_detailed}
\caption{Structure of the proposed compression-scaling system (the main components are emphasized).}
\label{Fig:compression_scaling_system_structure}
\end{figure*}

In this work, the analysis proposed in \cite{RefWorks:2} for still images is adapted to video signals. 
A comprehensive spatio-temporal analysis of the compression is proposed, and the optimal spatio-temporal down-scaling factors are examined. 
We show, both analytically and experimentally, that at low bit-rates, we benefit from applying a spatio-temporal down-scaling, i.e., reduction of frame-rate and frame-size, before the compression and a corresponding up-scaling afterwards (Fig. \ref{Fig:compression_scaling_system_structure}). The suggested procedure improves the compression efficiency at low bit-rates by means of estimating the optimal down-scaling using an analytic model of the compression, and performing scaling operations outside the codec; i.e., unlike in frame-skipping, we leave the codec itself unmodified. 

We provide a reasonable complete analytic model for the suggested compression-scaling system. In section \ref{sec:Signal Model for Multi-Resolution Analysis} we start with presenting a video signal model that is suitable for multi-resolution analysis.
Then, we analyze low bit-rate compression as carried out in H.264's baseline profile that serves as an example of a hybrid video codec. Specifically, the coding modes used, motion-compensated coding, transform coding and bit-allocation are studied in sections \ref{sec:Coding-Mode Usage at Low Bit-Rates}-\ref{sec:Overall Compression}.
In section \ref{sec:Compression-Scaling System}, we examine the overall compression-scaling system and formulate a bit-allocation optimization problem for finding the best down-scaling factor of a given video from its second-order statistics and a bit-budget. Numerical evaluations of the proposed model are also presented in section \ref{sec:Theoretical Estimations of the Model} as an alternative approach for finding the optimal down-scaling factor. Then, in section \ref{sec:Experimental Results}, we give experimental results for qualitative comparison with a real H.264 codec.

\section{Signal Model for Multi-Resolution Analysis}
\label{sec:Signal Model for Multi-Resolution Analysis}
\subsection{Continuous Signal Model}
Digital video signal is a sequence of 2D-images (i.e., frames) sampled from a continuous video signal at a constant frequency. In the coding procedure the frames are further partitioned into 2D-blocks.

Let us consider a video signal of one-second length. We assume it is defined on the unit cube $\left[ {0,1} \right] \times \left[ {0,1} \right] \times \left[ {0,1} \right]$, and represented by the function:
\begin{equation}
\label{eq:video signal as continuous function over the unit-cube}
{f_v}\left( {x,y,t} \right){\rm{ }}:{\rm{ }}\left[ {0,1} \right] \times \left[ {0,1} \right] \times \left[ {0,1} \right] \to \mathbb{R}
\end{equation}
A set of $T$ frames is defined in the unit cube as
\begin{equation}
\label{eq:video temporal slicing into frames definition}
\left\{ {{f_v}\left( {x,y,t} \right){\mkern 1mu} {\mkern 1mu} \left| {{\mkern 1mu} {\mkern 1mu} \left( {x,y} \right) \in \left[ {0,1} \right] \times \left[ {0,1} \right]{\mkern 1mu} {\mkern 1mu} {\mkern 1mu} ,{\mkern 1mu} {\mkern 1mu} {\mkern 1mu} t \in \left\{ {\frac{i}{T}} \right\}_{i = 0}^{T - 1}} \right.} \right\}
\end{equation}

A frame, $f _v \left( x, y, t=h \right)$, is assumed to be a realization of a 2D random process. Specifically, we assume it is wide-sense stationary with zero mean in the form of separable first-order Markov model; i.e., the spatial autocorrelation of a frame is 
\begin{equation}
\label{eq:spatial autocorrelation of a frame  - Markov model}
{R_v}\left( {{\tau _x},{\tau _y}} \right) = \sigma _v^2 \cdot {e^{ - {\alpha _x}\left| {{\tau _x}} \right|}}{e^{ - {\alpha _y}\left| {{\tau _y}} \right|}} .
\end{equation}

Since we study a block-based compression system, it is useful to consider partitioning of a frame into $M \cdot N$ equal-size 2D-blocks, i.e., the $h^{th}$ frame in the sequence is divided into the following set of 2D-regions defined as:
\begin{IEEEeqnarray}{rCl}
\label{eq:frame spatial slicing definition}
\Delta _{ij}^h \equiv \left[ {\frac{{i - 1}}{M},\frac{i}{M}} \right] \times \left[ {\frac{{j - 1}}{N},\frac{j}{N}} \right]
\\ \nonumber
for\,\,\,i = 1,...,M{\rm{ ; }}j = 1,...,N.
\end{IEEEeqnarray}
We refer to $M$ and $N$ as the spatial slicing parameters, and to $T$ as the temporal slicing parameter.

\subsection{Slices and Down-Scaling}
We assume block-based compression with a fixed block size denoted as ${W_{block}} \times {H_{block}}$; e.g., ${W_{block}} = {H_{block}} = 16$ for H.264 macroblocks. The block dimensions relate the spatial slicing parameters with the actual frame size, as follows:
\begin{IEEEeqnarray}{rCl}
\label{eq:spatial slicing parameters and frame size relation}
{W_{frame}} & = & {W_{block}} \cdot M
\label{eq:spatial slicing parameters and frame size relation - horizontal}
\\ 
{H_{frame}} & = & {H_{block}} \cdot N.
\label{eq:spatial slicing parameters and frame size relation - vertical}
\end{IEEEeqnarray}
The considered video is one-second length; hence, the frame-rate and the temporal-slicing parameter hold:
\begin{IEEEeqnarray}{rCl}
\label{eq:temporal slicing parameters and frame rate relation}
{F_{rate}} & = & T
\end{IEEEeqnarray}

We define the down-scaling factor as the ratio between dimension values of the original and the down-scaled videos. $D_M$, $D_N$ and $D_T$ denote the spatial-horizontal, spatial-vertical and temporal down-scaling factors. Using (\ref{eq:spatial slicing parameters and frame size relation - horizontal})-(\ref{eq:temporal slicing parameters and frame rate relation}) we write:
\begin{IEEEeqnarray}{rCl}
\label{eq:down-scaling factors definitions}
{D_M} & = & \frac{{W_{frame}^{original}}}{{W_{frame}^{scaled}}} = \frac{{{M^{original}}}}{{{M^{scaled}}}}
\label{eq:horizontal down-scaling factors definition}
\\ 
{D_N} & = & \frac{{H_{frame}^{original}}}{{H_{frame}^{scaled}}} = \frac{{{N^{original}}}}{{{N^{scaled}}}}
\label{eq:vertical down-scaling factors definition}
\\
{D_T} & = & \frac{{F_{rate}^{original}}}{{F_{rate}^{scaled}}} = \frac{{{T^{original}}}}{{{T^{scaled}}}}.
\label{eq:temporal down-scaling factors definition}
\end{IEEEeqnarray}

Let us demonstrate the last definitions by an example. Consider a video with frame size of 720x720 pixels. According to (\ref{eq:spatial slicing parameters and frame size relation - horizontal}), the horizontal slicing parameter is ${M^{original}} = \frac{{{W_{frame}}}}{{{W_{block}}}} = \frac{{720}}{{16}} = 45$. Assume that some analytic optimization procedure advice us to use ${M^{scaled}} = 15$. Using (\ref{eq:horizontal down-scaling factors definition}), we calculate the horizontal down-scaling factor: ${D_M} = \frac{{{M^{original}}}}{{{M^{scaled}}}} = \frac{{45}}{{15}} = 3$.

\subsection{Measuring Bit-Rate}
The compression process addresses the classic rate-distortion trade-off. The distortion is defined as the video quality after reconstruction from the compressed data, and it is measured in Peak-Signal-to-Noise-Ratio (PSNR). The rate is the amount of bits allocated for the signal compression, and also referred as bit-rate. Usually, video bit-rate is expressed in bits per time-unit, i.e., bits per second, kilo-bits per second, or mega-bits per second (abbreviated forms: bps, kbps and mbps, respectively).

In this work, we examine results of performing compression of spatio-temporal down-scaled video, and then up-scaling to the original resolution (Fig. \ref{Fig:compression_scaling_system_structure}); hence, the compared results involve compression of videos in different frame rates and sizes. The spatio-temporal resolution difference results in diverse amount of slices (or macroblocks) in the videos, and varying bit-budget per slice. The proposed model for the entire compression process evaluates performance at a constant bit-budget measured in bits-per-time-unit. However, some building blocks of the model treat macroblock-level procedures affected from the macroblock-bit-budget, which is different for various spatio-temporal resolutions at a given bit-rate per time-unit. These macroblock-level procedures (such as the coding-mode selection and the motion-compensation) are fairly compared for various spatio-temporal resolutions by analyses for bit-rates given in a bits-per-slice units (denoted as $B_{slice}$). Note that the bit-per-slice unit is an indicative representation of the bit-rate rather than the exact bit amount allocated for the coding of each slice in the sequence.

Bit-rate in bits-per-slice can be calculated from a bits-per-second value according to our spatio-temporal slicing parameters, $M$, $N$ and $T$:
\begin{equation}
\label{eq:bits-per-slice definition}
{B_{slice}} = \frac{B_{second}}{{M \cdot N \cdot T}}
\end{equation}
Where $B_{second}$ is the bit-rate in bits-per-second units.
For a compression block size of $W_{block} \times H_{block}$ the following relation to bits-per-pixel, $B_{pixel}$, holds:
\begin{equation}
\label{eq:bits-per-slice relation to bit-per-pixel}
{B_{slice}} = W_{block} \cdot H_{block} \cdot B_{pixel}.
\end{equation}

Let us demonstrate the last definitions by an example. 
Bit-rate of 1 mbps is considered as low-bit-rate for coding an HD video of 720p (i.e., frames of 1280x720 pixels) at 50 frames-per-second (fps), whereas it is high bit-rate for coding a QCIF (i.e., frames of 176x144 pixels) video at 15 fps. Whereas the bits-per-second units does not reflect this difference, expressing the bit-rate in bit-per-slice or bits-per-pixel will reflect it well. The QCIF bit-rates are:
\begin{IEEEeqnarray}{rCl}
\label{eq:calculation example of QCIF bit-rate}
&& B_{pixel}^{QCIF@15fps} = \frac{{{{10}^6}}}{{176 \times 144 \times 15}} = 2.6\left[ {bits/pixel} \right]
\\ \nonumber
&& B_{slice}^{QCIF@15fps} = 16 \times 16 \times B_{pixel}^{QCIF@15fps} = 673\left[ {bits/slice} \right],
\end{IEEEeqnarray}
and the 720p bit-rates are:
\begin{IEEEeqnarray}{rCl}
\label{eq:calculation example of 720p bit-rate}
&& B_{pixel}^{720p@50fps} = \frac{{{{10}^6}}}{{1280 \times 720 \times 50}} = 0.022\left[ {bits/pixel} \right]
\\ \nonumber
&& B_{slice}^{720p@50fps} = 16 \times 16 \times B_{pixel}^{720p@50fps} = 5.5\left[ {bits/slice} \right].
\end{IEEEeqnarray}
The bit-rate ratio between QCIF@15fps and 720p@50fps is around 120; this reflects the suitability of 1mbps to these distinct video resolutions.

\subsection{Reconstructed Signal Error}
The video signal reconstructed from the compressed signal $f_v$ is represented as a set of   $T$ reconstructed frames, where each frame is defined on the continuous spatial unit-square:
\begin{equation}
\label{eq:reconstructed frames definition}
\left\{ {{\hat f_v}\left( {x,y,t} \right){\mkern 1mu} {\mkern 1mu} \left| {{\mkern 1mu} {\mkern 1mu} \left( {x,y} \right) \in \left[ {0,1} \right] \times \left[ {0,1} \right]{\mkern 1mu} {\mkern 1mu} {\mkern 1mu} ,{\mkern 1mu} {\mkern 1mu} {\mkern 1mu} t \in \left\{ {\frac{i}{T}} \right\}_{i = 0}^{T - 1}} \right.} \right\}
\end{equation}
The mean-squared-error of reconstructing the signal $f$ over a 2D region, $\Delta$, is defined as:
\begin{IEEEeqnarray}{rCl}
\label{eq:MSE of 2D signal}
\nonumber && MSE_f\left( \Delta  \right) \equiv \frac{1}{{A\left( \Delta  \right)}} \int\!\!\!\int_{\Delta}{{{\left( {f\left( {x,y} \right) - \hat f\left( {x,y} \right)} \right)}^2}dxdy}
\\
\end{IEEEeqnarray}
where $A\left(  \cdot  \right)$ is the area of a given region.
The mean-squared-error of the reconstruction of the entire signal is denoted as $\varepsilon _v^2$ and it is calculated as follows:
\begin{IEEEeqnarray}{rCl}
\label{eq:reconstruction MSE}
\varepsilon _v^2 & = & \frac{1}{{T \cdot A\left( {\left[ {0,1} \right] ^ 2} \right)}} \cdot  \sum \limits_{ h = 1 }^{ T } \iint \limits_{{\left( {x,y} \right) \in \left[ {0,1} \right] ^ 2}} \left( {{f_v}\left( {x,y,t} \right) - \widehat {{f_v}}\left( {x,y,\frac{h-1}{T}} \right)} \right)^2 dxdy
\end{IEEEeqnarray}
where $A\left( {\left[ {0,1} \right] ^ 2} \right) = 1$.
Let us define ${D_{{f_v}}}\left( {x,y,t} \right) \equiv {f_v}\left( {x,y,t} \right) - \widehat {{f_v}}\left( {x,y,t} \right)$.
We assume that all the 2D-slices have the same second-order statistics and $f_v$ is a wide-sense stationary signal. The expected MSE in terms of the 2D-slices is:
\begin{IEEEeqnarray}{rCl}
\label{eq:reconstruction MSE - in 2D slices terms}
E\left[ {\varepsilon _v^2} \right] & = & E\left[ \frac{1}{T} \cdot \sum\limits_{h = 1}^T \sum\limits_{i = 1}^M \sum\limits_{j = 1}^N {\iint_{\Delta _{ij}^h} {{{D_{{f_v}}}\left( {x,y,\frac{{h - 1}}{T}} \right)^2}dxdy} } \right]
\\ \nonumber
& = & \frac{1}{T} \cdot M \cdot N \cdot T \cdot E\left[ {\iint_{\Delta _{11}^1} {{{D_{{f_v}}}\left( {x,y,\frac{{h - 1}}{T}} \right)^2}dxdy}} \right]
\\ \nonumber
& = & E\left[ {MSE_{f_v}\left( {\Delta _{11}^1} \right)} \right]
\end{IEEEeqnarray}

%
%
%
%
%

\section{Coding-Mode Usage at Low Bit-Rates}
\label{sec:Coding-Mode Usage at Low Bit-Rates}

Hybrid video coding combines predictive (including motion-compensation) and transform coding techniques. Modern block-based compression systems have several coding modes that are chosen blockwise by the encoder. The main difference between coding modes is the prediction method; e.g., inter prediction utilizes information from previously decoded frames, while intra prediction considers only the current frame. H.264's skip mode is an example for low bit-cost method that offers a simple motion-compensation prediction without any tranform coding of the prediction error.
Coding-mode selection depends on factors such as bit-rate, signal properties and run-time limitations.
In this section we study the coding mode usage at low bit-rates.

\subsection{Proposed Model}
We consider only one H.264-slice per frame; hence, the frames are classified into I and P frames by adapting the definitions for H.264-slices. An I-frame can contain only intra coded blocks and skipped blocks; whereas a P-frame contains at least one inter coded block and any number of intra coded or skipped blocks.

In this work, we assume a long sequence of P-frames as common in H.264, especially in low bit-rates. Therefore, we neglect I-frames in our compression analysis and consider only P-frames.

We assume that the encoder chooses a coding mode (i.e., intra, inter or skip) for each 2D block, $\Delta _{ij}^h$, independently of other blocks in the frame. This deviates from a real encoder, where frame types are assigned before the macroblock processing; e.g., many encoders divide the sequence into frame-groups, each begins with an I-frame and the rest are P-frames. Due to the low bit-rate scenario, we further assume that intra coding in P-frames has minor portion, and therefore can be neglected in rate-distortion analysis.

The macroblocks in H.264 are of 16x16 pixels size. Partitioning of macroblocks to smaller blocks results in better video quality; however, the bit-cost gets higher due to representation of finer block partitions. Therefore, we assume that in our low bit-rate scenario, all the encoded blocks are of 16x16 pixels; i.e., macroblocks are not split.

We represent a block's coding mode as a discrete random variable with probability mass function, ${P_{\Delta _{ij}^h\,\,{\rm{coding}}\,\,{\rm{mode}}}}\left(  \cdot  \right)$,that depends on the bit-rate in bits-per-slice units, ${B_{slice}}$, as follows:
\begin{IEEEeqnarray}{rCl}
\label{eq:probability mass function of block coding mode}
{P_{\Delta _{ij}^h{\kern 1pt} {\kern 1pt} {\text{coding}}{\kern 1pt} {\kern 1pt} {\text{mode}}}}\left( {\Delta _{ij}^h{\mkern 1mu} \,coding{\mkern 1mu} \,mode} \right) = \left\{ {\begin{array}{*{20}{c}}
  {inter}&{{\mkern 1mu} ,{\mkern 1mu} {\mkern 1mu} {\mkern 1mu} w.p.{\mkern 1mu} {\mkern 1mu} \,{P_{{\text{inter}}}}\left( {{B_{slice}}} \right)} \\ 
  {skip}&{{\mkern 1mu} ,{\mkern 1mu} {\mkern 1mu} {\mkern 1mu} w.p.{\mkern 1mu} \,{\mkern 1mu} {P_{{\text{skip}}}}\left( {{B_{slice}}} \right)} 
\end{array}} \right.
\end{IEEEeqnarray}
Where, for a given bit-rate, the total probability hold:
\begin{IEEEeqnarray}{rCl}
\label{eq:total probability for a given slice-bit-rate}
{P_{{\text{inter}}}}\left( {{B_{slice}}} \right) + {P_{{\text{skip}}}}\left( {{B_{slice}}} \right) = 1
\end{IEEEeqnarray}

\begin{figure*}[!t]
\centering
\subfloat[]{\includegraphics[width=3.25in]{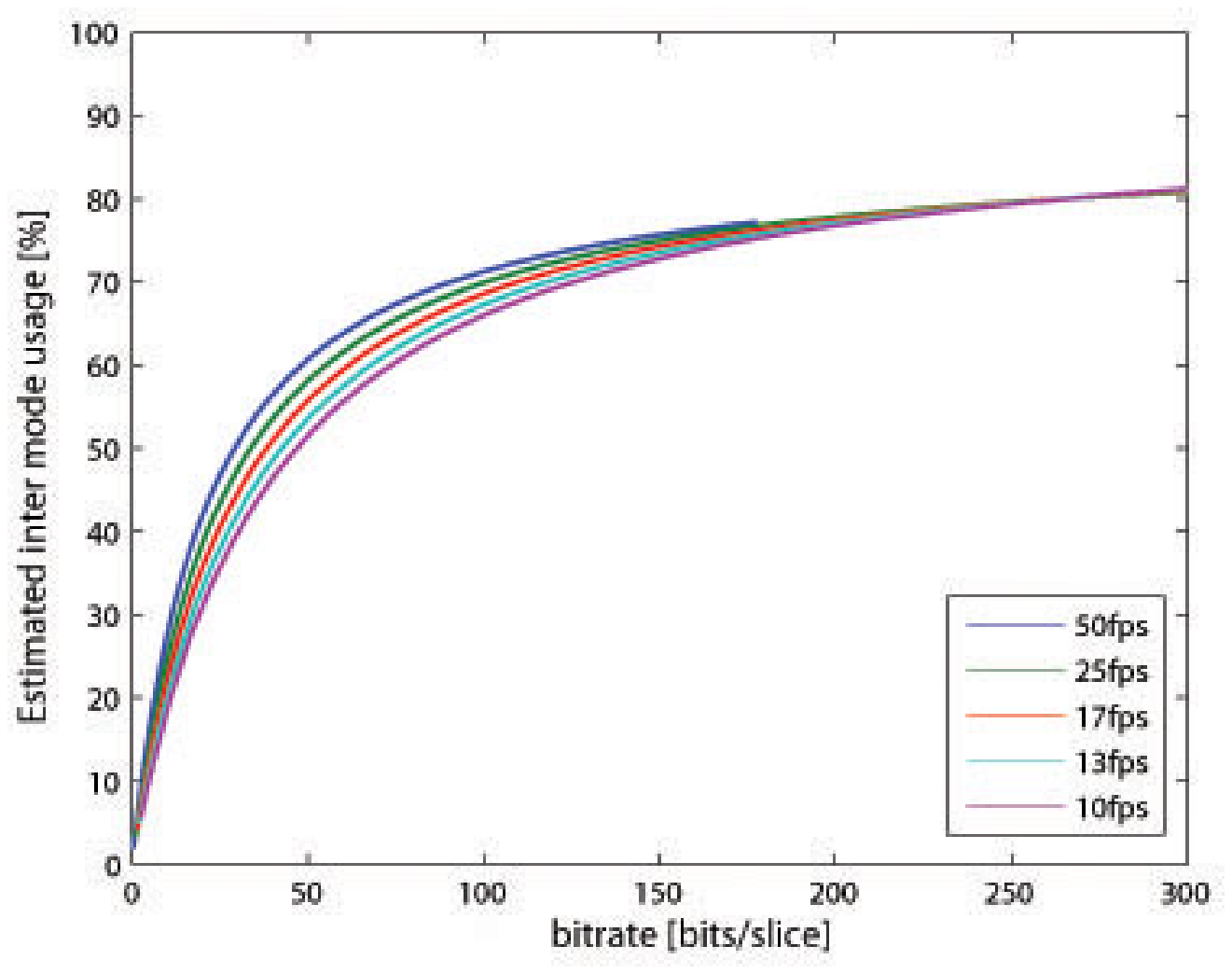}
\label{Fig:inter_mode_usage_linear_fractional_model_qvarbasic225_gammaC_0_3}}
\subfloat[]{\includegraphics[width=3.25in]{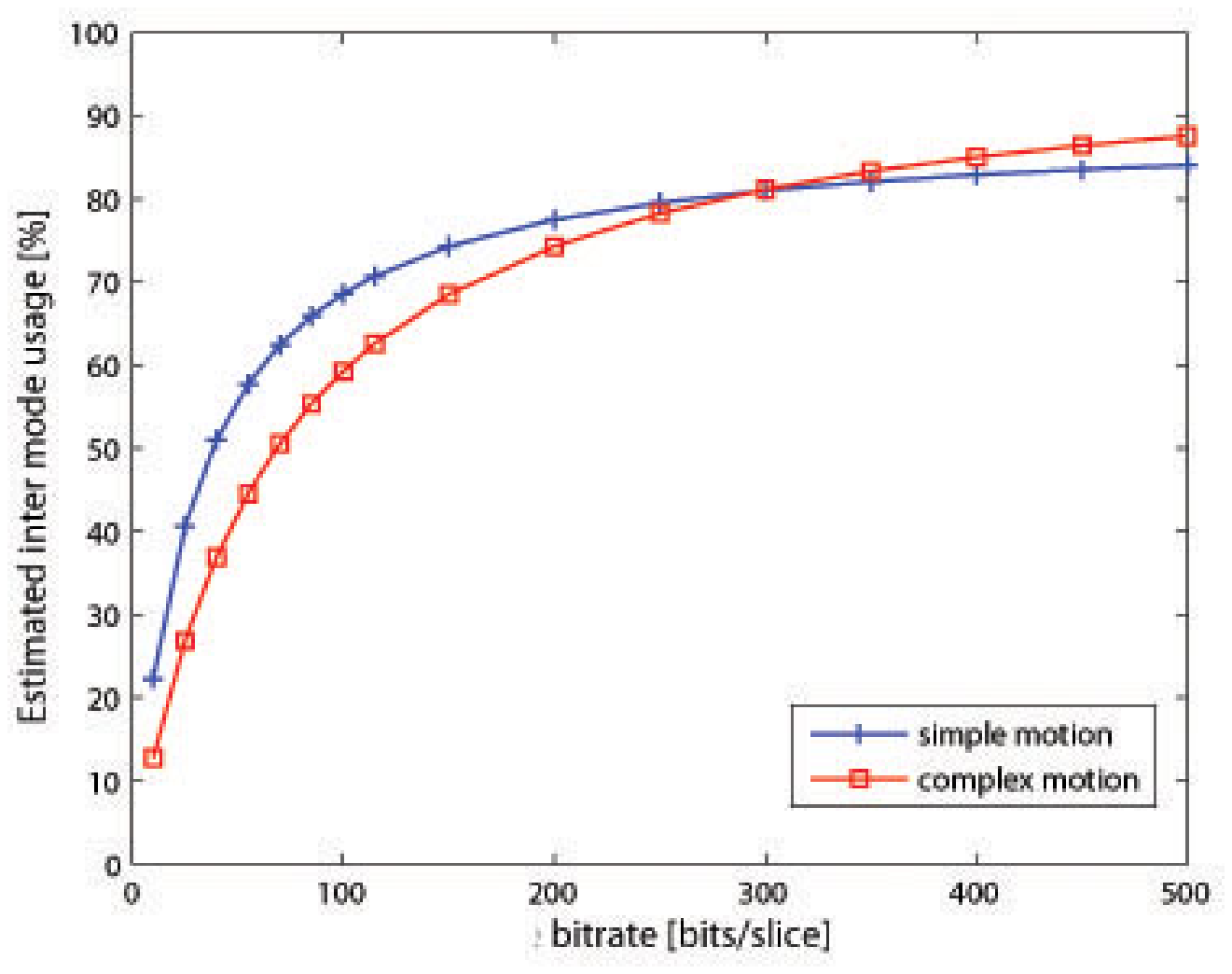}
\label{Fig:inter_mode_usage_linear_fractional_model_gammaC_0_3__50fps_comparison_qvarbasic225and800}}
\caption{Estimated inter mode usage according to a linear-fractional model ($P_{inter}^{asymp,\min } = 85$, ${\gamma _c}=0.3$ and ${\gamma _d}=20$). (a) a video in various frame-rates (${\tilde \sigma _q^2} = 225$) (b) different videos at 50fps (${\tilde \sigma _q^2} = 225$ and ${\tilde \sigma _q^2} = 800$).}
\label{Fig:Estimated inter mode usage according to a linear-fractional model.}
\end{figure*}

We claim that as the bit-rate decreases and approaches to be very low, more blocks are coded in skip mode instead of inter coding. We assume this process is a linear-fractional function of bit-rate $B_{slice}$, and write it as:
\begin{IEEEeqnarray}{rCl}
\label{eq:inter and skip mode probabilities as piecewise-linear function}
\nonumber && {P_{inter}}\left( {{B_{slice}},{F_{rate}}} \right) = \frac{{{B_{slice}}}}{{{c_m}\left( {{F_{rate}}} \right)\cdot{B_{slice}} + {d_m}\left( {{F_{rate}}} \right)}}
\\ 
&& {P_{skip}}\left( {{B_{slice}}} \right) = 1 - {P_{inter}}\left( {{B_{slice}}} \right)
\end{IEEEeqnarray}
Where $c_m$ and $d_m$ control the asymptotic value of the function and the convergence rate, respectively (Fig. \ref{Fig:inter_mode_usage_linear_fractional_model_qvarbasic225_gammaC_0_3}). $c_m$ and $d_m$ are affected by the frame-rate, $F_{rate}$, and the motion-characteristics of the video, $\tilde \sigma _q^2$ (as was defined in \cite{RefWorks:68}):
\begin{IEEEeqnarray}{rCl}
\label{eq:inter and skip mode probabilities as piecewise-linear function}
{c_m}\left( {{F_{rate}}} \right) & = & \frac{{100}}{{P_{inter}^{asymp,\min } + {\gamma _c} \cdot \frac{{\tilde \sigma _q^2}}{{{F_{rate}}}}}}
\\ \nonumber
{d_m}\left( {{F_{rate}}} \right) & = & \gamma _d + \frac{{\tilde \sigma _q^2}}{{{F_{rate}}}}
\end{IEEEeqnarray}
Where, $P_{inter}^{asymp,\min }$ is the minimal inter mode percentage (e.g., as in a video with simple motion); $\gamma _c$ and $\gamma _d$ are for normalization of the motion characteristics part and the convergence rate, respectively.

When the frame-rate is higher, the motion-compensated prediction residual has reduced energy and inter coding is more advantageous. Therefore, the inter coding percentage grows with the frame-rate (Fig.\ref{Fig:inter_mode_usage_linear_fractional_model_qvarbasic225_gammaC_0_3}). 
As the motion in the video is more complex (i.e., higher $\tilde \sigma _q^2$), the skip mode performance degrades. Hence, for high bit-rates the inter mode percentage is higher; however, for low bit-rates the percentage is lower due to the increased bit-cost for a given inter coding quality (Fig. \ref{Fig:inter_mode_usage_linear_fractional_model_gammaC_0_3__50fps_comparison_qvarbasic225and800}).

\subsection{Experimental Results}
\begin{figure*}
\centering
\subfloat[]{\includegraphics[height=2.5in]{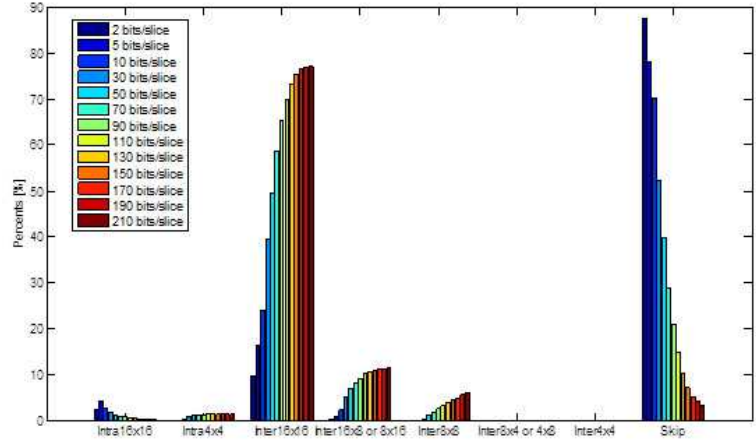}
\label{Fig:coding_mode_usage_parkrun}}
\\
\subfloat[]{\includegraphics[height=2.7in]{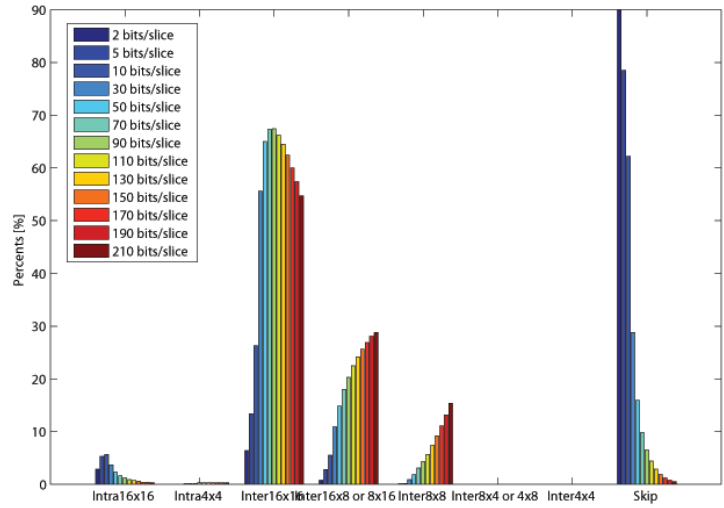}
\label{Fig:coding_mode_usage_ducks}}
\caption{Coding mode usage in real H.264 P-frame. (a) 'Parkrun' (b) ‘Ducks take off', both grayscale 720p@50fps}
\label{Fig:Coding mode usage in real H.264 P-frame}
\end{figure*}

The proposed model for coding mode usage was validated by analyzing data of a real H.264 codec. Fig. \ref{Fig:Coding mode usage in real H.264 P-frame} shows the coding modes and macroblock partitioning statistics for various bit-rates. The data was taken from P-frames. The experiments covered the entire bit-rate range: from very low (2bits/slice), through low (around 30 bits/slice), and up to high (210 bits/slice). Translating the range boundaries into bits-per-second units for 720x720 pixels frames at 50fps, gives $2\frac{{bits}}{{slice}} \approx 200\frac{{Kbits}}{{\sec }}$ and $210\frac{{bits}}{{slice}} \approx 20.7\frac{{Mbits}}{{\sec }}$. The experimental results support our model by showing the growing dominancy of encoding block size of 16x16 pixels (inter16x16 and skip modes) as bit-rate reduces. Moreover, the intra-coding percentage at low bit-rates is very low; thus, our disregard of intra mode was justified. The inter coding and skip mode percentages increase and decrease, respectively, as the bit-rate grows to the very high bit-rate area; furthermore, for 'Parkrun' (Fig. \ref{Fig:coding_mode_usage_parkrun}) they seem to converge to high and low values as in our model, whereas for 'Ducks take off' (Fig. \ref{Fig:coding_mode_usage_ducks}) more inter macroblock are partitioned at the expense of the inter 16x16 mode. Note that the shown avoidance from block partition smaller than 8x8 pixels is due to technical considerations of the codec implementers; however, this is negligible since these partition sizes are expected to have minor usage at the expense of 8x8 pixels inter coded blocks. 

In addition, the experiments show that H.264 uses long sequences of P-frames, even at high bit-rates. E.g., only 2 out of 500 frames were I-frames when coding the 'Old town cross' sequence at the high rate of 20.7 Mbit/sec. This further justifies our treatment of only P-frames in our performance analysis.

\begin{figure*}
\centering
\subfloat[]{\includegraphics[height=2.5in]{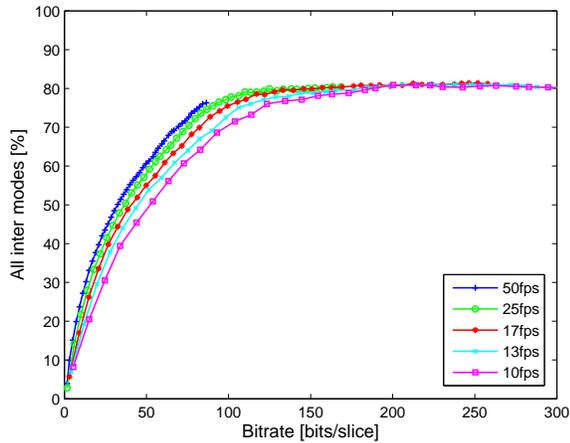}
\label{Fig:all_inter_modes_usage_old_town_cross}}
\subfloat[]{\includegraphics[height=2.5in]{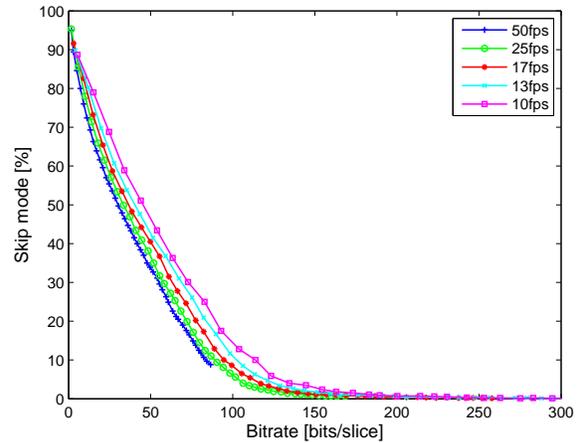}
\label{Fig:skip_mode_usage_old_town_cross}}
\caption{Coding mode usage in real H.264 P-frames of 'Old town cross' (720x720, grayscale at various frame-rates). (a) overall inter modes (b) skip mode}
\label{Fig:Skip and inter mode usage in real H.264 P-frame for various frame-rates. Old town cross.}
\end{figure*}

\begin{figure*}
\centering
\subfloat[]{\includegraphics[height=2.5in]{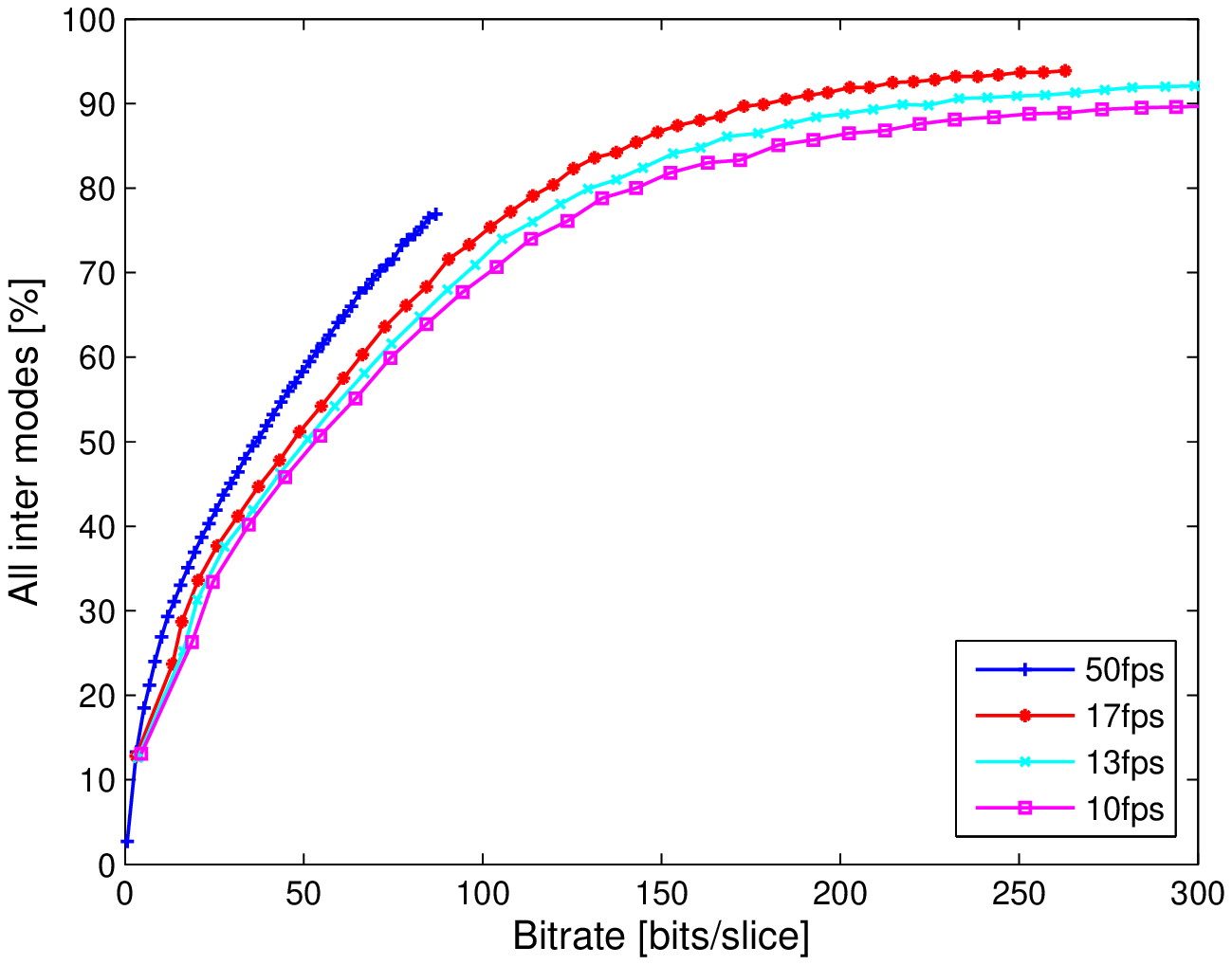}
\label{Fig:all_inter_modes_usage_parkrun}}
\subfloat[]{\includegraphics[height=2.5in]{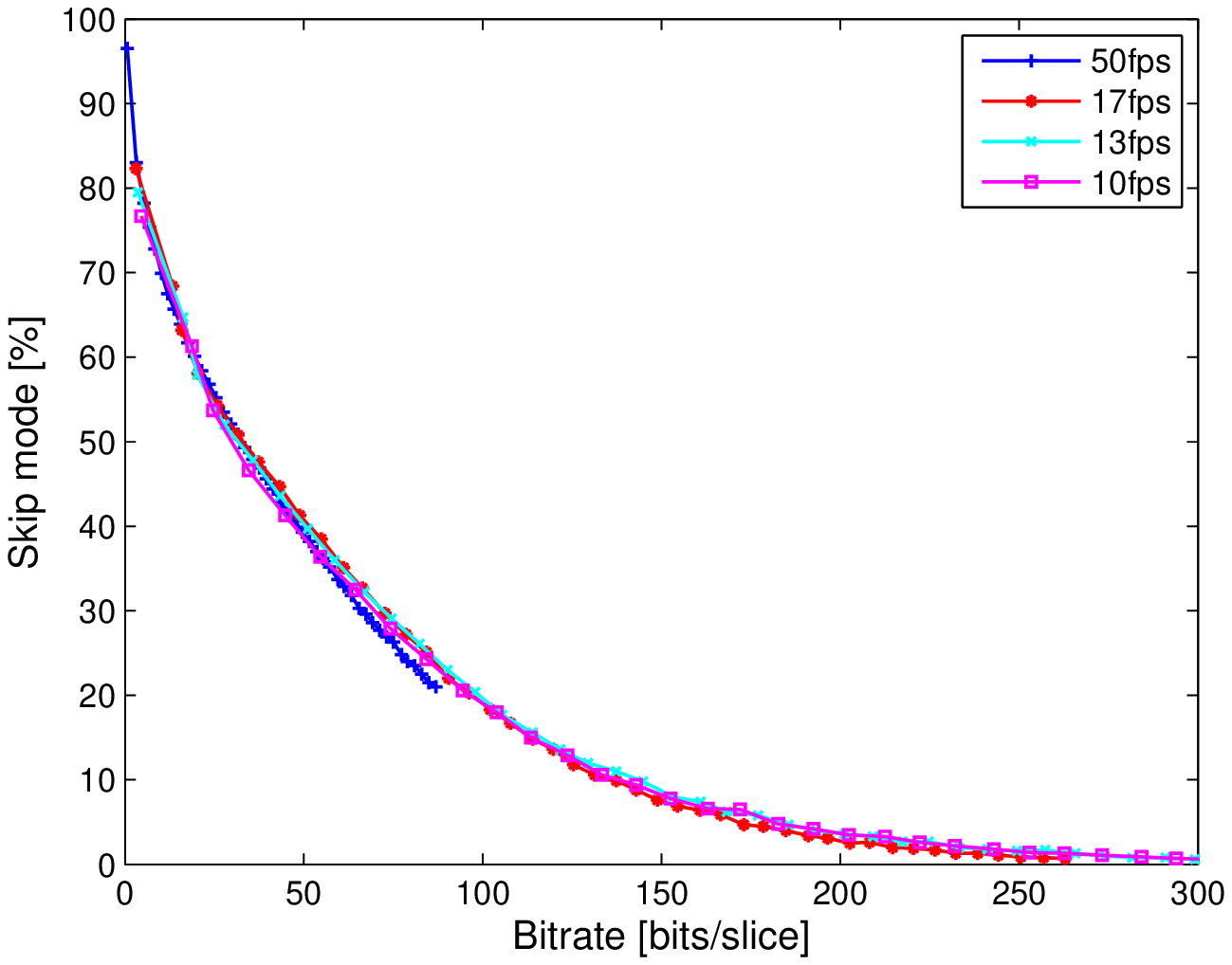}
\label{Fig:skip_mode_usage_parkrun}}
\caption{Coding mode usage in real H.264 P-frames of 'Parkrun' (720x720, grayscale at various frame-rates). (a) overall inter modes (b) skip mode}
\label{Fig:Skip and inter mode usage in real H.264 P-frame for various frame-rates. Parkrun.}
\end{figure*}

\begin{figure}
\centering
\includegraphics[width=3.5in]{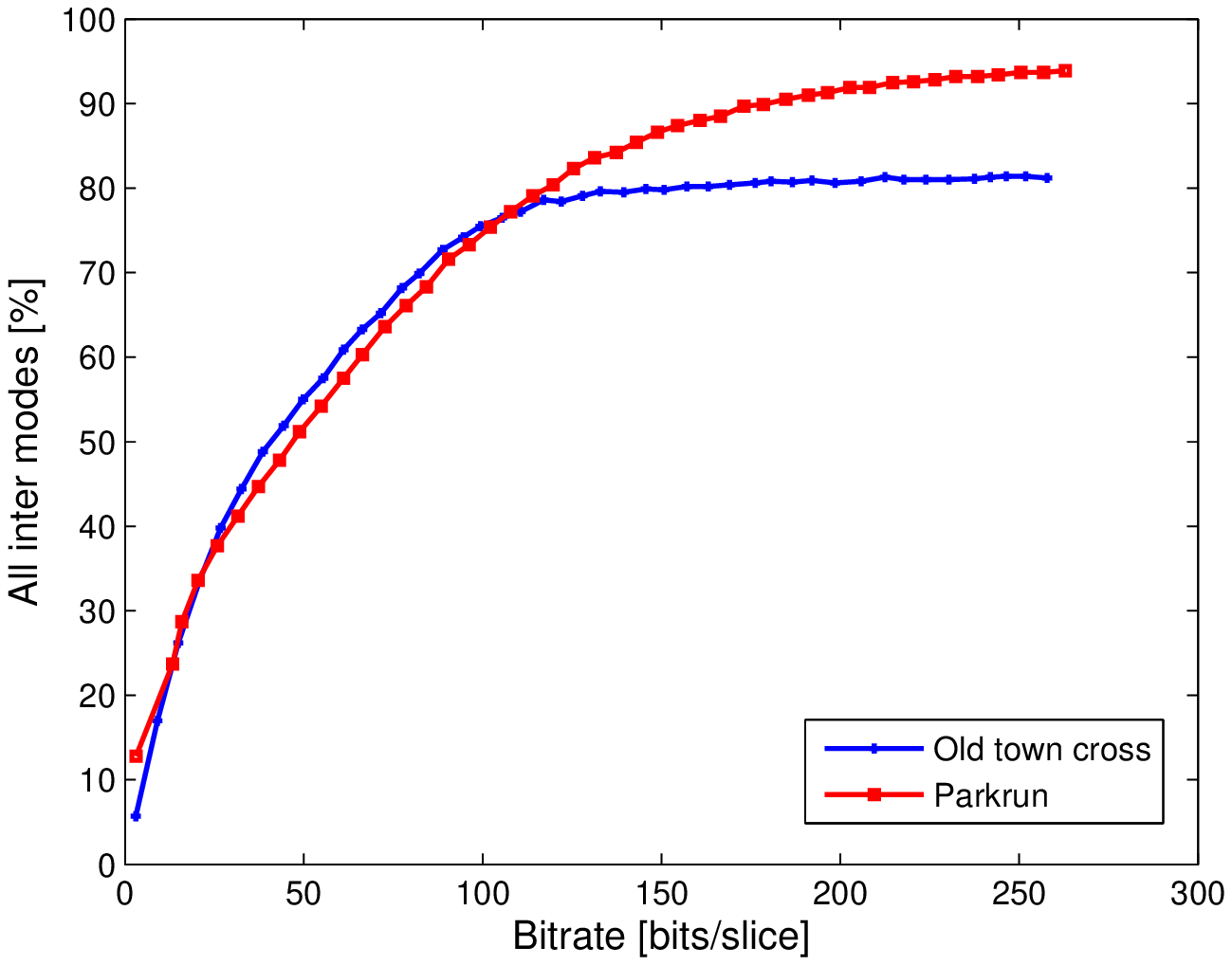}
\caption{Coding mode usage in real H.264 P-frames of 'Old town cross' and 'Parkrun' (720x720, grayscale at 17fps).}
\label{Fig:all_inter_modes_usage_old_town_cross_vs_parkrun_at_17fps}
\end{figure}

Measurements of inter and skip mode percentage (Fig. \ref{Fig:Skip and inter mode usage in real H.264 P-frame for various frame-rates. Old town cross.},\ref{Fig:Skip and inter mode usage in real H.264 P-frame for various frame-rates. Parkrun.}) show a linear-fractional behavior. Moreover, the dependency of our model on motion characteristics of the video is approved by the convergence of percentage of inter coding usage in 'Parkrun' is higher than in 'Old town cross' (Fig. \ref{Fig:all_inter_modes_usage_old_town_cross_vs_parkrun_at_17fps}).

Further comparison with model estimation is possible using calculating model parameters for the tested sequences, as follows.
$\tilde \sigma _q^2$ is calculated as follows \cite{RefWorks:68}:
\begin{IEEEeqnarray}{rCl}
\label{eq:Calculation of basic q variance}
\tilde \sigma _q^2 = \frac{{\hat \sigma _{1,2}^2 - 2\left( {\sigma _{\Delta x}^2 + \sigma _{\Delta y}^2} \right)\sigma _v^2\cdot\left( {1 - {\rho _v}} \right)}}{{2\left[ {\left( {\sigma _{\Delta x}^2 + \sigma _{\Delta y}^2} \right)L + 1} \right] \cdot \frac{1}{{{F_{rate}}}}}},
\end{IEEEeqnarray}
where $\hat \sigma _{1,2}^2$ is an empirically measurement of MC-prediction error between two frames; $\sigma _{\Delta x}^2$ and $\sigma _{\Delta y}^2$ reflect accuracy errors in ME and are calculated according to uniform distribution over error interval of half-pel accuracy; $\sigma _v^2$ and $\rho _v$ are the variance and correlation coefficient of the frame pixels; and $L$ is a temporal memory factor that is set to 100. The values calculated for 'Old town cross' were $\sigma ^2 _v = 2352$ and $\tilde \sigma _q^2 = 253$. For 'Parkrun' the measurements showed $\sigma ^2 _v = 2682$ and $\tilde \sigma _q^2 = 851$.



\section{Motion-Compensated Coding}
\label{sec:Motion-Compensated Coding}
According to section \ref{sec:Coding-Mode Usage at Low Bit-Rates}, we can study compression at low bit-rates by considering the inter and skip modes only. These two modes rely on motion-compensated prediction. 
In this section we examine the motion-compensated prediction error, also denoted as MC-prediction residual. In a previous work \cite{RefWorks:68} we study and modeled the motion-compensation procedure. Here we use results from \cite{RefWorks:68} and adapt them to the considered compression-scaling system.

\subsection{Autocorrelation of MC-Prediction Residual}
We assume the MC-prediction error is a wide-sense stationary process; hence, modeling its second-order statistics is sufficient.
In \cite{RefWorks:68} we presented a model for the motion-compensation procedure, and an autocorrelation function for its prediction residual. The autocorrelation has the simple form of a separable first-order Markov model:
\begin{IEEEeqnarray}{rCl}
\label{eq:motion compensated prediction residual autocorrelation - separable Markov model - correlation coefficients}
&& {R_{{f_r}}}\left( {{\tau _x},{\tau _y}} \right) = \sigma _{{f_r}}^2\cdot{\rho _{f_v , x}^{ \left| {{\tau _x}} \right|}}{\rho _{f_v , y}^{ \left| {{\tau _y}} \right|}} .
\end{IEEEeqnarray}
We define 
\begin{IEEEeqnarray}{rCl}
\label{eq:motion compensated prediction residual autocorrelation - separable Markov model - correlation parameters definition}
&& {\alpha _x} = -W \cdot \log \left( \rho _{f_v , x} \right)
\\ \nonumber
&& {\alpha _y} = -H \cdot \log \left( \rho _{f_v , y} \right)
\end{IEEEeqnarray}
where $W$ and $H$ are the frame width and height, respectively, in pixels. Then, alternative form of (\ref{eq:motion compensated prediction residual autocorrelation - separable Markov model - correlation coefficients}) is available by using (\ref{eq:motion compensated prediction residual autocorrelation - separable Markov model - correlation parameters definition}):
\begin{IEEEeqnarray}{rCl}
\label{eq:motion compensated prediction residual autocorrelation - separable Markov model - exponential}
&& {R_{{f_r}}}\left( {{\tau _x},{\tau _y}} \right) = \sigma _{{f_r}}^2\cdot{e^{ - {\alpha _x}\left| {{\tau _x}} \right|}}{e^{ - {\alpha _y}\left| {{\tau _y}} \right|}} .
\end{IEEEeqnarray}

The following expressions for the variance was given in \cite{RefWorks:68}:
\begin{IEEEeqnarray}{rCl}
\label{eq:inter coding - R00}
&& \sigma _{{f_r}}^2 = 2\left( {\sigma _{\Delta x}^2 + \sigma _{\Delta y}^2} \right)\cdot\left[ {\sigma _v^2\cdot\left( {1 - {\rho _v}} \right) + {\frac{L}{{{F_{rate}}}}\tilde \sigma _q^2 + \sigma _{w,ref}^2}} \right] + 2\tilde \sigma _q^2{d_t} + \sigma _{w,current}^2 + \sigma _{w,ref}^2
\end{IEEEeqnarray}
where $\sigma _{\Delta x}^2$ and $\sigma _{\Delta y}^2$ reflect accuracy errors in ME, $\sigma _v^2$ and $\rho _v$ are the variance and correlation coefficient of the frame pixels, $L$ is a temporal memory factor, $\sigma _q^2$ is the motion energy in the video, $F_{rate}$ is the frame rate, $\sigma _{w,current}^2$ and $\sigma _{w,ref}^2$ are temporally-local noise energy of the coded frame and the reference frame, respectively. Temporally-local noise reflects distortions due to compression or spatial processing.

We neglect here temporally-local noises other than compression and spatial down-scaling. The noise energy values of compression and spatial down-scaling are denoted as $\sigma _{compression} ^2$ and $\sigma _{spatial-scaling} ^2$, respectively. We assume the compression and scaling noise processes are independent. The coded frame is affected only by the spatial down-scaling, hence
\begin{IEEEeqnarray}{rCl}
\label{eq:current coded frame temporall-local noise}
\sigma _{w,current}^2 = \sigma _{spatial-scaling} ^2 .
\end{IEEEeqnarray}
In contrast, the used reference frame is reconstructed from compression. Therefore, it is affected by both compression and scaling noises; their independence yields
\begin{IEEEeqnarray}{rCl}
\label{eq:coding reference frame temporall-local noise}
\sigma _{w,ref}^2 = \sigma _{spatial-scaling} ^2 + \sigma _{compression} ^2.
\end{IEEEeqnarray}

Expression (\ref{eq:inter coding - R00}) was derived in \cite{RefWorks:68} for a spatially discrete signal. Here we model the signal to be spatially continuous. Therefore, the discrete derivative approximation from \cite{RefWorks:68}:
\begin{IEEEeqnarray}{rCl}
\label{eq:Discrete derivation approximation}
{\left. {\frac{\partial }{{\partial \tilde x}}f\left( {\tilde x,\tilde y} \right)} \right|_{\left( {\tilde x,\tilde y} \right) = \left( {x,y} \right)}} \approx f\left( {x + 1,y} \right) - f\left( {x,y} \right),
\end{IEEEeqnarray}
where the pixel width is defined to be 1, is replaced with the continuous derivative approximation:
\begin{IEEEeqnarray}{rCl}
\label{eq:Continuous derivation approximation}
{\left. {\frac{\partial }{{\partial \tilde x}}f\left( {\tilde x,\tilde y} \right)} \right|_{\left( {\tilde x,\tilde y} \right) = \left( {x,y} \right)}} = \frac{{f\left( {x + \varepsilon ,y} \right) - f\left( {x,y} \right)}}{\varepsilon }
\end{IEEEeqnarray}
where $\varepsilon$ is the pixel width, in the derivation direction, defined on the continuous unit-square according to the original frame size. I.e., for original frame size $H_0 \times W_0$, the horizontal and vertical pixel widths are $\varepsilon_x={1 \mathord{\left/{\vphantom {1 {{W_0}}}} \right.\kern-\nulldelimiterspace} {{W_0}}}$ and $\varepsilon_y={1 \mathord{\left/{\vphantom {1 {{H_0}}}} \right.\kern-\nulldelimiterspace} {{H_0}}}$, respectively.
This yields a continuous form for (\ref{eq:inter coding - R00}):
\begin{IEEEeqnarray}{rCl}
\label{eq:motion-compensated coding - Markov model - variance}
\sigma _{{f_r}}^2 = 2\left( {\frac{\sigma _{\Delta x}^2}{\varepsilon ^2_x} + \frac{\sigma _{\Delta y}^2}{\varepsilon ^2_y}} \right)\cdot\left[ {\sigma _v^2\cdot\left( {1 - {\rho _v}} \right) + {\frac{L}{{{F_{rate}}}}\tilde \sigma _q^2 + \sigma _{w,ref}^2}} \right] + 2\tilde \sigma _q^2{d_t} + \sigma _{w,current}^2 + \sigma _{w,ref}^2
\end{IEEEeqnarray}

The following expression for correlation coefficients were also given in \cite{RefWorks:68}:
\begin{IEEEeqnarray}{rCl}
\label{eq:motion-compensated coding - Markov model - correlation coefficient definition}
{\rho _{{f_v},x}} \cdot \sigma _{{f_r}}^2 = 2\left[ {\sigma _{\Delta x}^2 + \sigma _{\Delta y}^2} \right]\cdot\sigma _v^2\cdot{\rho _v} - \sigma _{\Delta x}^2\cdot\left[ {\sigma _v^2 \cdot \left( {1 + \rho _v^2} \right) + \frac{L}{{{F_{rate}}}}\tilde \sigma _q^2 + \sigma _{w,ref}^2} \right] - 2\sigma _{\Delta y}^2\sigma _v^2\rho _v^2 .
\end{IEEEeqnarray}
Equivalent expression for ${\rho _{{f_v},y}}$ is available by replacing $x$ and $y$ in (\ref{eq:motion-compensated coding - Markov model - correlation coefficient definition}).

\subsection{Temporal Down-scaling Effect}
Our compression-scaling system examines preceding the compression with temporal down-scaling, i.e., frame-rate reduction. Lower frame-rate implies increased temporal-distance between frames; hence, affecting the motion estimation and compensation procedures.
The autocorrelation of MC-prediction residual (\ref{eq:motion compensated prediction residual autocorrelation - separable Markov model - exponential}) expresses the quality reduction of ME and MC as the frame-rate gets lower. The variance (\ref{eq:motion-compensated coding - Markov model - variance}), which is also the prediction-error energy, increases as the frame-rate decreases. The behavior of the model as function of frame-rate is further discussed in \cite{RefWorks:68}.

\subsection{Spatial Down-scaling Effect}
Previous work treated spatial down-scaling before compression of image \cite{RefWorks:2} and video (\cite{RefWorks:64, RefWorks:65}) signals. Unlike \cite{RefWorks:2, RefWorks:64}, we consider here a predictive coding system; hence, the original signal, which is down-scaled, is not the transform-coded one. In our analysis, the transform-coded signal is the MC-prediction residual that is modeled according its second-order statistics (\ref{eq:motion compensated prediction residual autocorrelation - separable Markov model - exponential}); whereas in (\ref{eq:motion-compensated coding - Markov model - variance})-(\ref{eq:motion-compensated coding - Markov model - correlation coefficient definition}), we express the effect of spatial down-scaling of the original video as a part of the temporally-local noise signals of the coded and reference frames, as in (\ref{eq:current coded frame temporall-local noise}) and (\ref{eq:coding reference frame temporall-local noise}), respectively. Here we analyze the error introduced by the spatial scaling and calculate $\sigma _{spatial-scaling} ^2$ for a given spatial statistics of a video frame and a down-scaling factor.

Recall that we model a video frame, $f_v \left( x, y \right)$, as a continuous realization of a 2D WSS random process with zero-mean and autocorrelation $R_v \left( \tau _x, \tau _y \right)$ defined in (\ref{eq:spatial autocorrelation of a frame  - Markov model}). The power-spectral-density (PSD) function of a frame is calculated as follows.
\begin{IEEEeqnarray}{rCl}
\label{eq:PSD of a frame - continuous}
{S_{vv}}\left( {{\omega _x},{\omega _y}} \right) & = & {F_{2D}}\left\{ {{R_v}\left( {{\tau _x},{\tau _y}} \right)} \right\}\left( {{\omega _x},{\omega _y}} \right)
\\ \nonumber
& = & \sigma _v^2 \cdot {F_{1D}}\left\{ {{e^{ - {\alpha _x}\left| {{\tau _x}} \right|}}} \right\}\left( {{\omega _x}} \right) \cdot {F_{1D}}\left\{ {{e^{ - {\alpha _y}\left| {{\tau _y}} \right|}}} \right\}\left( {{\omega _y}} \right) 
\\ \nonumber
& = & \sigma _v^2 \cdot \frac{{2{\alpha _x}}}{{\alpha _x^2 + \omega _x^2}} \cdot \frac{{2{\alpha _y}}}{{\alpha _y^2 + \omega _y^2}} 
\\ \nonumber
& = & \frac{{4\sigma _v^2{\alpha _x}{\alpha _y}}}{{\left( {\alpha _x^2 + \omega _x^2} \right)\left( {\alpha _y^2 + \omega _y^2} \right)}} .
\end{IEEEeqnarray}
The last result shows that $S_{vv}$ is not band-limited; hence, according to the sampling theorem, sampling the continuous signal $f_v \left( x, y \right)$ will introduce an error. 

As part of our multi-resolution model, the video signal is considered as a continuous function. However, the studied compression-scaling system processes the video as a discrete signal with two respective spatial resolutions. First, the original frame size, denoted as $W_0 \times H_0$. Second is the spatially-downsampled frame size, denoted as $W_d \times H_d$ and defined as
\begin{IEEEeqnarray}{rCl}
\label{eq:down-scaled frame size definition}
{W_d} = \frac{{{W_0}}}{{{D_M}}}\,\,\,\,\,\,,\,\,\,\,\,\,{H_d} = \frac{{{H_0}}}{{{D_N}}}
\end{IEEEeqnarray}
Where $D_M$ and $D_N$ are the spatial down-scaling factors defined in (\ref{eq:down-scaling factors definitions}).
The original frame is sampled from the continuous signal $f_v \left( x, y \right)$ in horizontal and vertical sampling intervals of $T ^{0} _{s,x} = \frac{1}{W_0}$ and $T ^{0} _{s,y} = \frac{1}{H_0}$, respectively. The down-scaled frame can be viewed as the output of a downsampling procedure where the continuous frame is sampled in larger intervals $T ^{d} _{s,x} = \frac{1}{W_d}$ and $T ^{d} _{s,y} = \frac{1}{H_d}$. Since $D_M, D_N \geqslant 1$ then $T ^{d} _{s,x} \geqslant T ^{0} _{s,x}$ and $T ^{d} _{s,y} \geqslant T ^{0} _{s,y}$. The sampling frequencies are defined as
\begin{IEEEeqnarray}{rCl}
\label{eq:sampling frequency definition}
\omega _{s,x} = \frac{2\pi}{T _{s,x}}\,\,\,\,\,\,,\,\,\,\,\,\,\omega _{s,y} = \frac{2\pi}{T _{s,y}} .
\end{IEEEeqnarray}
Hence, $\omega ^{0} _{s,x} \geqslant \omega ^{d} _{s,x}$ and $\omega ^{0} _{s,y} \geqslant \omega ^{d} _{s,y}$.

\begin{figure*}[!t]
\centering
\subfloat[]{\includegraphics[width=3in]{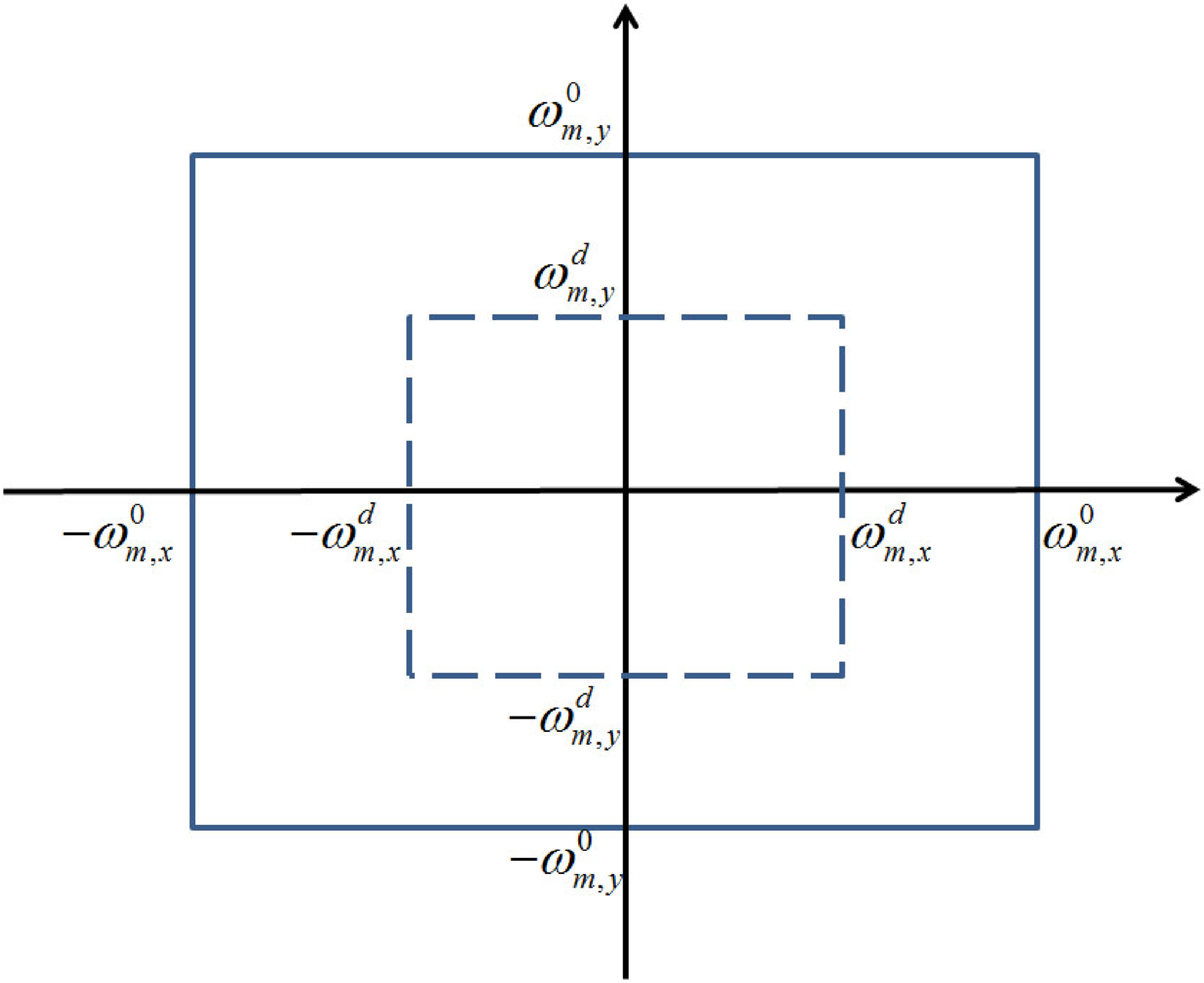}
\label{Fig:PSD_support_of_sampled_frame}}
\subfloat[]{\includegraphics[width=3in]{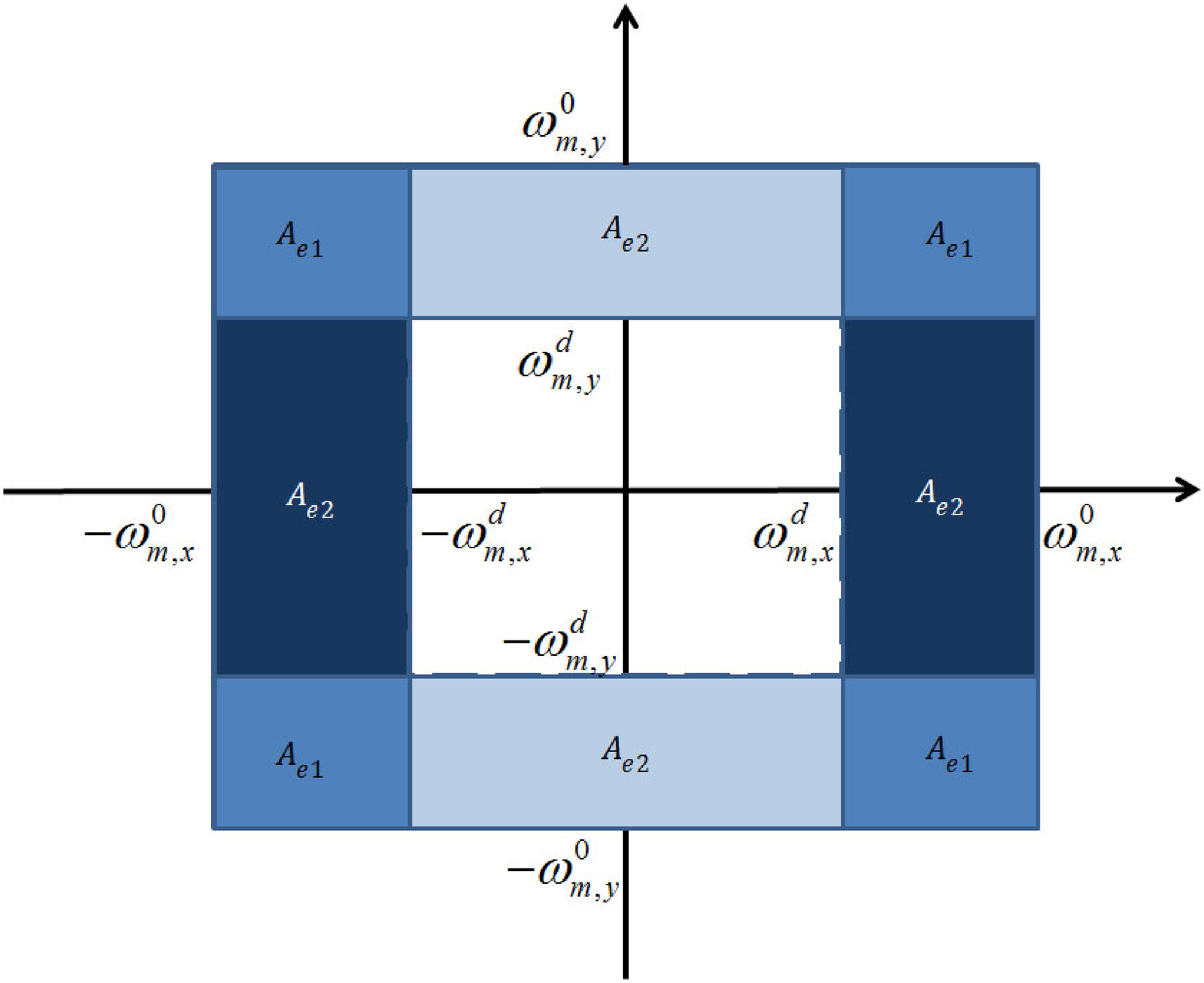}
\label{Fig:PSD_support_of_spatial_downsampling_error}}
\caption{(a) Support of a frame's PSD function after spatial sampling to original frame size (solid) and down-scaled (dashed) frame sizes. (b) The support of the down-scaling error and its decomposition into calculation areas.}
\label{Fig:spatial downsampling - PSD support demonstration}
\end{figure*}

According to the sampling theorem for random processes \cite{RefWorks:69}, a random signal sampled at $\omega _s$ frequency should have a PSD function that is band-limited to $\left| \omega  \right| \leqslant {\omega _m}$ ,where ${\omega _m} = \frac{{{\omega _s}}}{2}$, in order to avoid aliasing. The support of the PSD of the original frame is
\begin{IEEEeqnarray}{rCl}
\label{eq:PSD support of original frame}
{A_0} & = & \left\{ {\left| {{\omega _x}} \right| \leqslant \omega _{m,x}^0\,\,,\,\,\,\,\left| {{\omega _y}} \right| \leqslant \omega _{m,y}^0\,} \right\} 
\end{IEEEeqnarray}
where $\omega _{m,x}^0 = \pi {W_0}$ and $\omega _{m,y}^0 = \pi {H_0}$; and the corresponding support for the downsampled frame is
\begin{IEEEeqnarray}{rCl}
\label{eq:PSD support of down-scaled frame}
{A_d} & = & \left\{ {\left| {{\omega _x}} \right| \leqslant \omega _{m,x}^d\,\,,\,\,\,\,\left| {{\omega _y}} \right| \leqslant \omega _{m,y}^d\,} \right\} 
\end{IEEEeqnarray}
where $\omega _{m,x}^d = \frac{{\pi {W_0}}}{{{D_M}}}$ and $\omega _{m,y}^d = \frac{{\pi {H_0}}}{{{D_N}}}$.
Hence, $A_d$ is contained in $A_0$ (Fig. \ref{Fig:PSD_support_of_sampled_frame}).
Truncating the PSD function $S_{vv}$ (\ref{eq:PSD of a frame - continuous}), gives the PSD functions of the original and down-scaled frames:
\begin{IEEEeqnarray}{rCl}
\label{eq:PSD of original frame}
S_{vv}^0\left( {{\omega _x},{\omega _y}} \right) = \left\{ {\begin{array}{*{20}{c}}
  {S_{vv}^{}\left( {{\omega _x},{\omega _y}} \right)} \\ 
  0 
\end{array}} \right.\begin{array}{*{20}{c}}
  {\,\,\,\,\,,}&{\left( {{\omega _x},{\omega _y}} \right) \in {A_0}} \\ 
  {\,\,\,\,\,\,,}&{otherwise} 
\end{array}
\end{IEEEeqnarray}
and
\begin{IEEEeqnarray}{rCl}
\label{eq:PSD of down-scaled frame}
S_{vv}^d\left( {{\omega _x},{\omega _y}} \right) = \left\{ {\begin{array}{*{20}{c}}
  {S_{vv}^{}\left( {{\omega _x},{\omega _y}} \right)} \\ 
  0 
\end{array}} \right.\begin{array}{*{20}{c}}
  {\,\,\,\,\,,}&{\left( {{\omega _x},{\omega _y}} \right) \in {A_d}} \\ 
  {\,\,\,\,\,\,,}&{otherwise} 
\end{array}
\end{IEEEeqnarray},
respectively.

\begin{figure}
\centering
\includegraphics[height=2in]{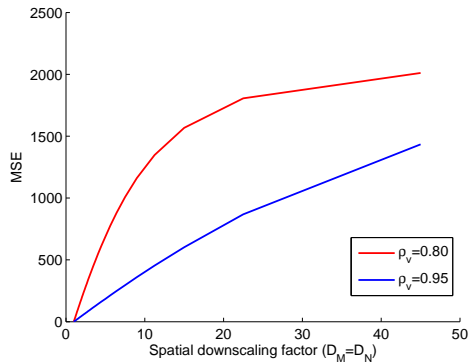}
\caption{Spatial downsampling error as function of down-scaling factor ($D_M=D_N$) and for various pixel correlation values $\rho _v$ ($\rho _v=\rho _{v,x}=\rho _{v,y}$).}
\label{Fig:Spatial_downsampling_MSE}
\end{figure}

While the original frame has a sampling error, we consider here the downsampling error as the additional error due to the smaller frame size. The PSD function of the downsampling error is
\begin{IEEEeqnarray}{rCl}
\label{eq:PSD of downsampling error}
S_{ee}^{spatial}\left( {{\omega _x},{\omega _y}} \right) & = & S_{vv}^0\left( {{\omega _x},{\omega _y}} \right) - S_{vv}^d\left( {{\omega _x},{\omega _y}} \right) 
\\ \nonumber
& = & \left\{ {\begin{array}{*{20}{c}}
  {S_{vv}^{}\left( {{\omega _x},{\omega _y}} \right)} \\ 
  0 
\end{array}} \right.\begin{array}{*{20}{c}}
  {\,\,\,,}&{\left( {{\omega _x},{\omega _y}} \right) \in {A_0}\backslash {A_d}} \\ 
  {\,\,\,,}&{otherwise} 
\end{array} ,
\end{IEEEeqnarray}
where its support is the subtraction of $A_d$ from $A_0$. The mean-squared-error of the downsampling is therefore
\begin{IEEEeqnarray}{rCl}
\label{eq:MSE of spatial down-scaling - definition}
\sigma _{spatial-scaling} ^2 & = & \frac{1}{{4{\pi ^2}}}\int\limits_{{\omega _x}}^{} {} \int\limits_{{\omega _x}}^{} {S_{ee}^{spatial}\left( {{\omega _x},{\omega _y}} \right)d{\omega _x}d{\omega _y}} 
\\ \nonumber
& = & \frac{1}{{4{\pi ^2}}}\iint\limits_{\left( {{\omega _x},{\omega _y}} \right) \in {A_0}\backslash {A_d}} {S_{vv}^{}\left( {{\omega _x},{\omega _y}} \right)d{\omega _x}d{\omega _y}}
\end{IEEEeqnarray}
The area ${A_0}\backslash {A_d}$ is a union of three regions (Fig. \ref{Fig:PSD_support_of_spatial_downsampling_error}), i.e.,
\begin{IEEEeqnarray}{rCl}
\label{eq:MSE of spatial down-scaling - error area as union of 3}
{A_e} \triangleq {A_0}\backslash {A_d} = {A_{e1}} \cup {A_{e2}} \cup {A_{e3}}
\end{IEEEeqnarray}
where
\begin{IEEEeqnarray}{rCl}
\label{eq:MSE of spatial down-scaling - 3 areas}
&& {A_{e1}} = \left\{ {\omega _{m,x}^d \leqslant \left| {{\omega _x}} \right| \leqslant \omega _{m,x}^0\,\,\,\,and\,\,\,\,\omega _{m,y}^d \leqslant \left| {{\omega _y}} \right| \leqslant \omega _{m,y}^0} \right\}
\nonumber \\ 
&& {A_{e2}} = \left\{ {\omega _{m,x}^d \leqslant \left| {{\omega _x}} \right| \leqslant \omega _{m,x}^0\,\,\,\,and\,\,\,\,\left| {{\omega _y}} \right| \leqslant \omega _{m,y}^d} \right\}
\\ \nonumber
&& {A_{e2}} = \left\{ {\left| {{\omega _x}} \right| \leqslant \omega _{m,x}^d\,\,\,\,and\,\,\,\,\omega _{m,y}^d \leqslant \left| {{\omega _y}} \right| \leqslant \omega _{m,y}^0} \right\} .
\end{IEEEeqnarray}
Using the last decomposition, (\ref{eq:MSE of spatial down-scaling - definition}) becomes
\begin{IEEEeqnarray}{rCl}
\label{eq:MSE of spatial down-scaling - area decomposition}
\sigma _{spatial - scaling}^2 & = & \frac{1}{{4{\pi ^2}}}\left[ \iint\limits_{\left( {{\omega _x},{\omega _y}} \right) \in {A_{e1}}} {S_{vv}^{}\left( {{\omega _x},{\omega _y}} \right)d{\omega _x}d{\omega _y}} + \iint\limits_{\left( {{\omega _x},{\omega _y}} \right) \in {A_{e2}}} {S_{vv}^{}\left( {{\omega _x},{\omega _y}} \right)d{\omega _x}d{\omega _y}} \right.
\\ \nonumber
&& \qquad {} \left. + \iint\limits_{\left( {{\omega _x},{\omega _y}} \right) \in {A_{e3}}} {S_{vv}^{}\left( {{\omega _x},{\omega _y}} \right)d{\omega _x}d{\omega _y}} \right] .
\end{IEEEeqnarray}
Let us define the following integral
\begin{IEEEeqnarray}{rCl}
\label{eq:MSE of spatial down-scaling - auxiliary integral definition}
&& I\left( {{\omega _{x1}},{\omega _{x2}},{\omega _{y1}},{\omega _{y2}}} \right) \triangleq \int\limits_{{\omega _{x1}}}^{{\omega _{x2}}} {} \int\limits_{{\omega _{y1}}}^{{\omega _{y2}}} {\frac{{2{\alpha _x}}}{{\alpha _x^2 + \omega _x^2}} \cdot \frac{{2{\alpha _y}}}{{\alpha _y^2 + \omega _y^2}}d{\omega _x}} d{\omega _y},
\end{IEEEeqnarray}
where its solution is
\begin{IEEEeqnarray}{rCl}
\label{eq:MSE of spatial down-scaling - auxiliary integral solution}
I\left( {{\omega _{x1}},{\omega _{x2}},{\omega _{y1}},{\omega _{y2}}} \right) & = & 4 \cdot \int\limits_{{\omega _{x1}}}^{{\omega _{x2}}} {\frac{{{\alpha _x}}}{{\alpha _x^2 + \omega _x^2}}d{\omega _x}} \int\limits_{{\omega _{y1}}}^{{\omega _{y2}}} {\frac{{{\alpha _y}}}{{\alpha _y^2 + \omega _y^2}}} d{\omega _y} 
\\ \nonumber
& = & 4 \cdot \left[ {\left. {\arctan \left( {\frac{{{\omega _x}}}{{{\alpha _x}}}} \right)} \right|_{{\omega _x} = {\omega _{x1}}}^{{\omega _{x2}}}} \right] \cdot \left[ {\left. {\arctan \left( {\frac{{{\omega _y}}}{{{\alpha _y}}}} \right)} \right|_{{\omega _y} = {\omega _{y1}}}^{{\omega _{y2}}}} \right] 
\\ \nonumber
& = & 4 \cdot \arctan \left( {\frac{{\frac{1}{{{\alpha _x}}} \cdot \left( {{\omega _{x2}} - {\omega _{x1}}} \right)}}{{1 + \frac{{{\omega _{x1}}{\omega _{x2}}}}{{\alpha _x^2}}}}} \right) \cdot \arctan \left( {\frac{{\frac{1}{{{\alpha _y}}} \cdot \left( {{\omega _{y2}} - {\omega _{y1}}} \right)}}{{1 + \frac{{{\omega _{y1}}{\omega _{y2}}}}{{\alpha _y^2}}}}} \right) .
\end{IEEEeqnarray}
We express (\ref{eq:MSE of spatial down-scaling - area decomposition})'s components as follows:
\begin{IEEEeqnarray}{rCl}
\label{eq:MSE of spatial down-scaling - expressions using aux integral}
\nonumber && \iint\limits_{\left( {{\omega _x},{\omega _y}} \right) \in {A_{e1}}} {S_{vv}^{}\left( {{\omega _x},{\omega _y}} \right)d{\omega _x}d{\omega _y}} = 4 \cdot I\left( {\omega _{m,x}^d,\omega _{m,x}^0,\omega _{m,y}^d,\omega _{m,y}^0} \right)
\\ \nonumber
&& \iint\limits_{\left( {{\omega _x},{\omega _y}} \right) \in {A_{e2}}} {S_{vv}^{}\left( {{\omega _x},{\omega _y}} \right)d{\omega _x}d{\omega _y}} = 4 \cdot I\left( {\omega _{m,x}^d,\omega _{m,x}^0,0,\omega _{m,y}^d} \right)
\\ \nonumber
&& \iint\limits_{\left( {{\omega _x},{\omega _y}} \right) \in {A_{e3}}} {S_{vv}^{}\left( {{\omega _x},{\omega _y}} \right)d{\omega _x}d{\omega _y}} = 4 \cdot I\left( {0,\omega _{m,x}^d,\omega _{m,y}^d,\omega _{m,y}^0} \right) .
\\
\end{IEEEeqnarray}
Substituting (\ref{eq:MSE of spatial down-scaling - expressions using aux integral}) in (\ref{eq:MSE of spatial down-scaling - area decomposition}) yields the final expression for spatial down-scaling MSE:
\begin{IEEEeqnarray}{rCl}
\label{eq:MSE of spatial down-scaling - final expression}
\sigma _{spatial - scaling}^2 & = & \frac{{\sigma _v^2}}{{{\pi ^2}}}\left[  I\left( {\omega _{m,x}^d,\omega _{m,x}^0,\omega _{m,y}^d,\omega _{m,y}^0} \right) +  I\left( {\omega _{m,x}^d,\omega _{m,x}^0,0,\omega _{m,y}^d} \right) +  I\left( {0,\omega _{m,x}^d,\omega _{m,y}^d,\omega _{m,y}^0} \right) \right]
\end{IEEEeqnarray}
where $\omega _{m,x}^0$, $\omega _{m,y}^0$, $\omega _{m,x}^d$ and $\omega _{m,y}^d$ values are given in (\ref{eq:PSD support of original frame}) and (\ref{eq:PSD support of down-scaled frame}).
The MSE (\ref{eq:MSE of spatial down-scaling - final expression}) increases together with the spatial down-scaling factor (Fig. \ref{Fig:Spatial_downsampling_MSE}). Moreover, as the frame pixels are more correlated, i.e. higher $\rho _v$, then the MSE increases more slowly due to the decreasing energy of high frequency components.

\subsection{Compression Effect}
Expressions (\ref{eq:motion-compensated coding - Markov model - variance})-(\ref{eq:motion-compensated coding - Markov model - correlation coefficient definition}) model the second-order statistics of the MC-prediction residual.
Motion-compensated coding avoids encoder-decoder mismatch by using reference frames that were reconstructed from compression. Therefore, the quality of the compression affects the statistics of the MC-prediction residual. Specifically, the bit-rate is influencing. Consequently, the temporally-local noise signal of the reference frame (\ref{eq:coding reference frame temporall-local noise}) contains the compression noise independently together with the spatial down-scaling noise, with respectively $\sigma _{compression} ^2$ and $\sigma _{spatial-scaling} ^2$ noise energies. 

In our previous work we studied the compression effect on MC-prediction residual statistics and offered two alternatives for estimating its variance for the application of coding. Firstly, an empirical rate-distortion curve was suggested:
\begin{IEEEeqnarray}{rCl}
\label{eq:Compression effect - w-compression variance - as rate distortion function}
\sigma ^2 _{w, compression} = \beta \cdot r ^ {-\alpha} 
\end{IEEEeqnarray}
where $\alpha$ and $\beta$ are curve parameters, and $r$ is the bit-rate.
Secondly, a theoretic rate-distortion estimation for memoryless Gaussian source was considered:
\begin{IEEEeqnarray}{rCl}
\label{eq:Compression effect - w-compression variance - theoretical estimation}
\sigma ^2 _{w, compression} = \sigma _v^2\cdot{2^{ - 2r}}
\end{IEEEeqnarray}
where $\sigma _v^2$ is the variance of the Gaussian source, and $r$ is the bit-rate.

\subsection{MC-Prediction Error in Inter-Coding and Skip Modes}
As we explained earlier, we consider two MC-based coding modes: inter-coding and skip modes. In inter-coding the prediction is fairly done and the prediction error is transform coded for the reconstruction. Hence, we consider statistical model defined in this section to represent the MC-prediction residual in inter-coding; i.e.,
\begin{IEEEeqnarray}{rCl}
\label{eq:MC-coding residual autocorrelation}
R_{{f_r}}^{inter}\left( {{\tau _x},{\tau _y}} \right) \equiv {R_{{f_r}}}\left( {{\tau _x},{\tau _y}} \right)
\end{IEEEeqnarray}
where $R_{{f_r}}$ was defined in (\ref{eq:motion compensated prediction residual autocorrelation - separable Markov model - correlation coefficients}). Calculating the reconstruction error for inter-coding mode requires analysis of predictive and transform coding, therefore it will be given after studying these methods in the following sections.

In contrast, skip mode is a low bit-cost mode. First, it offers inferior MC-prediction by constructing it from its spatial neighbors; then, the prediction error is not transmitted to the decoder. We assume the former property alone leads us to a proportional-increment in prediction-error energy relative to inter-coding; i.e., 
\begin{IEEEeqnarray}{rCl}
\label{eq:skip mode MC-prediction residual variance - definition}
R_{{f_r}}^{skip}\left( 0, 0 \right) \triangleq  \gamma \cdot {R_{{f_r}}}\left( 0, 0 \right) = \gamma \cdot \sigma _{{f_r}}^2
\end{IEEEeqnarray}
where $\gamma \geqslant 1$, and $\sigma _{{f_r}}^2$ was defined in (\ref{eq:motion-compensated coding - Markov model - variance}).
Let us consider a block ${\Delta _{ij}}$ that is encoded in skip mode. The original block information and its reconstruction are denoted as $f_v$ and $\hat f_v^{skip}$, respectively. 
The reconstruction MSE is calculated as follows.
\begin{IEEEeqnarray}{rCl}
\label{eq:skip mode - MSE}
E\left\{ {MSE_{{f_v}}\left( {{\Delta _{ij}}} \right)} \right\} & = & \frac{1}{{A\left( {{\Delta _{ij}}} \right)}}\int  \int_{{\Delta _{ij}}} {E\left\{ {{{\left( {{f_v}\left( {x,y} \right) - \hat f_v^{skip}\left( {x,y} \right)} \right)}^2}} \right\}dxdy} 
\\ \nonumber
& = & \frac{1}{{A\left( {{\Delta _{ij}}} \right)}} A\left( {{\Delta _{ij}}} \right) \cdot R_{{f_r}}^{skip}\left( 0, 0 \right) = \gamma \cdot \sigma _{{f_r}}^2
\end{IEEEeqnarray}
The use of motion-compensation in skip mode yields a dependency of the this mode's error on frame-rate and bit-rate as expressed in (\ref{eq:motion-compensated coding - Markov model - variance}).

\section{Predictive Coding Analysis}
\label{sec:Predictive Coding Analysis}
\subsection{Basic Error Expression}
In this section we analyze the compression error for a general predictive coding method. The specific case of inter coding is addressed in the next section. Let us consider a 2D-slice, $\Delta _{ij}^h$, in the video. This block is encoded by a block-based predictive-coding technique.
The prediction results in the encoder and decoder are identical, and the prediction error is lossy-coded. Hence, the overall error is the coding error of the prediction-residual.
The prediction-residual signal is represented by the function $f_r$. The residual of the $\Delta _{ij}^h$ slice is ${f_r}:\left[ {\frac{{i - 1}}{M},\frac{i}{M}} \right] \times \left[ {\frac{{j - 1}}{N},\frac{j}{N}} \right] \to \mathbb{R}$.

The prediction-residual, $f_r$, depends on the prediction method (e.g., intra, inter, etc.). In this section, we consider a general $f_r$ signal, and describe it by properties that are assumed to hold for any relevant prediction method. We model $f_r$ as a wide-sense stationary process with zero mean and autocorrelation function ${R_{{f_r}}}\left( {{\tau _x},{\tau _y}} \right)$. The reconstructed residual is denoted as $\hat f_r$. According to (\ref{eq:MSE of 2D signal}), the reconstruction MSE of $f_r$ is
\begin{IEEEeqnarray}{rCl}
\label{eq:reconstruction MSE of prediction residual}
\nonumber && MSE_{f_r}\left( {\Delta _{ij}^h} \right) = M N \iint\limits_{\Delta _{ij}^h} {{{\left( {{f_r}\left( {x,y} \right) - \hat f_r \left( {x,y} \right)} \right)}^2}dxdy}
\\
\end{IEEEeqnarray}

\subsection{Transformation of Prediction-Residual}
The residual block $f_r$ is represented using an orthonormal basis of functions. 
Previous video encoders have utilized transform on the entire macroblock, e.g., MPEG2 used 8x8 transform while its macroblocks were 8x8 pixels. However, H.264 and its extensions support transformation of sub-blocks of 4x4 or 8x8 pixels \cite{RefWorks:3, RefWorks:67}.

\begin{figure}
\centering
\includegraphics[height=1.5in]{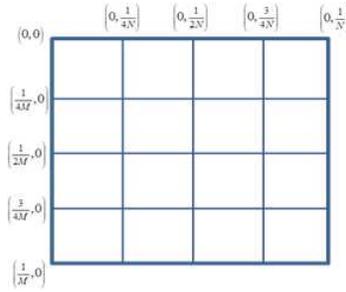}
\caption{Inner partition of a slice into transformed sub-slices ($\beta = 4$).}
\label{Fig:transform_subslices}
\end{figure}

Our assumption that all coded-blocks at low bit-rates are of 16x16 pixels size means that all block-predictions are made on 16x16 pixels blocks. Hence, for a given transform block size (e.g., 4x4) the ratio between the dimensions of the prediction block and the transform block are fixed. The prediction-transform dimension ratio is denoted as $\beta$:
\begin{IEEEeqnarray}{rCl}
\label{eq:prediction-transform dimension ratio}
\beta = \frac{DIM_{prediction}}{DIM_{transform}}
\end{IEEEeqnarray}
where $DIM_{prediction}$ and $DIM_{transform}$ are the dimensions of the prediction and transform blocks, respectively. We assume equal width and height, therefore the dimension equals to both of them. 
We further assume that $\beta$ is always an integer. For prediction-blocks of 16x16 pixels, $\beta$ values are 4 and 2 for 4x4 and 8x8 transforms, respectively. Consequently, if our slice to code is defined on $\left[ {0,\frac{1}{M}} \right] \times \left[ {0,\frac{1}{N}} \right]$ then the transformation will be applied separately on $\beta^2$ equal-sized square sub-slices of this slice (Fig. \ref{Fig:transform_subslices}); i.e., on
\begin{IEEEeqnarray}{rCl}
\label{eq:transformed sub-slices}
&& \left[ {\left( {p - 1} \right) \cdot \frac{1}{{\beta M}},p \cdot \frac{1}{{\beta M}}} \right] \times \left[ {\left( {q - 1} \right) \cdot \frac{1}{{\beta N}},q \cdot \frac{1}{{\beta N}}} \right]
\\ \nonumber
&& for \,\,\,p = 1,...,\beta \,\, {\rm{ ; }} \,\, q = 1,...,\beta
\end{IEEEeqnarray}

Let us denote the $\left( {p,q} \right)$ sub-slice of the $\left( {i,j} \right)$ slice as ${\Delta _{ij,pq}}$, where $i \in \left\{ {1,...,M} \right\}{\text{ , }}j \in \left\{ {1,...,N} \right\}{\text{ , }}p,q \in \left\{ {1,...,\beta} \right\}$.
The residual signal defined on the region of the sub-slice ${\Delta _{ij,pq}}$ is denoted as the function ${f_{r,{\Delta _{ij,pq}}}}\left( {x,y} \right)$. The basis defined over a sub-slice is denoted as $\left\{ {{\Phi _{kl}}\left( {x,y} \right){\text{ }},{\text{ }}k,l = 0,1,2,....} \right\}$ and due to its orthonormality, the following equation holds: 
\begin{IEEEeqnarray}{rCl}
\label{eq:transform basis orthonormality}
\iint\limits_{{\Delta _{ij,pq}}} {{\Phi _{kl}}{\Phi _{k'l'}}dxdy = {\delta _{k - k'}}{\delta _{l - l'}}} = \left\{ {\begin{array}{*{20}{c}}
  {1,{\text{if}}\left( {k,l} \right) = \left( {k',l'} \right)} \\ 
  {0,{\text{otherwise}}} 
\end{array}} \right.
\end{IEEEeqnarray}
The sub-slice of the residual signal can be represented using the basis as:
\begin{IEEEeqnarray}{rCl}
\label{eq:sub-slice representation using transform basis}
{f_{r,{\Delta _{ij,pq}}}}\left( {x,y} \right) = \sum\limits_{k = 0}^\infty  {\sum\limits_{l = 0}^\infty  {{{\left\langle {{f_r}\left( {x,y} \right),{\Phi _{kl}}\left( {x,y} \right)} \right\rangle }_{{\Delta _{ij,pq}}}} \cdot {\Phi _{kl}}\left( {x,y} \right)} }
\end{IEEEeqnarray}
The coefficients $\left\{ {{F_{kl}}{\text{ , }}k,l = 0,1,2,...} \right\}$ are calculated by 
\begin{IEEEeqnarray}{rCl}
\label{eq:transform coefficient calculation}
{F_{kl}} & \equiv & {\left\langle {{f_r}\left( {x,y} \right),{\Phi _{kl}}\left( {x,y} \right)} \right\rangle _{{\Delta _{ij,pq}}}} = \iint_{{\Delta _{ij,pq}}} {{f_r}\left( {x,y} \right){\Phi _{kl}}\left( {x,y} \right)dxdy}
\end{IEEEeqnarray}
We assume $E\left[ {{f_r}\left( {x,y} \right)} \right] = 0$; hence,
\begin{IEEEeqnarray}{rCl}
\label{eq:Mean of coefficient is zero}
E\left[ {{F_{kl}}} \right] & = & \iint_{{\Delta _{ij,pq}}} {E\left[ {{f_r}\left( {x,y} \right)} \right]{\Phi _{kl}}\left( {x,y} \right)dxdy} = 0
\end{IEEEeqnarray}
Therefore, $\operatorname{var} \left\{ {{F_{kl}}} \right\} = E\left[ {F_{kl}^2} \right]$.

In the compression process the residual block function ${f_r}\left( {x,y} \right)$ is approximated over ${\Delta _{ij,pq}}$ by using only a finite set $\Omega$ of the orthonormal basis functions $\left\{ {{\Phi _{kl}}\left( {x,y} \right)} \right\}$, i.e.,
\begin{IEEEeqnarray}{rCl}
\label{eq:residual approximation using finite set of basis functions}
{\hat f_{r,{\Delta _{ij,pq}}}}\left( {x,y} \right) = \sum\limits_{}^{} {\sum\limits_{\left( {k,l} \right) \in \Omega }^{} {{F_{kl}} \cdot {\Phi _{kl}}\left( {x,y} \right)} }
\end{IEEEeqnarray}
The optimal coefficients in the approximation are the ${F_{kl}}$ coefficients that were defined in (\ref{eq:transform coefficient calculation}), hence
\begin{IEEEeqnarray}{rCl}
\label{eq:residual approximation using finite set of basis functions - explicit}
{\hat f_{r,{\Delta _{ij,pq}}}}\left( {x,y} \right) = \sum\limits_{}^{} {\sum\limits_{\left( {k,l} \right) \in \Omega }^{} {{{\left\langle {{f_r}\left( {x,y} \right),{\Phi _{kl}}\left( {x,y} \right)} \right\rangle }_{{\Delta _{ij,pq}}}} \cdot {\Phi _{kl}}\left( {x,y} \right)} }
\end{IEEEeqnarray}

In appendix \ref{appendix:Calculation of The Expected MSE of a slice} we calculate the following expected MSE of a slice:
\begin{IEEEeqnarray}{rCl}
\label{eq:MSE calculation of residual block}
E\left[ {MSE_{{f_r}}\left( {{\Delta _{11}}} \right)} \right] = {R_{{\Delta _{ij}}}}\left( {0,0} \right) - {\beta ^2} M N \cdot \sum\limits_{}^{} {\sum\limits_{\left( {k,l} \right) \in \Omega }^{} {E\left[ {F_{kl}^2} \right]} } ,
\end{IEEEeqnarray}
where we used (\ref{eq:residual approximation using finite set of basis functions - explicit}), and the WSS property of the signal over the sub-slices.

\subsection{Quantization}
The prediction residual approximation in (\ref{eq:residual approximation using finite set of basis functions - explicit}) includes the coefficients $\left\{ {{F_{kl}}} \right\}$. While these coefficients take values in $\mathbb{R}$, they are represented using a finite number of bits. Hence, quantization is applied on the coefficients. We denote the quantized coefficients as ${\left\{ {F_{kl}^Q} \right\}_{\left( {k,l} \right) \in \Omega }}$, and $b_{kl}$ denotes the number of bits dedicated for representing the $\left( {k,l} \right)$ coefficient. The representation of $f_r$ over the sub-slice $\Delta _{ij,pq}$ using the basis functions and quantized coefficients is
\begin{IEEEeqnarray}{rCl}
\label{eq:residual approximation using quantized coefficients}
\hat f_{r,{\Delta _{ij,pq}}}^Q\left( {x,y} \right) = \sum{\sum\limits_{\left( {k,l} \right) \in \Omega } {F_{kl}^Q \cdot {\Phi _{kl}}\left( {x,y} \right)} }
\end{IEEEeqnarray}

The quantization error of the $\left( {k,l} \right)$ coefficient is given by
\begin{IEEEeqnarray}{rCl}
\label{eq:coefficient quantization error}
\Gamma _{kl}^2 = {\left( {{F_{kl}} - F_{kl}^Q} \right)^2} .
\end{IEEEeqnarray}
In appendix \ref{appendix:Calculation of The Expected MSE of a slice with Quantization}, we calculate the expected MSE of a residual block with quantized coefficients, and get
\begin{IEEEeqnarray}{rCl}
\label{eq:expected MSE of the quantized representation over a sub-slice}
E\left[ {MSE_{{f_r}}^Q\left( {{\Delta _{11}}} \right)} \right] = {R_{{\Delta _{ij}}}}\left( {0,0} \right) - {\beta ^2}MN \cdot \sum {\sum\limits_{\left( {k,l} \right) \in \Omega } {\left( {E\left[ {F_{kl}^2} \right] - E\left[ {{{\left( {F_{kl} - F_{kl}^Q} \right)}^2}} \right]} \right)} } .
\end{IEEEeqnarray}

As in \cite{RefWorks:2}, we assume Gaussian distribution for the coefficients and model the quantization MSE as
\begin{IEEEeqnarray}{rCl}
\label{eq:quantization MSE model for Gaussian distribution}
E\left[ {{{\left( {{F_{kl}} - F_{kl}^Q} \right)}^2}} \right] \sim K \cdot \frac{{\operatorname{var} \left\{ {{F_{kl}}} \right\}}}{{{2^{2{b_{kl}}}}}}
\end{IEEEeqnarray}
Where $K \in \left[ {1,3} \right]$, and $b_{kl}$ is the number of bits for representing $F_{kl}$.

Inserting (\ref{eq:quantization MSE model for Gaussian distribution}) into the MSE expression (\ref{eq:expected MSE of the quantized representation over a sub-slice}) yields
\begin{IEEEeqnarray}{rCl}
\label{eq:predictive coding MSE}
E\left[ {MSE_{{f_r}}^Q\left( {{\Delta _{11}}} \right)} \right] = {R_{{\Delta _{ij}}}}\left( {0,0} \right) - {\beta ^2}MN\cdot\sum {\sum\limits_{\left( {k,l} \right) \in \Omega } {\operatorname{var} \left\{ {{F_{kl}}} \right\} \cdot \left( {1 - \frac{K}{{{2^{2{b_{kl}}}}}}} \right)} } .
\end{IEEEeqnarray}

\subsection{Separable Cosine Bases}
H.264 applies an integer transform that approximates DCT. Accordingly,
we adapt the transform model from \cite{RefWorks:2} to our scenario and choose the separable cosine basis for transforming the sub-slices in our analysis. For each sub-slice in $\left\{ {{\Delta _{ij,pq}} | p = 1,...,\beta{\text{ , q = 1,}}...{\text{,}\beta}} \right\}$ the following basis functions are defined over the corresponding region:
\begin{IEEEeqnarray}{rCl}
\label{eq:separable cosine bases - basis functions}
&& {\Phi _{kl}}\left( {x,y} \right) = {\varphi _k}\left( x \right){\varphi _l}\left( y \right)
\\ \nonumber
&& where
\\ \nonumber
&& \qquad{} {\varphi _k}\left( x \right) = \sqrt {{\beta}M\left( {2 - {\delta _k}} \right)} \cos \left( {k \cdot {\beta}M\pi x} \right)
\\ \nonumber
&& \qquad{} {\varphi _l}\left( y \right) = \sqrt {{\beta}N\left( {2 - {\delta _l}} \right)} \cos \left( {l \cdot {\beta}N\pi y} \right)
\\ \nonumber
&& for \,\,\, k = 0,...,{\beta}-1 \,;\, l = 0,...,{\beta}-1
\end{IEEEeqnarray}
The $F_{kl}$ coefficient is calculated by
\begin{IEEEeqnarray}{rCl}
\label{eq:coefficient calculation for separable basis}
{F_{kl}} = \iint_{{\Delta _{ij,pq}}} {{f_r}\left( {x,y} \right){\varphi _k}\left( x \right){\varphi _l}\left( y \right)dxdy}
\end{IEEEeqnarray}
The second moment is calculated as follows:
\begin{IEEEeqnarray}{rCl}
\label{eq:separable cosine bases - coefficient second moment}
E\left[ {F_{_{kl}}^2} \right] & = & E\left[ \iint_{{\Delta _{11,11}}} \iint_{{\Delta _{11,11}}} {f_r}\left( {x,y} \right){f_r}\left( {\xi ,\eta } \right)  \cdot {\varphi _k}\left( x \right){\varphi _l}\left( y \right){\varphi _k}\left( \xi  \right){\varphi _l}\left( \eta  \right)dxdyd\xi d\eta  \right]
\\ \nonumber
& = & \iint_{{\Delta _{11,11}}} {\iint_{{\Delta _{11,11}}} {}}{R_{{\Delta _{ij}}}}\left( {x - \xi ,y - \eta } \right) \cdot {\beta}M\left( {2 - {\delta _k}} \right)
\\ \nonumber
&& \times \cos \left( {{\beta}Mk\pi x} \right)\cos \left( {{\beta}Mk\pi \xi } \right) \cdot {\beta}N\left( {2 - {\delta _l}} \right)\cos \left( {{\beta}Nl\pi y} \right)\cos \left( {{\beta}Nl\pi \eta } \right)dxdyd\xi d\eta
\end{IEEEeqnarray}

\subsection{The Case of Inter Coding}
Here we develop (\ref{eq:separable cosine bases - coefficient second moment}) further for inter coding.
Let us substitute the inter prediction autocorrelation (\ref{eq:motion compensated prediction residual autocorrelation - separable Markov model - exponential}) in the expression (\ref{eq:separable cosine bases - coefficient second moment}) of the coefficient second moment:
\begin{IEEEeqnarray}{rCl}
\label{eq:inter coding - }
\nonumber E\left[ {F_{_{kl}}^2} \right] & = & \iint_{{\Delta _{11,11}}} {\iint_{{\Delta _{11,11}}} {}\sigma _{{e_i}}^2\left( {{F_{rate}},B} \right)\cdot{e^{ - {\alpha _x}\left| {x - \xi } \right|}}{e^{ - {\alpha _y}\left| {y - \eta } \right|}}}
\\ \nonumber
&& \times {\beta}M\left( {2 - {\delta _k}} \right)\cos \left( {{\beta}Mk\pi x} \right)\cos \left( {{\beta}Mk\pi \xi } \right)
\\ \nonumber
&& \times {\beta}N\left( {2 - {\delta _l}} \right)\cos \left( {{\beta}Nl\pi y} \right)\cos \left( {{\beta}Nl\pi \eta } \right)dxdyd\xi d\eta 
\\
\end{IEEEeqnarray}
We use separability and change the integration variables to
\begin{IEEEeqnarray}{rCl}
\label{eq:inter coding - integration variables transformation}
&& {\tilde x = {\beta}Mx} \,\,\,,\,\,\, {\tilde \xi  = {\beta}M\xi } \,\,\,,\,\,\, {\tilde y = {\beta}Ny} \,\,\,,\,\,\, {\tilde \eta  = {\beta}N\eta }
\\ \nonumber
&& {\tilde x,\tilde \xi,\tilde y,\tilde \eta \in \left[ {0,1} \right]}
\end{IEEEeqnarray}
and get
\begin{IEEEeqnarray}{rCl}
\label{eq:inter coding - coefficient second moment after integration variables transformation}
E\left[ {F_{_{kl}}^2} \right] & = & \sigma _{{e_i}}^2\left( {{F_{rate}},B} \right) \cdot \left( {2 - {\delta _k}} \right)\left( {2 - {\delta _l}} \right)
\\ \nonumber
&& \times \frac{1}{{\beta M}}\int\limits_{\tilde x = 0}^1 {\int\limits_{\tilde \xi  = 0}^1 {} } {e^{ - \frac{{{\alpha _x}}}{{\beta M}}\left| {\tilde x - \tilde \xi } \right|}}\cos \left( {k\pi \tilde x} \right)\cos \left( {k\pi \tilde \xi } \right)d\tilde xd\tilde \xi
\\ \nonumber
&&  \times \frac{1}{{\beta N}}\int\limits_{\tilde y = 0}^1 {\int\limits_{\tilde \eta  = 0}^1 {} {e^{ - \frac{{{\alpha _y}}}{{\beta N}}\left| {\tilde y - \tilde \eta } \right|}}} \cos \left( {l\pi \tilde y} \right)\cos \left( {l\pi \tilde \eta } \right)\cdot d\tilde yd\tilde \eta .
\end{IEEEeqnarray}
The following integral was defined in \cite{RefWorks:2}:
\begin{IEEEeqnarray}{rCl}
\label{eq:inter coding - integral M definition} 
Y\left( {A;k,l} \right) \triangleq \int\limits_{0}^{1} {\int\limits_{0}^{1} {\exp \left\{ { - A\left| {x - \xi } \right|} \right\} \cdot \cos \left( {k\pi x} \right)\cos \left( {l\pi \xi } \right)} } dxd\xi
\nonumber \\
\end{IEEEeqnarray}
where its solution \cite{RefWorks:2} is 
\begin{IEEEeqnarray}{rCl}
\label{eq:inter coding - integral M solution} 
Y\left( {A;k,l} \right) & = & \left[ {\frac{A}{{{A^2} + {{\left( {l\pi } \right)}^2}}} + \frac{A}{{{A^2} + {{\left( {k\pi } \right)}^2}}}} \right]\frac{1}{2}\left( {1 + {\delta _{k - or - l}}} \right){\delta _{\left| {k - l} \right|}} 
\\ \nonumber
&& - \frac{{{A^2}}}{{\left( {{A^2} + {{\left( {l\pi } \right)}^2}} \right)\left( {{A^2} + {{\left( {k\pi } \right)}^2}} \right)}} \cdot \left( {2 - {e^{ - A}}\left[ {{{\left( { - 1} \right)}^k} + {{\left( { - 1} \right)}^l}} \right]} \right) .
\end{IEEEeqnarray}
Thus, we can write (\ref{eq:inter coding - coefficient second moment after integration variables transformation}) as
\begin{IEEEeqnarray}{rCl}
\label{eq:inter coding - coefficient second moment simplified using M integral}
E\left[ {F_{_{kl}}^2} \right] & = & \sigma _{{e_i}}^2\left( {{F_{rate}},B} \right) \cdot \left( {2 - {\delta _k}} \right)\left( {2 - {\delta _l}} \right) \cdot \frac{1}{{\beta ^ 2 M N}} \cdot Y\left( {\frac{{{\alpha _x}}}{{\beta M}};k,k} \right) \cdot Y\left( {\frac{{{\alpha _y}}}{{\beta N}};l,l} \right) .
\end{IEEEeqnarray}

Note that the model for MC-prediction residual is assumed to include the effects of inaccurate motion-estimation, and quantization error in motion-vector coding. Therefore, motion-vectors are not treated directly in the distortion analysis.

\section{Overall Compression}
\label{sec:Overall Compression}

\subsection{Bit-Allocation}
\label{subsec:Bit-Allocation}
Practical transform-coding systems usually have an a-priori bit-allocation rule for dividing a given bit-budget among the transform coefficients. H.264's baseline profile applies uniform quantization on its 4x4 transform coefficients; whereas in the high profiles a weighted-quantization (i.e., non-uniform) is carried out on 8x8 transform coefficients. In this work we focus on modeling the compression process of the baseline profile; therefore, we will consider a uniform quantization among the coefficients. 
However, we give here a general analysis of a weighted-quantization matrix.

The a-priori bit-allocation among transform coefficients is modeled by relative bit-allocation. This is similar to the image compression model in \cite{RefWorks:2}; however, we present here few adaptations to treat the joint use of inter and skip modes.

Let us define the weighted-quantization matrix as $Q_{weight}$. Its dimensions equal the transform-block dimensions; i.e., for $D_{trans} \times D_{trans}$ transform, $Q_{weight}$ is a $D_{trans} \times D_{trans}$ matrix. The ${Q_{weight}}\left( {k,l} \right)$ is the quantization weight of the $\left( {k,l} \right)$ coefficient, and it should be considered relatively to all the other weights in $Q_{weight}$, i.e., it should be normalized to get its relative part from the joint bit-budget of the macroblock's coefficients. These normalized weights form the normalized weighted-quantization matrix $\tilde Q_{weight}$ as follows:
\begin{IEEEeqnarray}{rCl}
\label{eq:bit-allocation - elements of normalized weighted-quantization matrix}
{\tilde Q_{weight}}\left( {k,l} \right) = \frac{{\frac{1}{{{Q_{weight}}\left( {k,l} \right)}}}}{{\sum\limits_{q = 1}^{D_{trans}} {\sum\limits_{r = 1}^{D_{trans}} {\frac{1}{{{Q_{weight}}\left( {q,r} \right)}}} } }}.
\end{IEEEeqnarray}

Let us derive the amount of bits allocated for transform coefficients of a slice. For compactness of representation, we omit function notation of the coding mode probabilities and write them as $P_{inter}$ and $P_{skip}$. The total number of slices in the video is
\begin{IEEEeqnarray}{rCl}
\label{eq:bit-allocation - total number of slices in the video}
{S_{total}} = N \cdot M \cdot T.
\end{IEEEeqnarray}
The amounts of inter and skipped slices are calculated as follows:
\begin{IEEEeqnarray}{rCl}
\label{eq:bit-allocation - amounts of the intra, inter and skipped slices}
&& {S_{inter}} = {P_{inter}} \cdot {S_{total}}
\\ \nonumber
&& {S_{skip}} = {P_{skip}} \cdot {S_{total}}.
\end{IEEEeqnarray}

We exclude from our analysis two elements that affect the bit-cost. First are motion-vectors and coding-mode information. Second, entropy coding is also excluded from our scope. However, we assume these two untreated elements balance their overall effect on the bit-cost. Furthermore, their indirect effect on the distortion is considered through our proposed models.

The amount of bits invested in the coefficients of each intra or inter slice is
\begin{IEEEeqnarray}{rCl}
\label{eq:bit-allocation - total bits for transform coefficients}
B_{coeffs}^{slice} & = & \frac{{B_{total}}}{{{S_{{\text{inter}}}}}}
= \frac{{{B_{total}} }}{{{S_{total}} \cdot {P_{{\text{inter}}}}}}
= \frac{{{B_{total}} }}{{M \cdot N \cdot T \cdot {P_{{\text{inter}}}}}}
\end{IEEEeqnarray}

Recall that each slice consists of ${\beta ^2}$ sub-slices that are transformed separately. Therefore, we are interested in the bit-budget for the transform coefficients of a sub-slice:
\begin{IEEEeqnarray}{rCl}
\label{eq:bit-allocation - bit-cost for the transform coefficients of one sub-slice}
B_{coeffs}^{sub - slice} = \frac{1}{{{\beta ^2}}} \cdot B_{coeffs}^{slice}
\end{IEEEeqnarray}
The number of bits allocated for the $\left( {k,l} \right)$ coefficient as function of the slicing parameters $M$, $N$ and $T$ is
\begin{IEEEeqnarray}{rCl}
\label{eq:bit-allocation - bit-cost for the k l transform coefficient of one sub-slice}
{b_{kl}}\left( {{B_{total}},M,N,T,{P_{inter}},{P_{skip}}} \right) & = & {{\tilde Q}_{weight}}\left( {k,l} \right) \cdot B_{coeffs}^{sub - slice}
\\ \nonumber
& = & {\tilde Q_{weight}}\left( {k,l} \right)\cdot
\frac{{{B_{total}} }}{{ {\beta ^2} \cdot M \cdot N \cdot T \cdot {P_{{\text{inter}}}}}}
\end{IEEEeqnarray}
Recall that in H.264's baseline profile $D_{trans} = 4$ and ${\tilde Q_{weight}}\left( {k,l} \right) = \frac{1}{16}$ for $1 \leqslant k,l \leqslant 4$.

\subsection{Overall Distortion}
As discussed in previous sections, H.264 utilizes three macroblock coding modes: intra, inter and skip. In section ‎\ref{sec:Coding-Mode Usage at Low Bit-Rates}, we modeled the coding-mode usage as probabilities varying with the bit-rate while neglecting usage of intra-coding (\ref{eq:probability mass function of block coding mode})-(\ref{eq:inter and skip mode probabilities as piecewise-linear function}).
Moreover, we analyzed the distortion-rate behavior while the bit-cost of elements such as motion-vectors and coding-mode is considered indirectly by modeling the properties of the transform-coded signal as function of the total bit-rate.

Recall (\ref{eq:reconstruction MSE - in 2D slices terms}), where we got that the expected MSE of the entire signal reconstruction equals to the expected MSE of a slice, i.e., $E\left[ {\varepsilon _v^2} \right] = E\left[ {MSE_{{f_v}}\left( {\Delta _{11}^1} \right)} \right]$. However, the slice coding-mode affects the resulting reconstruction error. Moreover, the chosen coding-mode is a random-variable with a distribution function given in (\ref{eq:probability mass function of block coding mode}). Hence, we write
\begin{IEEEeqnarray}{rCl}
\label{eq:expected slice MSE - conditional in coding mode}
&& E\left[ {\left. {\varepsilon _v^2} \right|coding\,\,mode} \right] = E\left[ {\left. {MS{E_{{f_v}}}\left( {\Delta _{11}^1} \right)} \right|coding\,\,mode} \right] .
\end{IEEEeqnarray}
Applying the law of total expectation for the calculation of the expected MSE of a slice:
\begin{IEEEeqnarray}{rCl}
\label{eq:combined mode analysis - distortion}
E\left[ {\varepsilon _v^2} \right] & = & E\left[ {MSE_{{f_v}}\left( {\Delta _{11}^1} \right)} \right] 
\\ \nonumber
& = & E\left[ {E\left[ {\left. {MSE_{{f_v}}\left( {\Delta _{11}^1} \right)} \right|coding\,\,mode} \right]} \right]
\\ \nonumber
& = & {P_{{inter}}}\left( {{B_{slice}}} \right) \cdot E\left[ {\left. {MSE_{{f_v}}\left( {\Delta _{11}^1} \right)} \right|inter\,\,coding} \right]
\\ \nonumber
&& + {P_{skip}}\left( {{B_{slice}}} \right) \cdot E\left[ {\left. {MSE_{{f_v}}\left( {\Delta _{11}^1} \right)} \right|skip\,\,mode} \right]
\end{IEEEeqnarray}

\section{Compression-Scaling System}
\label{sec:Compression-Scaling System}
\subsection{Frame-Rate Up Conversion}
In our previous paper \cite{RefWorks:68}, we modeled the MC-FRUC procedure and derived the MSE of the $j^{th}$ interpolated frame:
\begin{IEEEeqnarray}{rCl}
\label{eq:MC-FRUC MSE}
&& MSE_{FRUC}\left( {D_{T},j} \right) = \frac{1}{2}\cdot\left[ {\tilde \sigma _q^2 \frac{D_{T}}{F_{rate}} + \sigma _{{w_0}}^2 + \sigma _{{w_j}}^2} \right] + \left( {\sigma _{\Delta {x^{abs}}}^2 + \sigma _{\Delta {y^{abs}}}^2} \right)\cdot\left[ {\left( {1 - {\rho _v}} \right)\cdot\sigma _v^2 + \frac{L}{F_{rate}} \tilde \sigma _q^2 + \sigma _{{w_0}}^2} \right]
\end{IEEEeqnarray}
where, 
$D_{T}$ is the frame-rate conversion factor. $j \in \left\{ {1,...,{D_T} - 1} \right\}$ is the index of the interpolated frame, which is assigned relative to the existing frames. 
$\sigma _{\Delta x ^{abs}}^2$ and $\sigma _{\Delta y ^{abs}}^2$ reflect accuracy errors in ME of absent frame, $\sigma _v^2$ and $\rho _v$ are the variance and correlation coefficient of the frame pixels, $L$ is a temporal memory factor, $\tilde \sigma _q^2$ is the motion energy in the video, $F_{rate}$ is the frame rate, $\sigma _{{w_0}}^2$ is a temporally-local noise energy of an available frame. Temporally-local noise reflects distortions due to compression or spatial processing; hence, we set
\begin{IEEEeqnarray}{rCl}
\label{eq:MC-FRUC MSE - compression effect}
\sigma _{{w_0}}^2 = MSE_{compression}
\end{IEEEeqnarray}
where $MSE_{compression}$ equals to $E\left[ {\varepsilon _v^2} \right]$ from (\ref{eq:combined mode analysis - distortion}). Recall that the spatial down-scaling error is included in the compression error through its effect on the MC-prediction residual.
We update (\ref{eq:MC-FRUC MSE}) accordingly:
\begin{IEEEeqnarray}{rCl}
\label{eq:MC-FRUC MSE - with compression MSE}
MSE_{FRUC}\left( {D_{T},j,MSE_{compression}} \right) & = & \frac{1}{2}\cdot\left[ {\tilde \sigma _q^2 \frac{D_{T}}{F_{rate}} + MSE_{compression} + \sigma _{{w_j}}^2} \right] 
\\ \nonumber
&& + \left( {\sigma _{\Delta {x^{abs}}}^2 + \sigma _{\Delta {y^{abs}}}^2} \right) \cdot \left[ {\left( {1 - {\rho _v}} \right)\cdot\sigma _v^2 + \frac{L}{F_{rate}} \tilde \sigma _q^2 + MSE_{compression}} \right]
\end{IEEEeqnarray}

\begin{figure}
\centering
\includegraphics[width=2.5in]{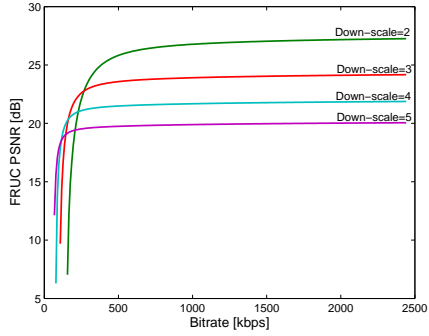}
\caption{FRUC PSNR prediction for a typical video ($\sigma ^2 _v = 2300$, $\rho _v = 0.95$, $\tilde \sigma _q^2 = 250$, $L=100$).}
\label{Fig:Estimated FRUC PSNR vs BitRates at Various T}
\end{figure}

The behavior of this model for FRUC PSNR (Fig. \ref{Fig:Estimated FRUC PSNR vs BitRates at Various T}) shows a convergence as the bit-rate increases, since at high bit-rates there is a minor distortion that does not affect the ME performance in FRUC. Moreover, 
the PSNR converges to an higher value for smaller up-scaling factors, as expected due to the unrecoverable loss of the discarded frames.

\subsection{Overall System Analysis}
Recall the structure of the investigated system for improved low bit-rate video coding (Fig. \ref{Fig:compression_scaling_system_structure}). This system suggests to compress a down-scaled video and to up-scale it to its original dimensions after decoding. The compression error of the down-scaled video was studied here over few sections resulting in error expression (\ref{eq:combined mode analysis - distortion}). The error of temporal interpolation by MC-FRUC techniques was given in (\ref{eq:MC-FRUC MSE - with compression MSE}). Here we combine the two aforementioned parts of the system and discuss on the complete system performance.

The spatio-temporal down-scaling operations are applied sequentially as decreasing the frame size and lowering the frame rate by discarding frames (assuming the temporal down-scaling factor is an integer). The down-scaling operation order is important only for computational efficiency, since reducing frame size of omitted frames is unnecessary. In contrast, the operation order of up-scaling after decoding is important for the quality of the result. Our proposed system includes MC-FRUC algorithm for frame interpolation; hence, motion-estimation is done and its performance affects the interpolation quality. Commonly, motion-estimation and compensation is performed on finer spatial resolution (e.g., half-pel or quarter-pel) where the reference frames are temporarily enlarged for better results \cite{RefWorks:17}. Therefore, spatial up-scaling before applying FRUC gives better results than in the opposite order.

\begin{figure}
\centering
\includegraphics[width=3.5in]{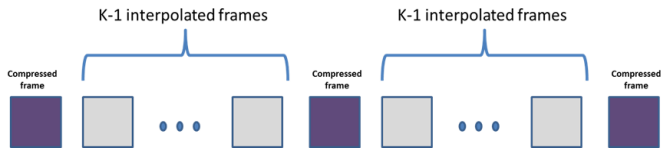}
\caption{Frame pattern of an output video that was temporally down-scaled in a factor of K.}
\label{Fig:frame_pattern_in_the_compression_scaling_system}
\end{figure}

The output video consists of two frame types. Firstly, frames that were encoded in the down-scaled video. These frames were spatially down-scaled and up-scaled, before and after compression, respectively; Therefore, the spatial scaling affects them directly, whereas the temporal scaling affects them only indirectly through the lower frame rate in the actually compressed video. The second frame type is the omitted frames that were interpolated after decoding. These frames are affected directly by the temporal scaling, while the spatial scaling affects them indirectly through its distortion on the frames of the first type that are used for the FRUC.
The frame types are arranged periodically according to the chosen temporal down-scaling factor (Fig. \ref{Fig:frame_pattern_in_the_compression_scaling_system}).

Since the frames are reconstructed from compressed data or by temporal-interpolation from this data, the overall mean-squared-error of the output video is a weighted-average of the compression and interpolation errors. The overall mean-squared-error is given by
\begin{IEEEeqnarray}{rCl}
\label{eq:overall system analysis - distortion}
MSE_{overall}\left( {M,N,T,B} \right) = \frac{1}{{{D_T}}} \cdot MSE_{spatial}\left( {M,N,T,B} \right) + \frac{{{D_T} - 1}}{{{D_T}}} \cdot MSE_{spatio-temporal}\left( {M,N,T,B} \right).
\end{IEEEeqnarray}
Where, ${D_T}$ is the frame-rate upsampling factor, defined according to (\ref{eq:temporal down-scaling factors definition}) as
\begin{IEEEeqnarray}{rCl}
\label{eq:overall system analysis - frame-rate upsampling factor definition}
{D_T} = \frac{{original\,\,\,frame\,\,\,rate}}{T},
\end{IEEEeqnarray}
where $T$ is the temporal slicing factor.
$MSE_{spatial}$ is the MSE of a frame that was only spatially scaled, it is defined as:
\begin{IEEEeqnarray}{rCl}
\label{eq:overall system analysis - MSE spatial}
MSE_{spatial}\left( {M,N,T,B} \right) = MSE_{compression}\left( {M,N,T,B} \right).
\nonumber \\
\end{IEEEeqnarray}
Note that the temporal-scaling affects these frames indirectly through the statistics of the MC-prediction residual that is spatially coded.
$MSE_{spatio-temporal}$ is the MSE of a frame that was discarded in temporal down-scaling procedure. This frame is reconstructed using frames that were only spatially down-scaled; therefore, we set $\sigma _{{w_0}}^2 = MSE_{spatial}\left( {M,N,T,B} \right)$ and define $MSE_{spatio-temporal}$ as the average MSE of interpolated frames:
\begin{IEEEeqnarray}{rCl}
\label{eq:overall system analysis - MSE spatio-temporal}
MSE_{spatio - temporal}\left( {M,N,T,B} \right) = \frac{1}{{{D_T} - 1}} \cdot \sum\limits_{j = 1}^{{D_T} - 1} {MSE_{FRUC}\left( {T,j,MSE_{spatial}\left( {M,N,T,B} \right)} \right)} .
\end{IEEEeqnarray}
$MSE_{compression}$ and $MSE_{FRUC}$ are the MSE of the frames of the compressed down-scaled video, and the temporally-interpolated frames, respectively.
The expressions for $MSE_{compression}$ and $MSE_{FRUC}$ were given in (\ref{eq:combined mode analysis - distortion}) and (\ref{eq:MC-FRUC MSE - with compression MSE}), respectively.

\subsection{Optimization of Signal Slicing}
The basic bit-allocation optimization problem is based on the probabilistic expressions for expected overall MSE of the compression-scaling system (\ref{eq:overall system analysis - distortion}), given a bit-budget denoted as $B_{total}$. 

We formulate an optimization problem that models the coding-mode choices actually taken in a real and unmodified H.264 encoder working in the baseline profile. This is done by setting a coding-mode usage according to the bit-rate as presented in section \ref{sec:Coding-Mode Usage at Low Bit-Rates}. Hence, $P_{inter}$ and $P_{skip}$ vary according to a defined model, rather than being directly optimized. The problem is formulated as follows:
\begin{IEEEeqnarray}{rCl}
\label{eq:overall system analysis - optimization problem of signal slicing}
\begin{gathered}
  \mathop {{\text{minimize}}}\limits_{M,N,T} {\text{ }}MSE_{overall}\left( {M,N,T,B} \right) \hfill \\
  {\text{subject to           }} \hfill \\
  \,\,\,\,\,\,\,\,\,\,\,\,\,\,\,\,\,{B_{slice}} = \frac{{{B_{total}}}}{{M \cdot N \cdot T}} \hfill \\
  \,\,\,\,\,\,\,\,\,\,\,\,\,\,\,\,\,{P_{inter}} = {P_{inter}}\left( {{B_{slice}}} \right) \hfill \\
  \,\,\,\,\,\,\,\,\,\,\,\,\,\,\,\,\,{P_{skip}} = {P_{skip}}\left( {{B_{slice}}} \right) \hfill \\
  \,\,\,\,\,\,\,\,\,\,\,\,\,\,\,\,\,\,{b_{kl}} = {b_{kl}}\left( {{B_{total}},M,N,T,{P_{inter}},{P_{skip}}} \right)\,\,\,,\,\,1 \leqslant k,l \leqslant 4 \hfill \\ 
\end{gathered}
\end{IEEEeqnarray}
The bit-allocation within each coding mode is assumed to model a real encoder, as presented in section \ref{subsec:Bit-Allocation}; specifically, ${b_{kl}}\left( {{B_{total}},M,N,T,{P_{inter}},{P_{skip}}} \right)$ is given in (\ref{eq:bit-allocation - bit-cost for the k l transform coefficient of one sub-slice}). 
The optimization can be applied on a subset of the slicing parameters. E.g., temporal-only optimization will keep $M$ and $N$ fixed, while optimizing $T$; spatial-only optimization will keep $T$ fixed, whereas $M$ and $N$ are optimized.

Recall we aim at finding the optimal down-scaling of a video given to an unmodified H.264 codec. Therefore, we optimize the down-scaling factors and not the encoder parameters; e.g., coding-mode assignment and bit-allocation are modeled as in a real H.264 compression rather than optimized.

\section{Theoretical Predictions of the Model}
\label{sec:Theoretical Estimations of the Model}
\subsection{Temporal Scaling}
\begin{figure*}[!t]
\centering
\subfloat[]{\includegraphics[width=3in]{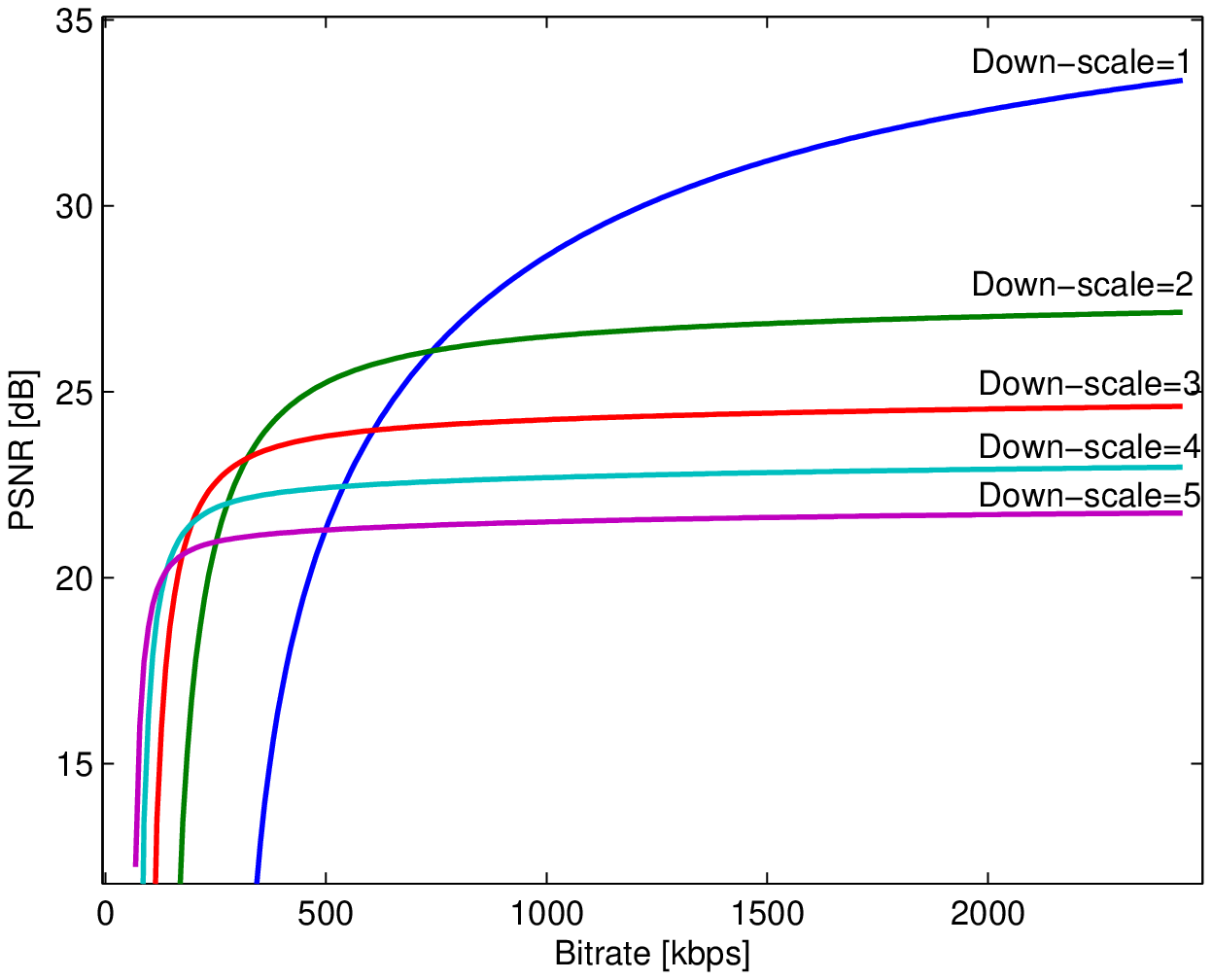}
\label{Fig:model_estimation_temporal_scaling}}
\subfloat[]{\includegraphics[width=3in]{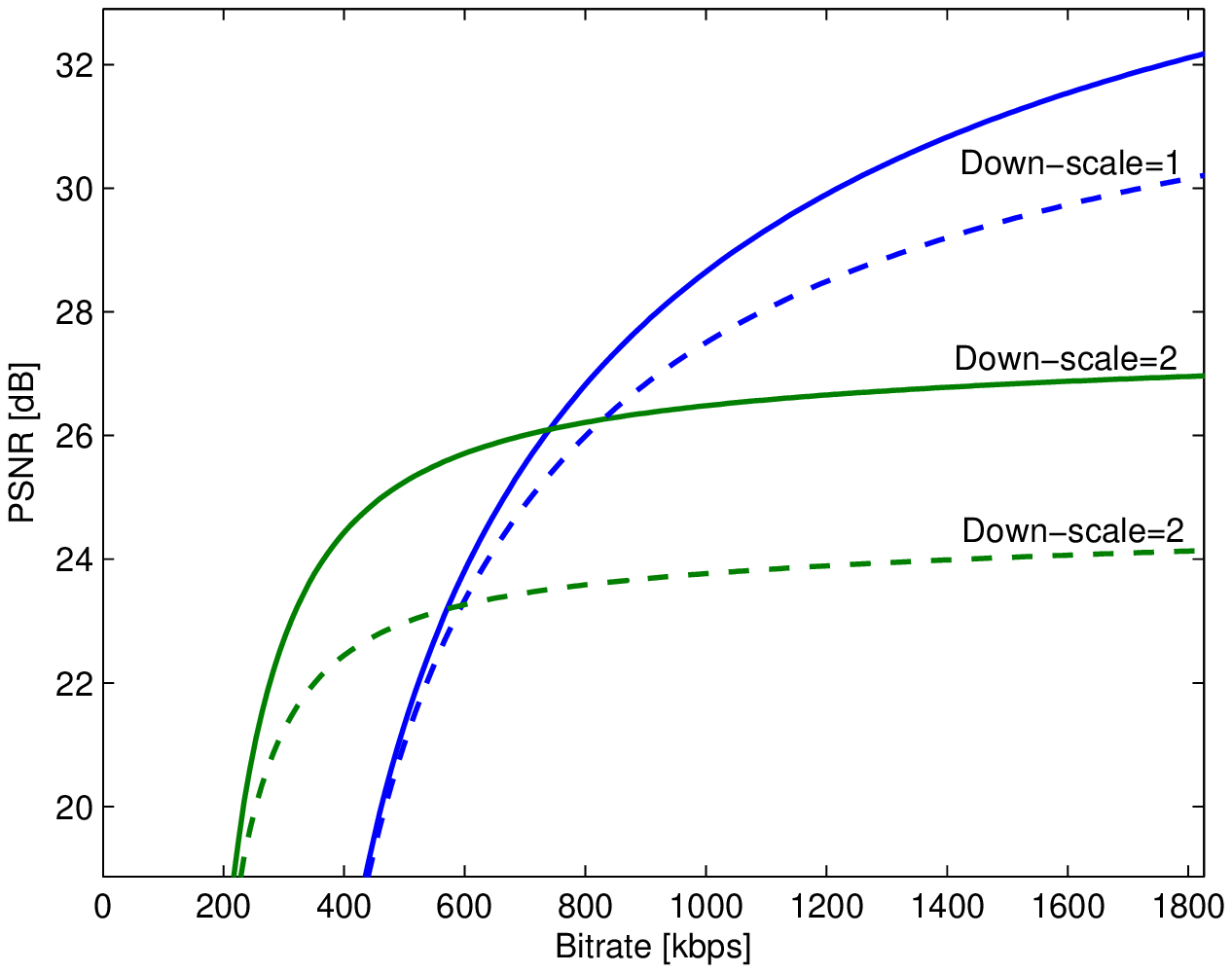}
\label{Fig:model_estimation_temporal_scaling_motion_complexity_comparison}}
\caption{Theoretical estimation of overall compression-scaling system PSNR for temporal scaling of a typical video ($\sigma ^2 _v = 2300$, $\rho _v = 0.95$ and $\tilde \sigma _q^2 = 250$, $L=100$). (b) comparison of results with video containing more complex motion  $\tilde \sigma _q^2 = 500$ (dashed lines).}
\label{Fig:model_estimation_temporal_scaling - including comparison of motion complexity effect}
\end{figure*}

Here we examine the overall compression-scaling system for temporal scaling only. We estimate PSNR for various temporal down-scaling factors at varying bit-rates (Fig. \ref{Fig:model_estimation_temporal_scaling - including comparison of motion complexity effect}) while having fixed spatial down-scaling factors that correspond to the original frame size (i.e., $M=N=45$ for $720\times720$ frame and $16\times16$ block). 

Our model behaves as follows. First, we got a pattern of decision regions (Fig. \ref{Fig:model_estimation_temporal_scaling}), 
where a decision region corresponding to a higher down-scaling factor is located in a lower bit-rate range.
Second, as the contained motion in video becomes more complex, the estimated PSNR is lower (Fig. \ref{Fig:model_estimation_temporal_scaling_motion_complexity_comparison}). Moreover, the intersection between scaling-curves occurs at lower bit-rates; hence, the advised temporal down-scaling factor is lower. These estimates are justified by the higher distortion expected from the temporal-interpolation due to unrecoverable information when the motion complexity increases.

\subsection{Spatial Scaling}
The compression-scaling system for spatial-scaling does not include FRUC, i.e., $D_T = 1$. Setting this in (\ref{eq:overall system analysis - distortion}) yields:
\begin{IEEEeqnarray}{rCl}
\label{eq:distortion expression for spatial scaling only}
MSE_{overall}\left( {M,N,T,B} \right) = MSE_{spatial}\left( {M,N,T,B} \right).
\nonumber \\
\end{IEEEeqnarray}
Therefore, the overall performance is evaluated by considering the PSNR of compression and spatial scaling.

\begin{figure*}[!t]
\centering
\subfloat[]{\includegraphics[width=2.4in]{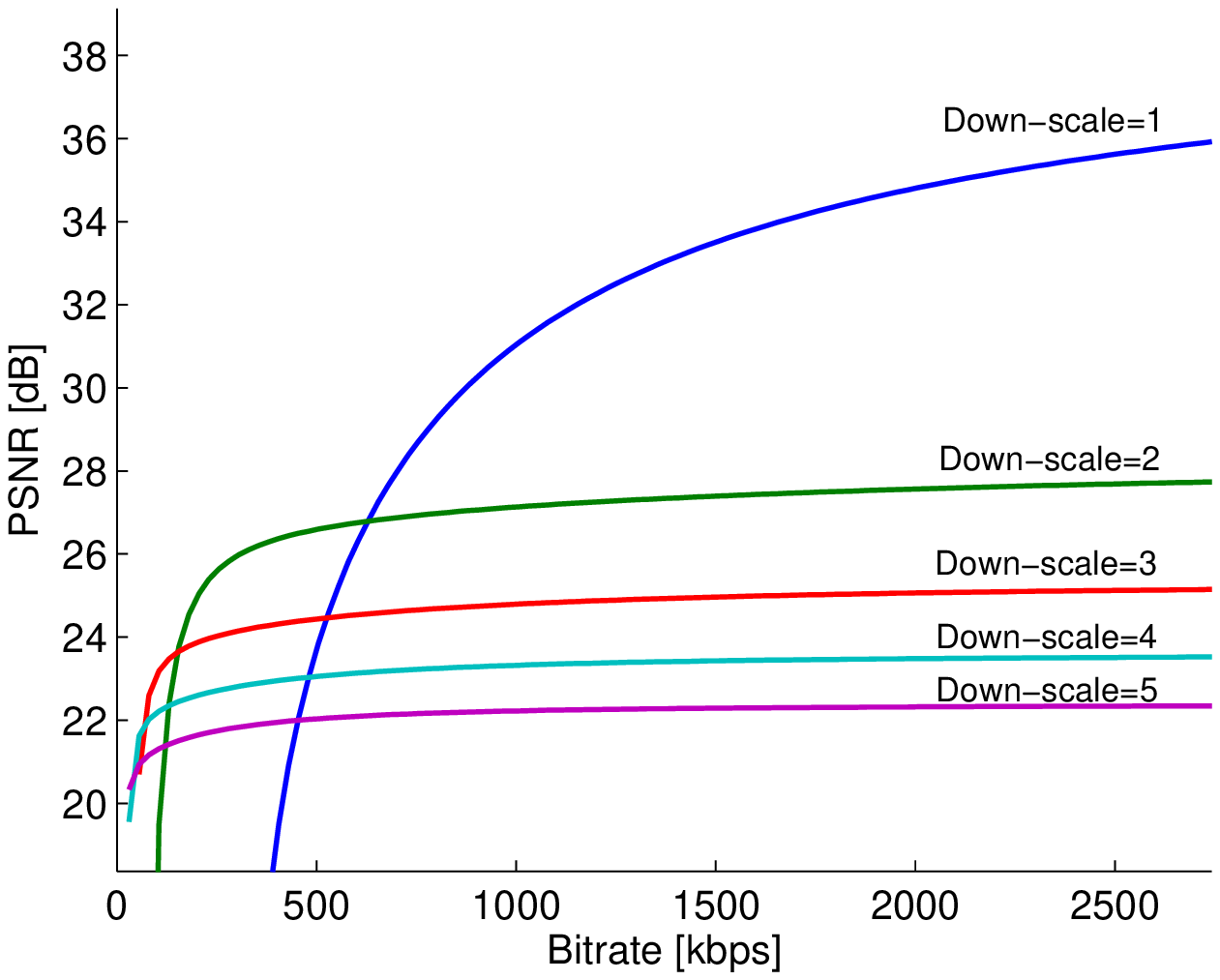}
\label{Fig:model_estimation_spatial_scaling__compression_psnr}}
\subfloat[]{\includegraphics[width=2.4in]{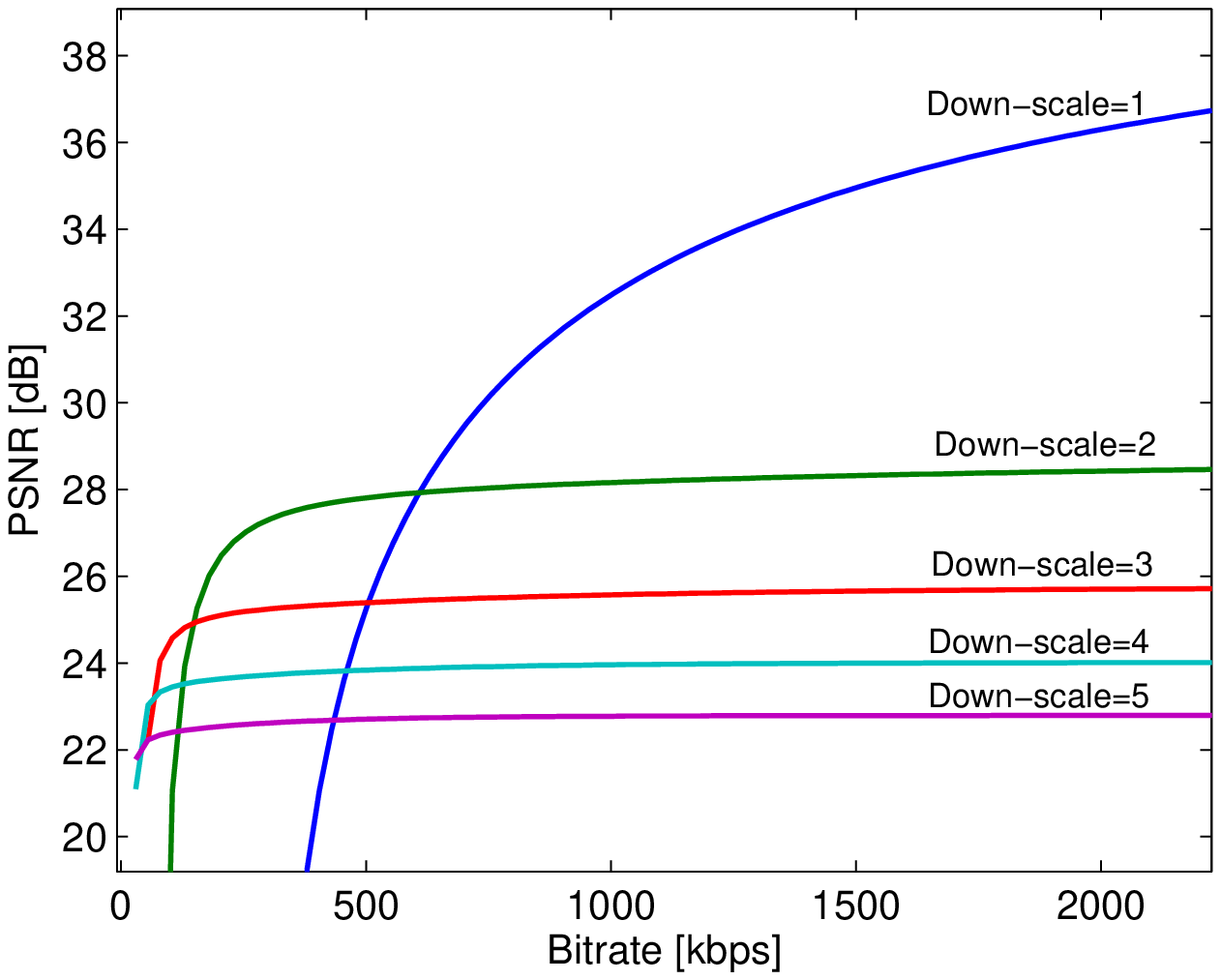}
\label{Fig:model_estimation_spatial_scaling__inter_coding_psnr}}
\subfloat[]{\includegraphics[width=2.4in]{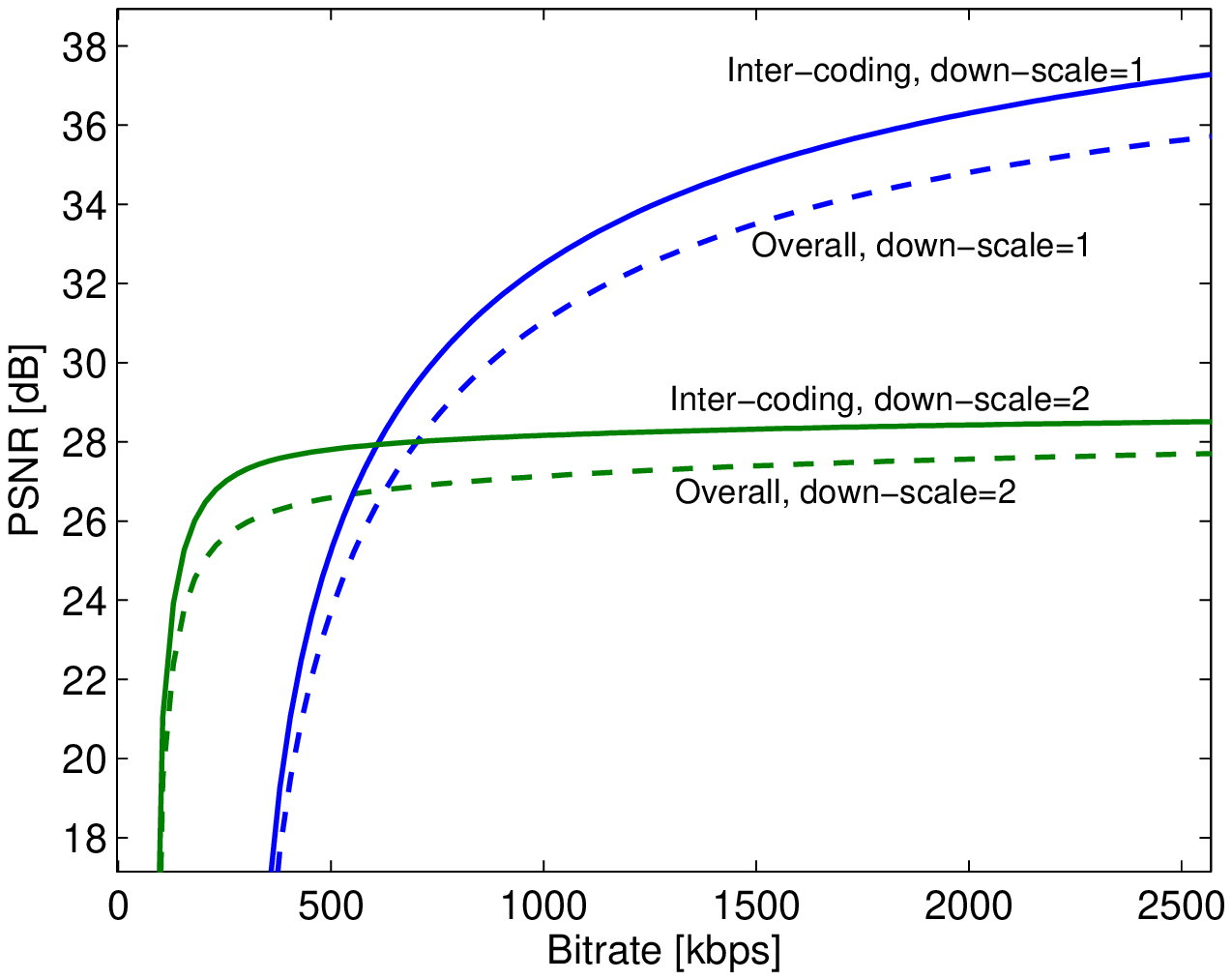}
\label{Fig:model_estimation_spatial_scaling__camparing_overall_to_inter_coding}}
\caption{Theoretical estimation of compression PSNR for spatial scaling of a typical video ($\sigma ^2 _v = 2300$, $\rho _v = 0.95$ and $\tilde \sigma _q^2 = 250$). (a) overall compression-scaling system, (b) inter coding of scaled video, (c) a comparison between overall compression and inter-coding.}
\label{Fig:model_estimation_spatial_scaling__compression_psnr and inter coding}
\end{figure*}

Theoretical PSNR prediction for the compression of a spatially-scaled video (Fig. \ref{Fig:model_estimation_spatial_scaling__compression_psnr}) shows that as the bit-rate reduces, PSNR can be improved by higher factor of spatial scaling.
Let us examine the effect of skip mode on the compression results. The PSNR of inter-coded blocks (Fig. \ref{Fig:model_estimation_spatial_scaling__inter_coding_psnr}) has similar behavior to the overall compression procedure. However, the PSNR values for inter-coding are higher than for the overall compression (Fig. \ref{Fig:model_estimation_spatial_scaling__camparing_overall_to_inter_coding}); moreover, the intersections between the PSNR graphs for the inter-coding only occur at lower bit-rates than for the overall compression that includes the skip mode. This is due to the increased amount of bits allocated per slice for higher down-scaling factors that reduces the use of skip mode, which have inferior reconstruction quality that increases the overall compression distortion.

\begin{figure*}[!t]
\centering
\includegraphics[width=2.6in]{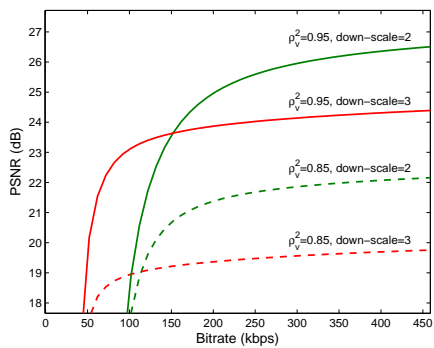}
\caption{Demonstration of texture complexity effect according to the theoretical estimations for spatial scaling. Texture level represented by the correlation coefficient.}
\label{Fig:model_estimation_spatial_scaling_texture_complexity_comparison}
\end{figure*}

Let us analyze the estimations for a varying texture level. As the video contains larger amount or higher complexity textured its pixel's correlation-coefficient decreases. Our estimations show that compression of more textured video results in lower quality (Fig. \ref{Fig:model_estimation_spatial_scaling_texture_complexity_comparison}), since representation of textured images require higher bit-budget. Moreover, texture information resides in high frequency components that are removed in spatial down-scaling; hence, lower spatial down-scaling factors are preferable for videos with increased texture content (Fig. \ref{Fig:model_estimation_spatial_scaling_texture_complexity_comparison}).

\subsection{Spatio-Temporal Scaling}
We studied above our compression-scaling system for spatial-only and temporal-only down-scaling. Optimal spatial and temporal scaling factors depend on the complexities of texture and motion, respectively. Here we examine the application of joint spatio-temporal down-scaling.
The optimal down-scaling factor gives the highest reconstruction quality among the considered factors, including compression at the original dimensions. Inevitably, any down-scaling of non-trivial signal will introduce information loss and distortions. However, for a given bit-rate and spatio-temporal characteristics of a video, the optimal choice of spatio-temporal down-scaling factors depends on the relation between the complexities of texture and motion. 

\begin{figure*}[!t]
\centering
\subfloat[]{\includegraphics[width=2.3in]{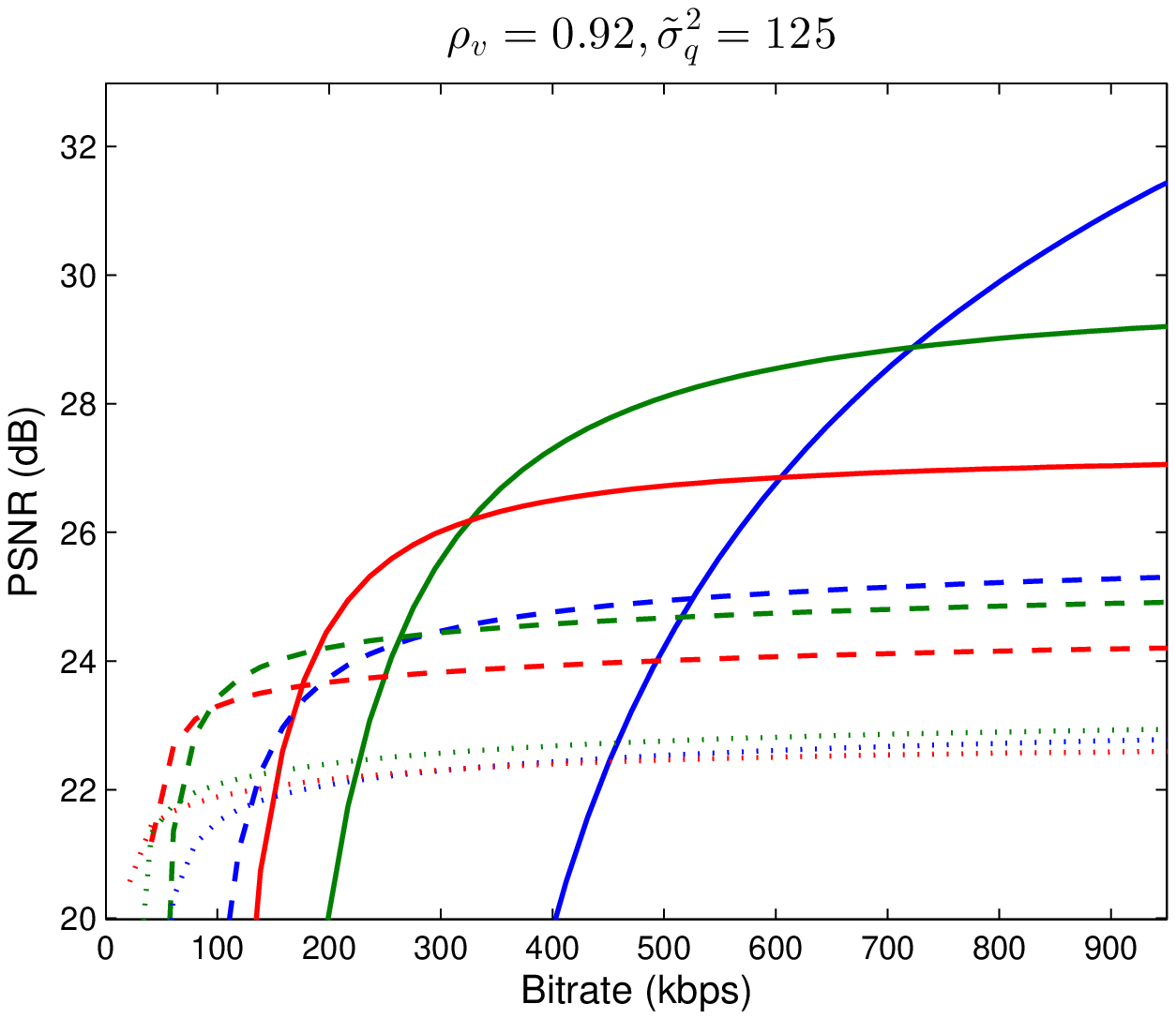}
\label{Fig:model_estimation_spatio_temporal_scaling_comparison_first}}
\subfloat[]{\includegraphics[width=2.3in]{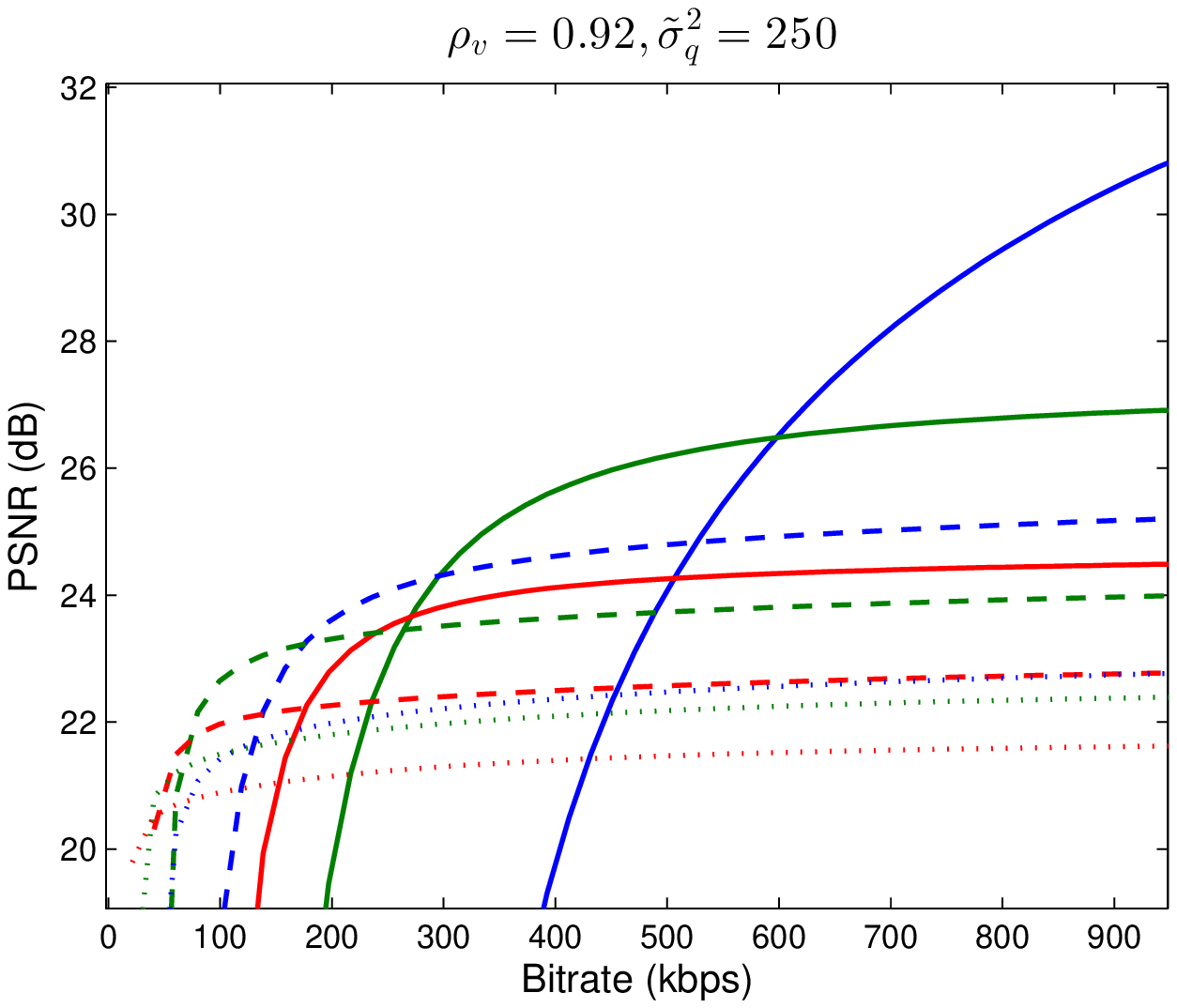}
\label{Fig:model_estimation_spatio_temporal_scaling_comparison_rho086_qvar250}}
\subfloat[]{\includegraphics[width=2.3in]{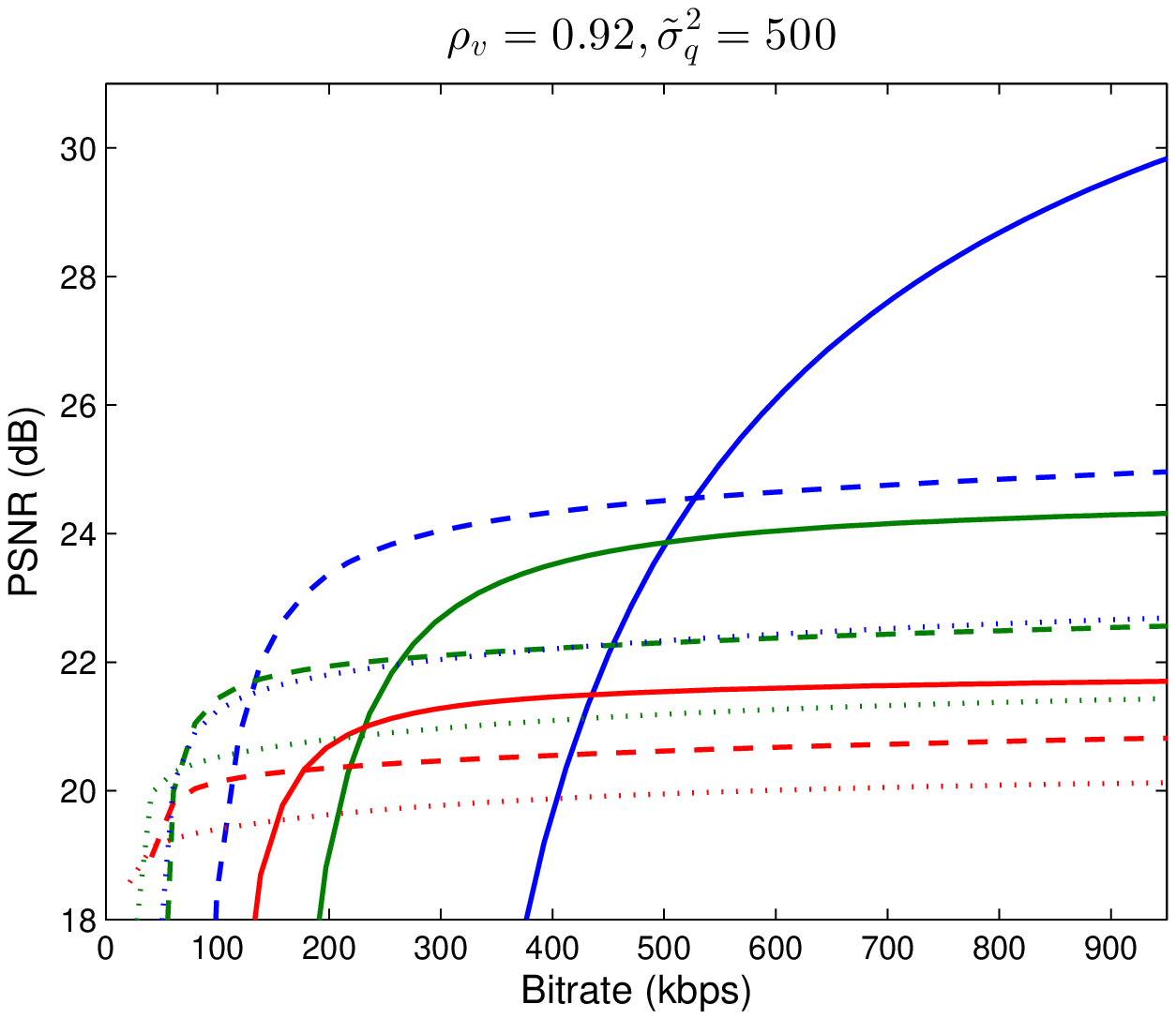}
\label{Fig:model_estimation_spatio_temporal_scaling_comparison_rho086_qvar500}}
\\
\subfloat[]{\includegraphics[width=2.3in]{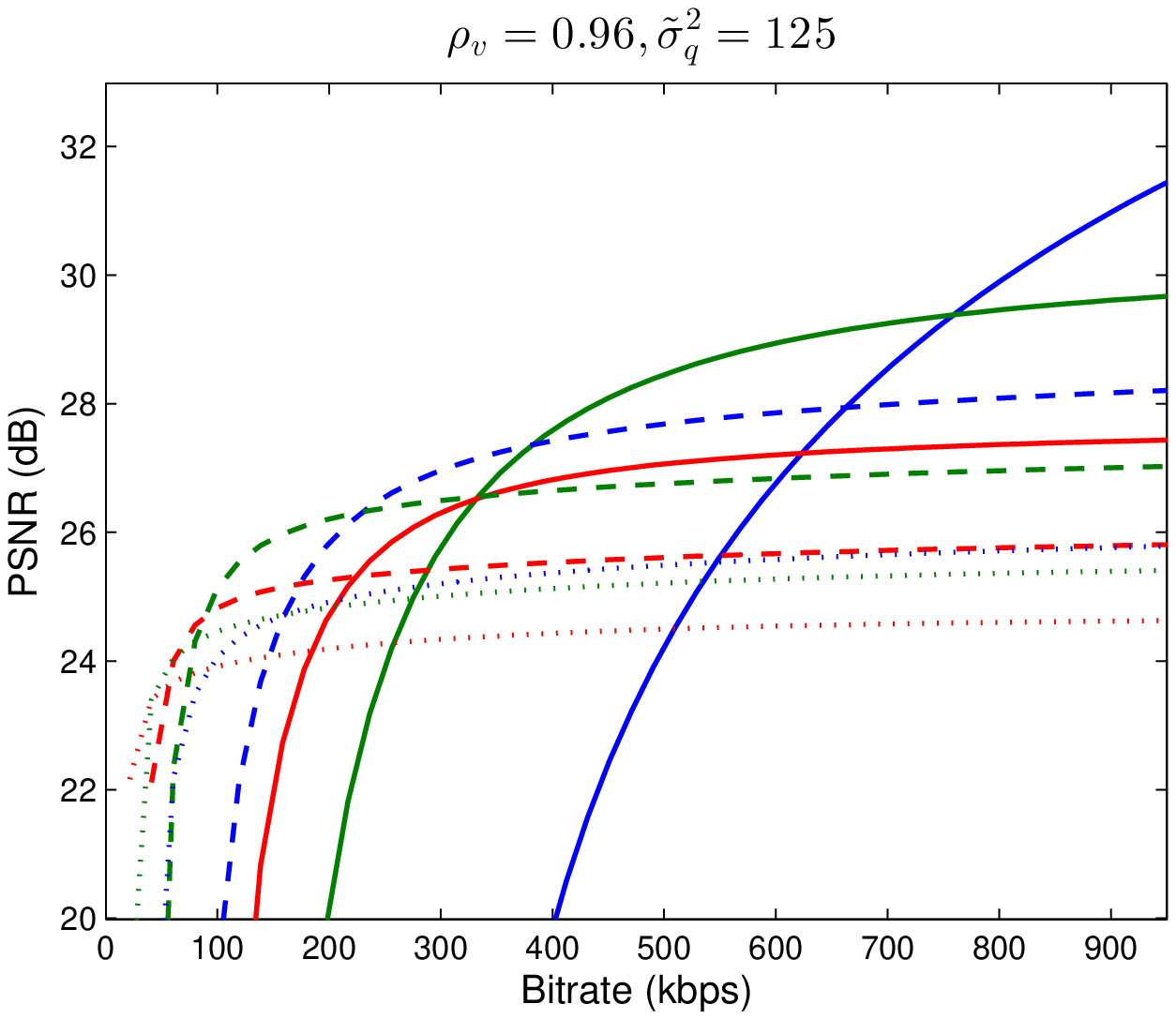}
\label{Fig:model_estimation_spatio_temporal_scaling_comparison_rho092_qvar125}}
\subfloat[]{\includegraphics[width=2.3in]{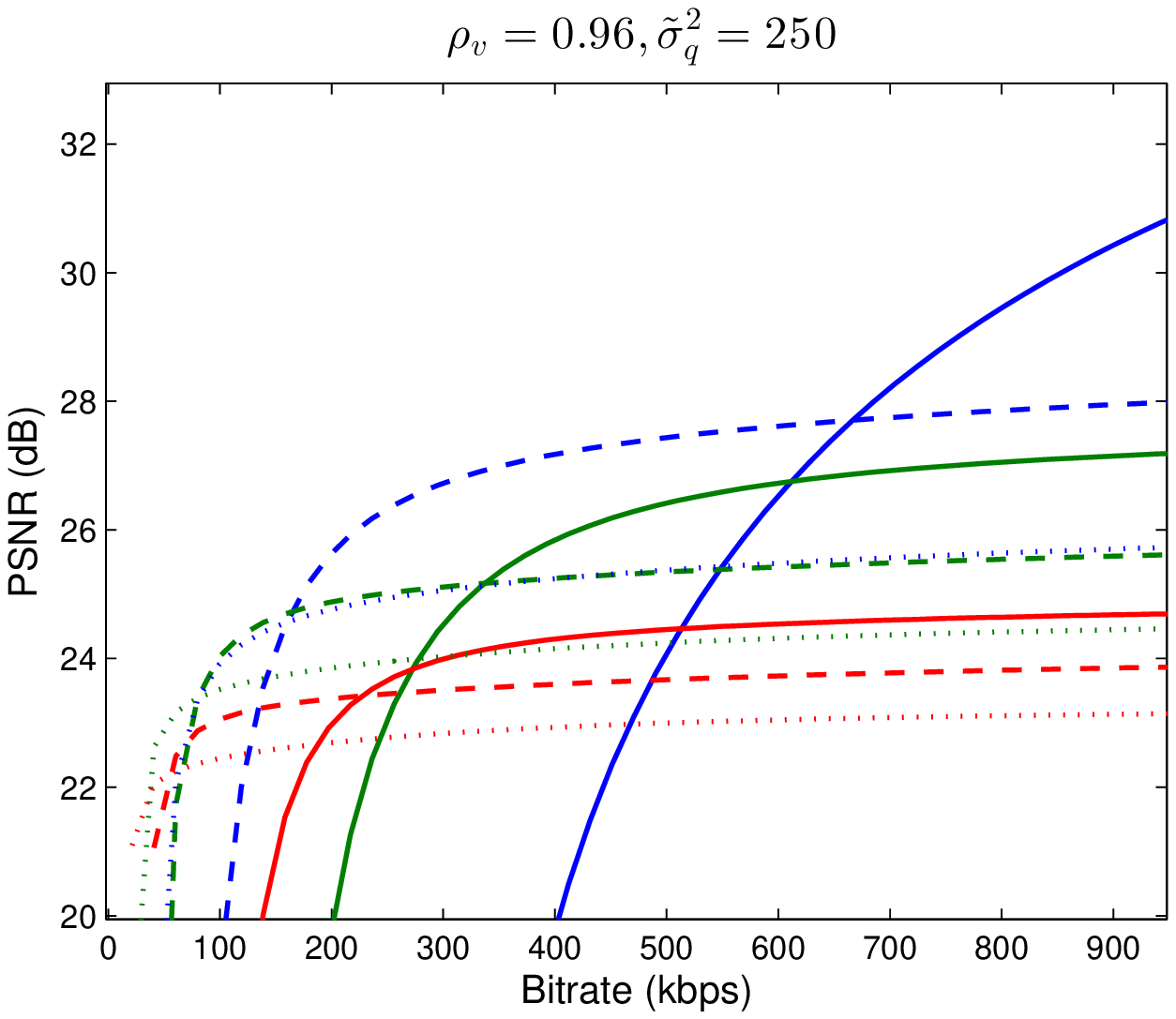}
\label{Fig:model_estimation_spatio_temporal_scaling_comparison_rho092_qvar250}}
\subfloat[]{\includegraphics[width=2.3in]{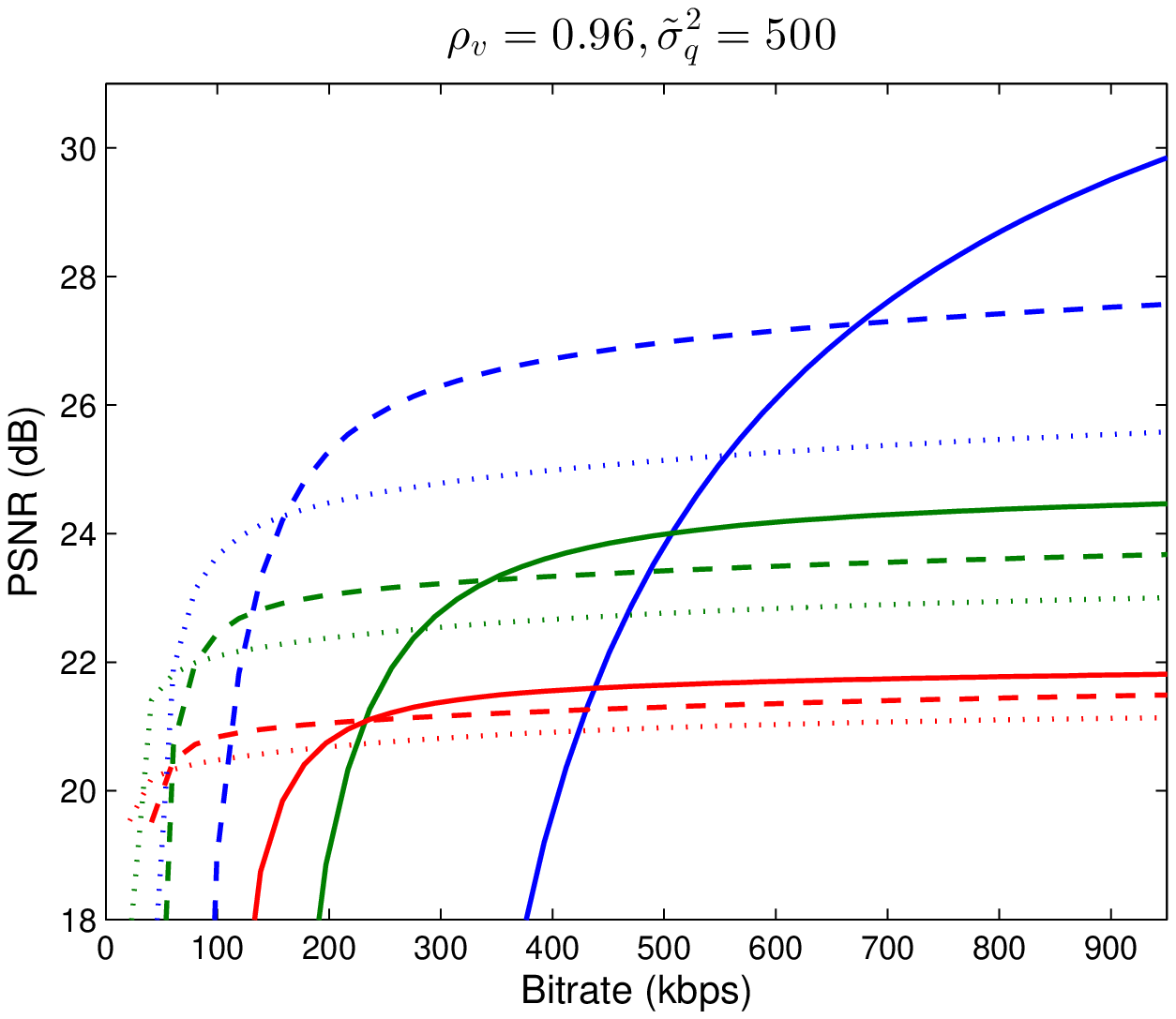}
\label{Fig:model_estimation_spatio_temporal_scaling_comparison_last}}
\caption{Theoretical estimation of compression PSNR for spatio-temporal scaling of a video signal with varying texture and motion levels (set using the correlation coefficient $\rho_v$, and the motion-complexity $\tilde \sigma _q^2$, respectively). Spatial down-scale factor is represented by line style: $D_M=D_N=$ 1 (solid), 2 (dashed), 3 (dotted). Temporal down-scale is represented by line color: $D_T=$ 1 (blue), 2 (green), 3 (red). (Fixed values: $\sigma ^2 _v = 2300$ and $L = 100$).}
\label{Fig:model_estimation_spatio_temporal_scaling_comparison}
\end{figure*}

Figure \ref{Fig:model_estimation_spatio_temporal_scaling_comparison} shows PSNR plots for varying levels of motion and texture complexities. Figures \ref{Fig:model_estimation_spatio_temporal_scaling_comparison_first}-\ref{Fig:model_estimation_spatio_temporal_scaling_comparison_last} are ordered as follows. As the figure located more right, than the motion-complexity, $\tilde \sigma _q ^2$, is higher. Additionally, as the figure location is lower, than the texture-complexity is lower and $\rho _v$ is higher. Each plot include 9 PSNR curves for combinations of spatio-temporal down-scaling factors, where ${D_T} \in \left\{ {1,2,3} \right\}$ and ${D_M} = {D_N} \in \left\{ {1,2,3} \right\}$. The intersections among the PSNR curves define the estimated optimal decision regions.  
Fig. \ref{Fig:model_estimation_spatio_temporal_scaling_comparison} shows that for higher motion-complexity (i.e. more right sub-figure), then spatial down-scaling in more beneficial than reducing the frame-rate. In addition, higher texture complexity, i.e. upper sub-figure location, makes the temporal down-scaling more preferable.

Many optimal combinations of down-scaling factors hold ${D_T} > 1$ and ${D_M} > 1$ (Fig. \ref{Fig:model_estimation_spatio_temporal_scaling_comparison}). Hence, in these cases both dimensions are down-scaled, implying better results than can be achieved in the spatial-only and temporal-only scaling systems.

\section{Experimental Results}
\label{sec:Experimental Results}
In this section we show experimental results of our compression-scaling system on the 'Old town cross' and 'Parkrun' sequences. The original videos are of 720x720 pixels frame-size at 50 frames per second. Our experimental setup consisted of an H.264 coded \cite{RefWorks:72} and spatio-temporal scaling (including MC-FRUC) implemented in Matlab.


\subsection{Temporal Scaling}
\begin{figure*}[!t]
\centering
\subfloat[]{\includegraphics[width=3.5in]{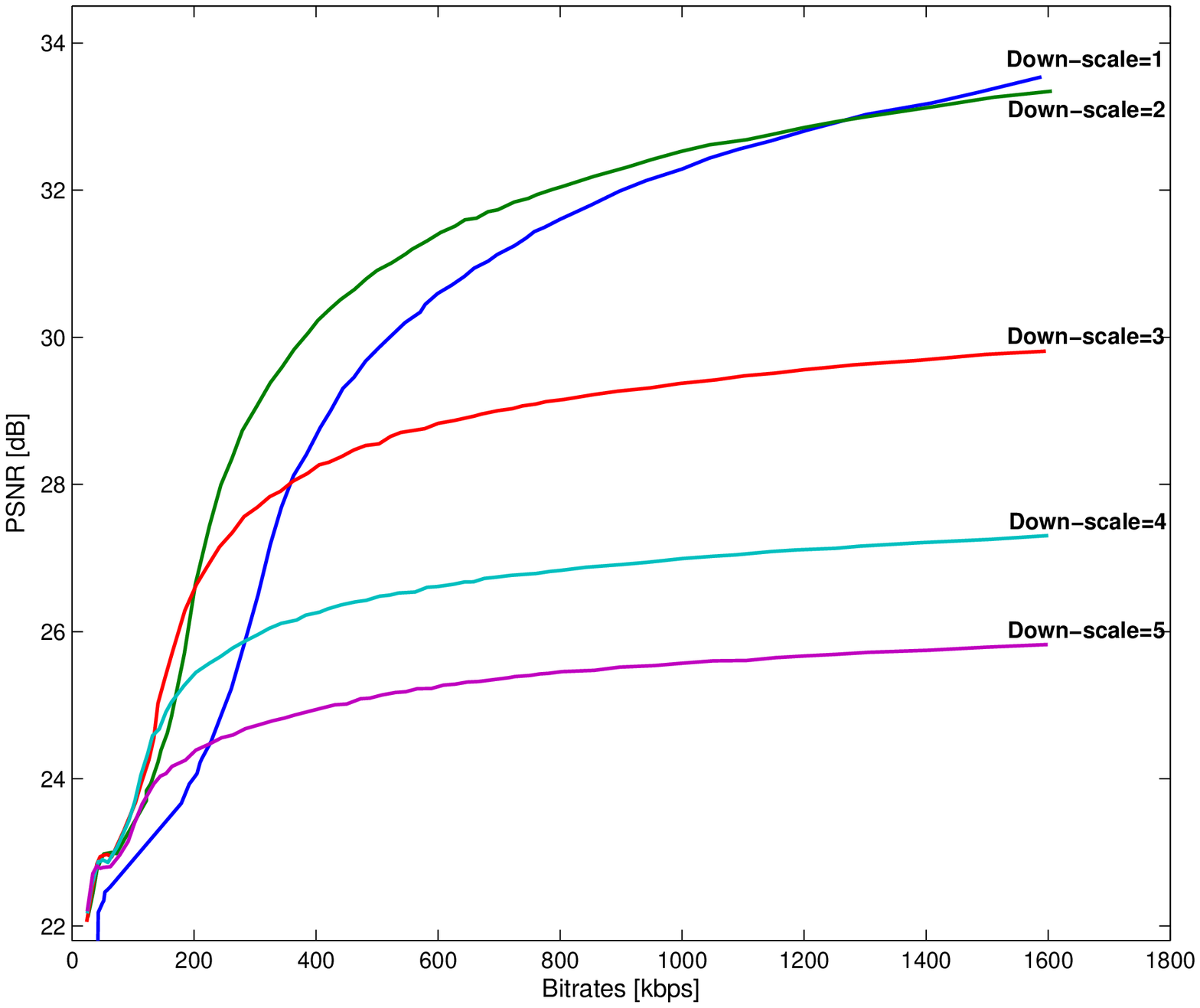}
\label{Fig:Compression-Upsampling_Results_For_Various_Temporal_Downsampling_Factors_old_town_cross}}
\subfloat[]{\includegraphics[width=3.5in]{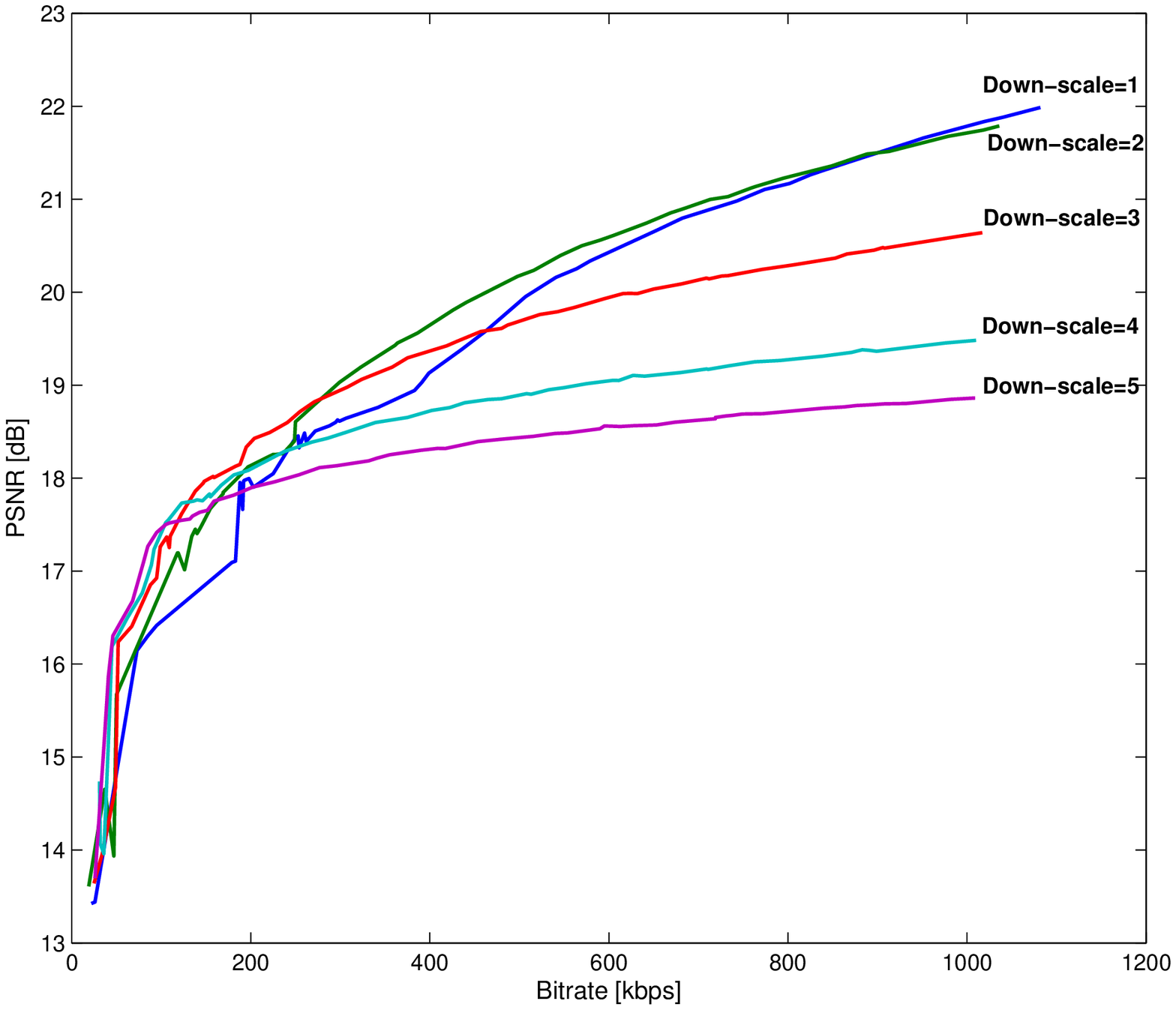}
\label{Fig:Compression-Upsampling_Results_For_Various_Temporal_Downsampling_Factors_parkrun}}
\caption{Output PSNR of compression-scaling system for temporal scaling only. (a) 'Old town cross' (b) 'Parkrun', both 720x720, grayscale.}
\label{Fig:Output PSNR of compression-scaling system - only temporal scaling}
\end{figure*}

Here we allow temporal scaling only. The results (Fig. \ref{Fig:Output PSNR of compression-scaling system - only temporal scaling}) show behavior that is similar to the theoretical prediction of the model. Specifically, the decision regions are constructed in a similar order of increasing down-scaling factors as the bit-rate decreases. Moreover, at a given bit-rate, the optimal down-scaling factor for sequence with more complex motion ('Parkrun') is equal or lower than for simpler motion ('Old town cross').

Our proposed compression-scaling system for temporal-scaling only improved the compression results for bit-rates lower than 1250kbps and 890kbps for 'Old town cross' (Fig. \ref{Fig:Compression-Upsampling_Results_For_Various_Temporal_Downsampling_Factors_old_town_cross}) and 'Parkrun' (Fig. \ref{Fig:Compression-Upsampling_Results_For_Various_Temporal_Downsampling_Factors_parkrun}), respectively. The difference in these threshold bit-rates is due to the disparate motion complexity levels of the videos.
The results for 'Old town cross' (Fig. \ref{Fig:Compression-Upsampling_Results_For_Various_Temporal_Downsampling_Factors_old_town_cross}) show a PSNR improvement of 2.6dB at 180kbps by reducing the frame-rate in a factor of 3; additionally, bit-savings of 34\% were achieved for fixed PSNR of 27dB by halving the frame-rate.


\subsection{Spatial Scaling}
\begin{figure*}
\centering
\subfloat[]{\includegraphics[width=3.5in]{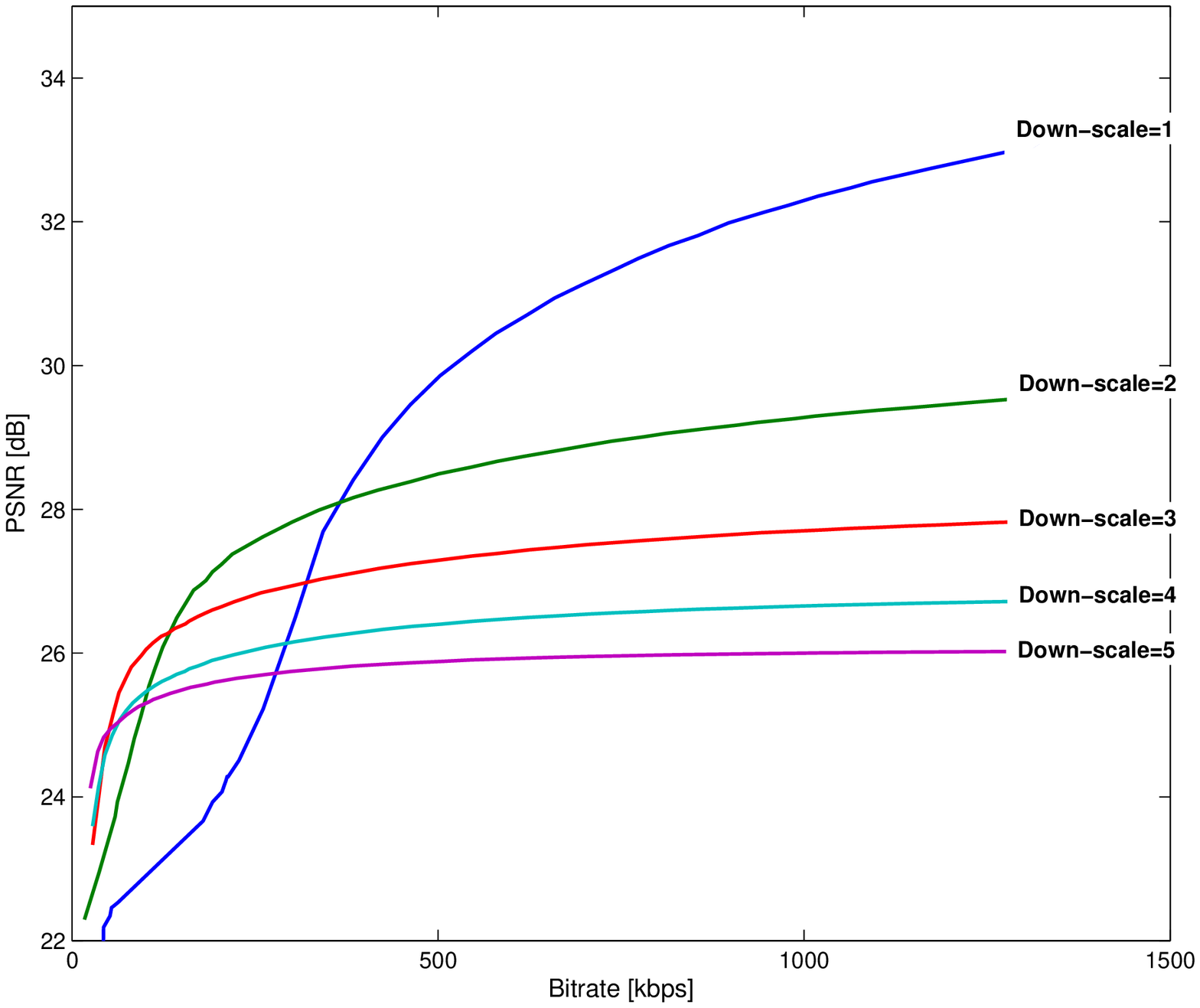}
\label{Fig:Spatial_Compression-Upsampling_Results_For_Various_Spatial_Downsampling_Factors_old_town_cross}}
\subfloat[]{\includegraphics[width=3.5in]{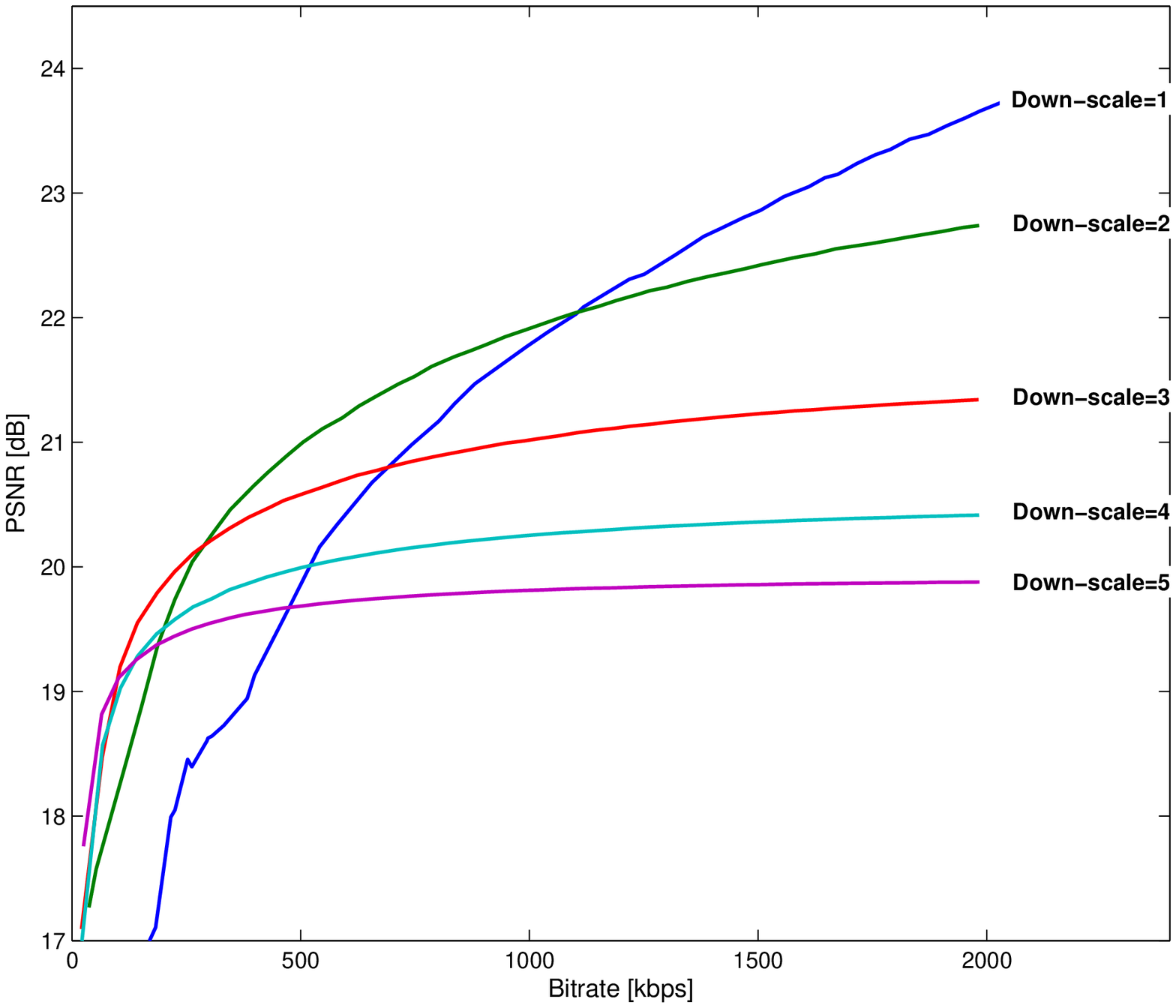}
\label{Fig:Spatial_Compression-Upsampling_Results_For_Various_Spatial_Downsampling_Factors_parkrun}}
\caption{Output PSNR of compression-scaling system for spatial scaling only. (a) 'Old town cross' (b) 'Parkrun', both 720x720, grayscale.}
\label{Fig:Output PSNR of compression-scaling system - only spatial scaling}
\end{figure*}

Let us consider spatial scaling only. The results (Fig. \ref{Fig:Output PSNR of compression-scaling system - only spatial scaling}) show behavior that is similar to the theoretical prediction of the model. Specifically, the decision regions are also constructed in an order of increasing down-scaling factors as the bit-rate decreases. 

Our proposed compression-scaling system for temporal-scaling only improved the compression results for bit-rates lower than 350kbps and 1100kbps for 'Old town cross' (Fig. \ref{Fig:Spatial_Compression-Upsampling_Results_For_Various_Spatial_Downsampling_Factors_old_town_cross}) and 'Parkrun' (Fig. \ref{Fig:Spatial_Compression-Upsampling_Results_For_Various_Spatial_Downsampling_Factors_parkrun}), respectively. 
For example, the results for 'Old town cross' (Fig. \ref{Fig:Spatial_Compression-Upsampling_Results_For_Various_Spatial_Downsampling_Factors_old_town_cross}) show a PSNR improvement of 3.3dB at 180kbps, and bit-savings of 43\% at 27dB both by halving the frame width and height. Visual demonstration of the result is given in Fig. \ref{Fig:oldtowncross_best_spatial_downscaling_at_180kbps}.

%

\subsection{Spatio-Temporal Scaling}
\begin{figure*}
\centering
\subfloat[]{\includegraphics[width=3.5in]{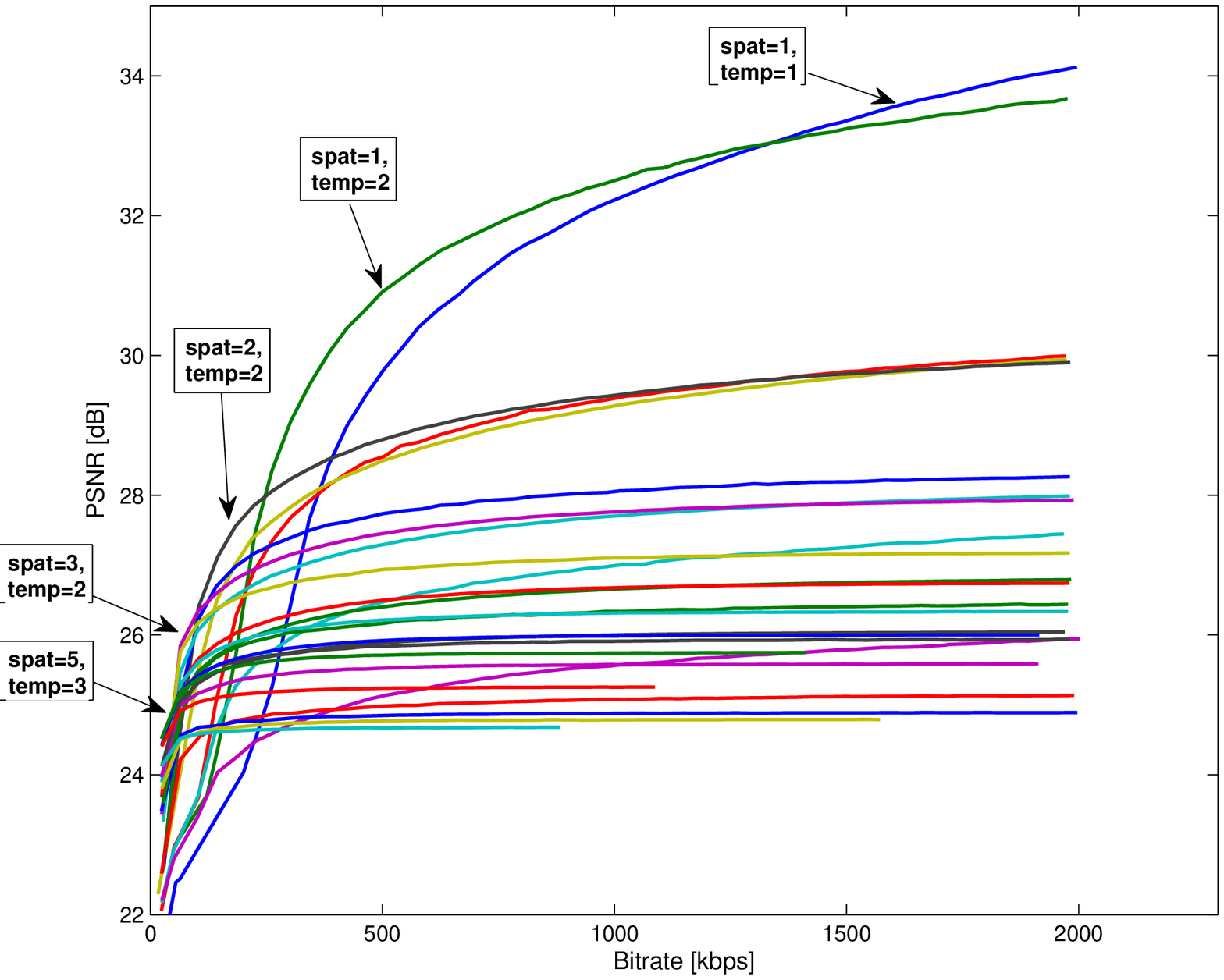}
\label{Fig:SpatioTemporal_Compression-Upsampling_Results_For_Various_SpatioTemporal_Downsampling_Factors_old_town_cross}}
\subfloat[]{\includegraphics[width=3.5in]{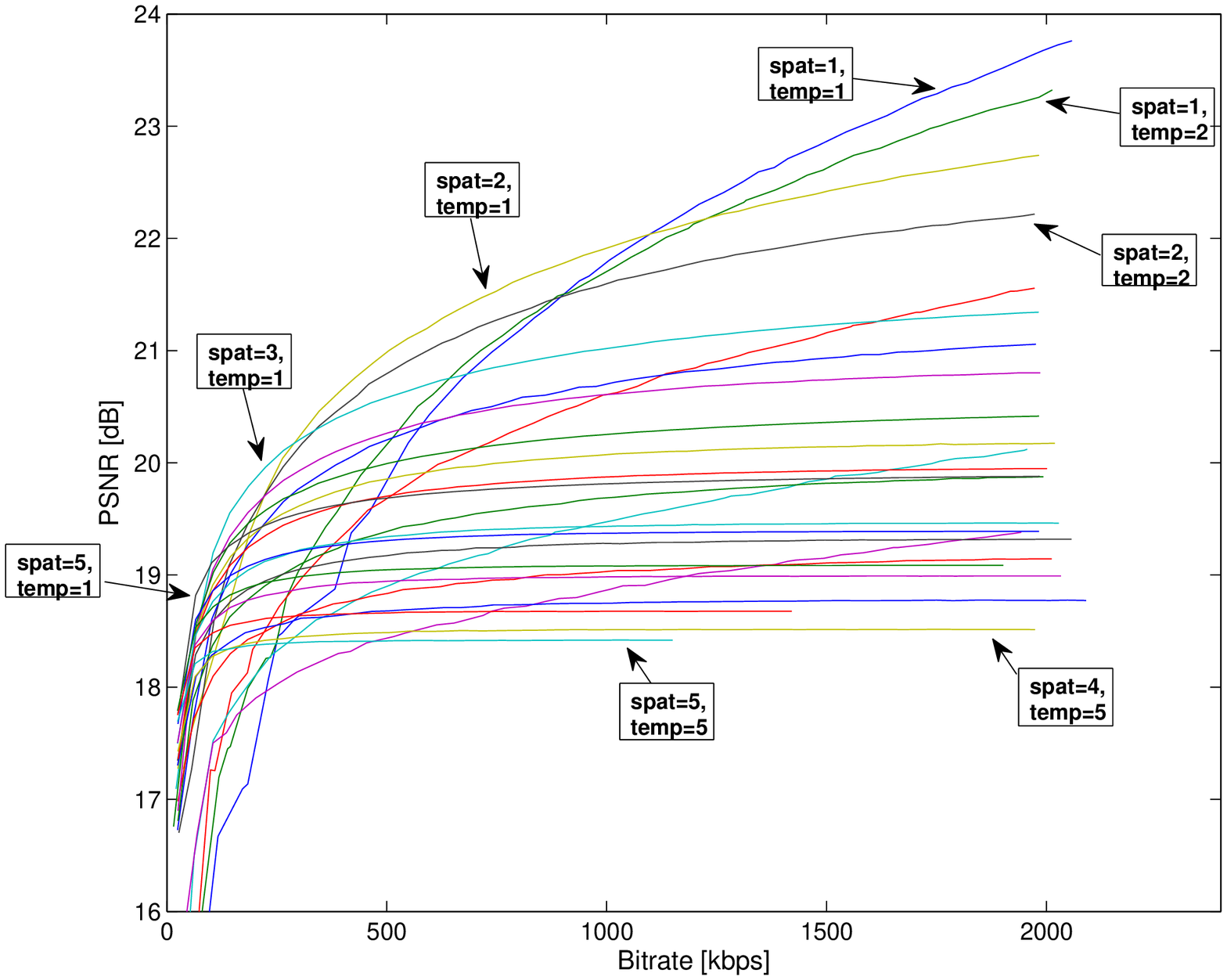}
\label{Fig:SpatioTemporal_Compression-Upsampling_Results_For_Various_SpatioTemporal_Downsampling_Factors_parkrun}}
\caption{Output PSNR of compression-scaling system for spatio-temporal scaling. (a) 'Old town cross' (b) 'Parkrun', both 720x720, grayscale.}
\label{Fig:Output PSNR of compression-scaling system - spatio-temporal scaling}
\end{figure*}

Parkrun has significant tendency to spatial scaling (Fig. \ref{Fig:SpatioTemporal_Compression-Upsampling_Results_For_Various_SpatioTemporal_Downsampling_Factors_parkrun}) due to the complex temporal properties (motion). Old town cross is more balanced and has mixed preferences for the spatio-temporal scaling (Fig. \ref{Fig:SpatioTemporal_Compression-Upsampling_Results_For_Various_SpatioTemporal_Downsampling_Factors_old_town_cross}).

Our proposed compression-scaling system for spatio-temporal scaling improved the compression results for bit-rates lower than 1250kbps and 1100kbps for 'Old town cross' (Fig. \ref{Fig:SpatioTemporal_Compression-Upsampling_Results_For_Various_SpatioTemporal_Downsampling_Factors_old_town_cross}) and 'Parkrun' (Fig. \ref{Fig:SpatioTemporal_Compression-Upsampling_Results_For_Various_SpatioTemporal_Downsampling_Factors_parkrun}), respectively. In these cases, the bit-rate threshold is the maximal value between the thresholds found for separately spatial and temporal scaling. This relation between threshold bit-rates for the joint and separated scaling should be common for non-trivial video signals.

\begin{figure*}[!t]
\centering
\subfloat[]{\includegraphics[width=5.5in]{oldtowncross_original_corresponding_to_direct_compression_at_180kbps}
\label{Fig:oldtowncross_original_corresponding_to_direct_compression_at_180kbps__2}}
\\
\subfloat[]{\includegraphics[width=5.5in]{oldtowncross_direct_compression_at_180kbps}
\label{Fig:oldtowncross_direct_compression_at_180kbps__2}}
\\
\subfloat[]{\includegraphics[width=5.5in]{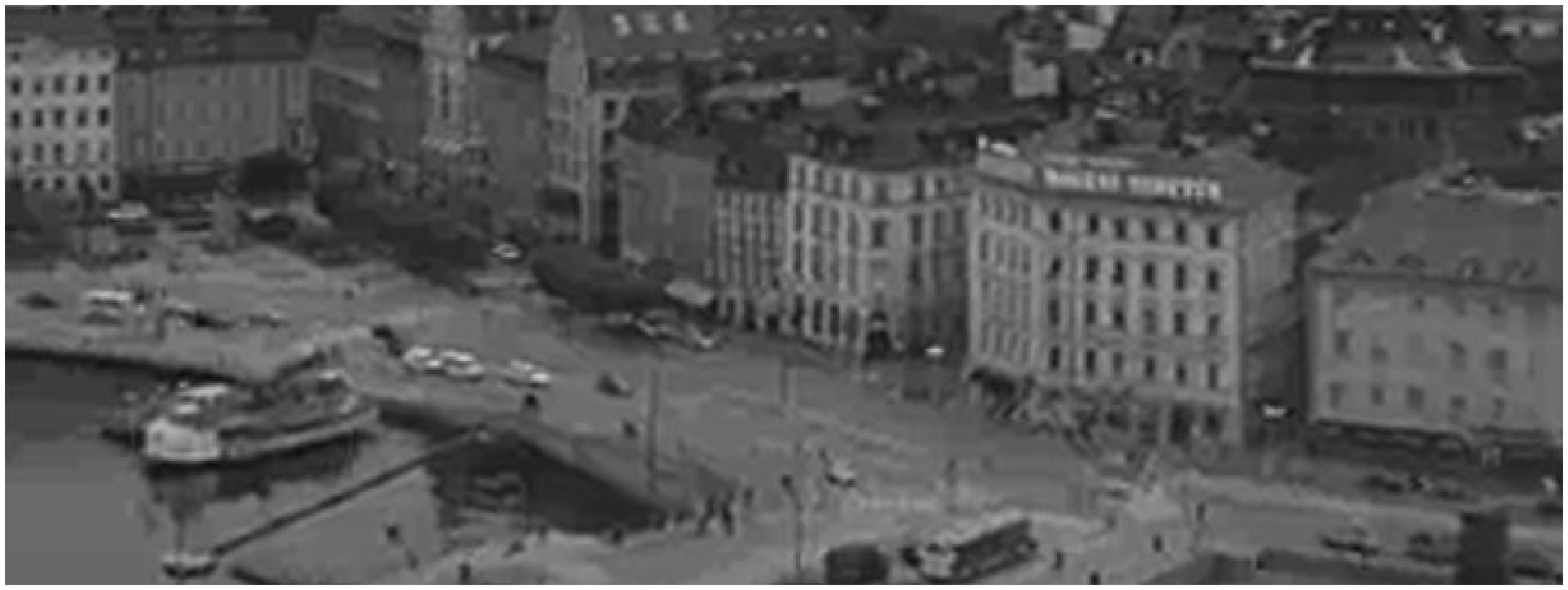}
\label{Fig:oldtowncross_best_spatial_downscaling_at_180kbps}}
\\
\subfloat[]{\includegraphics[width=5.5in]{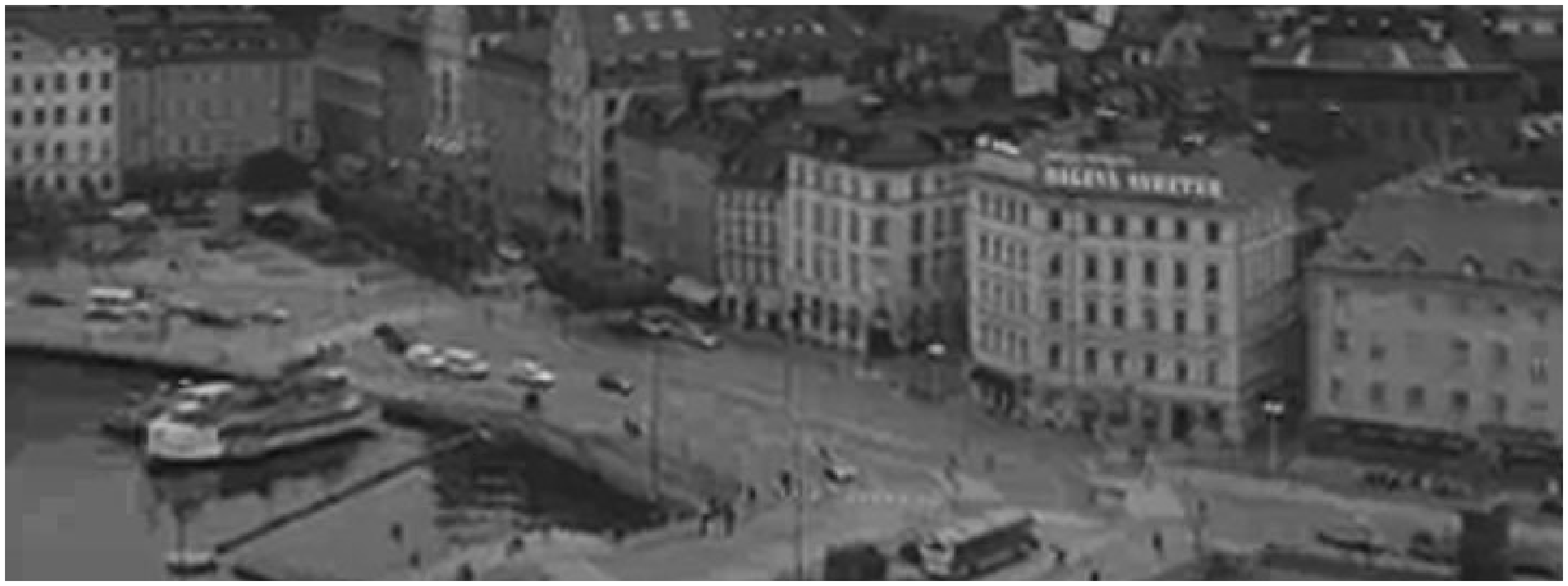}
\label{Fig:oldtowncross_best_spatio_temporal_downscaling_at_180kbps}}
\caption{Demonstration of video compression at low bit-rates. Part of a frame from 'Old town cross' (720p, 50fps). (a) original, (b) directly compressed at 180kbps, (c) spatial down-sampling by 2 before compression at 180kbps, and (d) spatio-temporal down-sampling by 2 before compression at 180kbps. }
\label{Fig:compression_at_low_bit_rates_comparison_with_spatio_temporal_scaling_results}
\end{figure*}

Halving the frame-rate, frame width and height of 'Old town cross' results in a PSNR gain of 3.9dB at 180kbps, and bit-savings of 56\% at fixed PSNR of 27dB (Fig. \ref{Fig:SpatioTemporal_Compression-Upsampling_Results_For_Various_SpatioTemporal_Downsampling_Factors_old_town_cross}). These improvements exceed those of spatial-only or temporal-only scaling and are observed in Fig. \ref{Fig:oldtowncross_best_spatio_temporal_downscaling_at_180kbps}.

%

%


\section{Conclusion}
In this work we examined spatio-temporal scaling operations for improving video compression at low bit-rates. We proposed an analytic model for video compression at low bit-rates. Moreover, a model for the entire compression-scaling system was also introduced. We analytically showed that we benefit from applying a spatio-temporal down-scaling when the given bit budget is low. The optimal spatio-temporal scaling factors were found to depend on the complexities of texture and motion and their mutual relation.

Future work can improve estimation efficiency for using in real-time applications. Moreover, the spatio-temporal characterization of the video signal can be further developed to practically describe longer and more dynamic video scenes, or to be adjusted to specific application needs. In addition, enhancing the model to support the subjective cost of down-scaling may be useful.


%

\appendices 
\section{The Expected MSE of a slice}
\label{appendix:Calculation of The Expected MSE of a slice}
We use (\ref{eq:residual approximation using finite set of basis functions - explicit}) to calculate the MSE of slice $\Delta _{11}$:
\begin{IEEEeqnarray}{rCl}
\label{eq:MSE calculation of residual block}
\nonumber MSE_{f_r}\left( {{\Delta _{11}}} \right) & = & \frac{1}{{{\beta ^2}}} \cdot \sum\limits_{p = 1}^{\beta} {\sum\limits_{q = 1}^{\beta} {MSE_{{f_r}}\left( {{\Delta _{11,pq}}} \right)} }
\\ \nonumber
& = & \frac{1}{{{\beta ^2}}} \sum\limits_{p = 1}^{\beta} \sum\limits_{q = 1}^{\beta} \frac{1}{{A\left( {{\Delta _{11,pq}}} \right)}} \iint\limits_{{\Delta _{11,pq}}} {{{\left( {{f_r}\left( {x,y} \right) - {{\hat f}_{r,{\Delta _{11,pq}}}}\left( {x,y} \right)} \right)}^2}dxdy}
\\ \nonumber
& = & \frac{1}{{{\beta ^2} A\left( {{\Delta _{11,11}}} \right)}} \sum\limits_{p = 1}^{\beta} {\sum\limits_{q = 1}^{\beta} {\iint\limits_{{\Delta _{11,pq}}} {{{\left( {{f_r}\left( {x,y} \right) - {{\hat f}_{r,{\Delta _{11,pq}}}}\left( {x,y} \right)} \right)}^2}dxdy}} }
\\ \nonumber
& = &  \frac{1}{{{\beta ^2} \cdot \frac{1}{{{\beta ^ 2} M N}} }} \cdot \sum\limits_{p = 1}^{\beta} \sum\limits_{q = 1}^{\beta} \left[ \iint_{{\Delta _{11,pq}}} {f_r^2\left( {x,y} \right)dxdy} \right.
\\ \nonumber
&& \qquad \left. - 2\iint_{{\Delta _{11,pq}}} {{f_r}\left( {x,y} \right){{\hat f}_{r,{\Delta _{11,pq}}}}\left( {x,y} \right)dxdy} + \iint_{{\Delta _{11,pq}}} {\hat f_{r,{\Delta _{11,pq}}}^2\left( {x,y} \right)dxdy} \right]
\\ \nonumber
& = & MN \cdot \sum\limits_{p = 1}^{\beta} \sum\limits_{q = 1}^{\beta} \left[ \iint_{{\Delta _{11,pq}}} {f_r^2\left( {x,y} \right)dxdy} \right.
\\ \nonumber
&& \qquad \left. - 2\sum \sum\limits_{\left( {k,l} \right) \in \Omega } {{\left\langle {{f_r}\left( {x,y} \right),{\Phi _{kl}}\left( {x,y} \right)} \right\rangle }_{{\Delta _{11,pq}}}} \times \iint_{{\Delta _{11,pq}}} {{f_r}\left( {x,y} \right){\Phi _{kl}}\left( {x,y} \right)dxdy}  \right.
\\ \nonumber
&& \qquad {} \left. + \sum\limits_{}^{} {\sum\limits_{\left( {k,l} \right) \in \Omega }^{} {\left\langle {{f_r}\left( {x,y} \right),{\Phi _{kl}}\left( {x,y} \right)} \right\rangle _{{\Delta _{11,pq}}}^2} } \right]
\\ \nonumber
& = & MN \cdot \sum\limits_{p = 1}^{\beta} \sum\limits_{q = 1}^{\beta} \left[ \iint_{{\Delta _{11,pq}}} {f_r^2\left( {x,y} \right)dxdy} - \sum\limits_{}^{} {\sum\limits_{\left( {k,l} \right) \in \Omega }^{} {\left\langle {{f_r}\left( {x,y} \right),{\Phi _{kl}}\left( {x,y} \right)} \right\rangle _{{\Delta _{11,pq}}}^2} }  \right]
\end{IEEEeqnarray}
Due to wide-sense stationarity of the residual signal over the sub-slices, we can expect the MSE for simplicity:
\begin{IEEEeqnarray}{rCl}
\label{eq:MSE calculation of residual block - appendix}
\nonumber E\left[ {MS{E_{{f_r}}}\left( {{\Delta _{11}}} \right)} \right] & = & MN \sum\limits_{p = 1}^{\beta} \sum\limits_{q = 1}^{\beta} E\left( \,\, \iint\limits_{{\Delta _{11,pq}}} {f_r^2\left( {x,y} \right)dxdy} - \sum\limits_{}^{} {\sum\limits_{\left( {k,l} \right) \in \Omega }^{} {\left\langle {{f_r}\left( {x,y} \right),{\Phi _{kl}}\left( {x,y} \right)} \right\rangle _{{\Delta _{11,pq}}}^2} }  \right) 
\\ \nonumber
& = & MN \cdot {\beta ^2} \cdot \left[ \iint_{{\Delta _{11,11}}} {E\left[ {f_r^2\left( {x,y} \right)} \right]dxdy} - \sum\limits_{}^{} {\sum\limits_{\left( {k,l} \right) \in \Omega }^{} {E\left[ {\left\langle {{f_r}\left( {x,y} \right),{\Phi _{kl}}\left( {x,y} \right)} \right\rangle _{{\Delta _{11,11}}}^2} \right]} }  \right]
\\ \nonumber
& = & {\beta ^2}MN \cdot \left[ {A\left( {{\Delta _{11,11}}} \right) \cdot {R_{{\Delta _{ij}}}}\left( {0,0} \right) - \sum\limits_{}^{} {\sum\limits_{\left( {k,l} \right) \in \Omega }^{} {E\left[ {F_{kl}^2} \right]} } } \right]
\\ \nonumber
& = & {\beta ^2}MN \cdot \left[ {\frac{1}{{{\beta ^2} M N}} \cdot {R_{{\Delta _{ij}}}}\left( {0,0} \right) - \sum\limits_{}^{} {\sum\limits_{\left( {k,l} \right) \in \Omega }^{} {E\left[ {F_{kl}^2} \right]} } } \right]
\\ \nonumber
& = & {R_{{\Delta _{ij}}}}\left( {0,0} \right) - {\beta ^2} M N \cdot \sum\limits_{}^{} {\sum\limits_{\left( {k,l} \right) \in \Omega }^{} {E\left[ {F_{kl}^2} \right]} } 
\end{IEEEeqnarray}

\section{The Expected MSE of a slice with Quantization}
\label{appendix:Calculation of The Expected MSE of a slice with Quantization}

Recall (\ref{eq:coefficient quantization error}), where the quantization error of the $\left( {k,l} \right)$ coefficient is given by $\Gamma _{kl}^2 = {\left( {{F_{kl}} - F_{kl}^Q} \right)^2}$.
The squared-error of the quantized representation over a sub-slice is
\begin{IEEEeqnarray}{rCl}
\label{eq:squared-error of the quantized representation over a sub-slice}
\nonumber && \iint_{{\Delta _{ij,pq}}} {{{\left( {f_{r,{\Delta _{ij,pq}}}^{}\left( {x,y} \right) - \hat f_{r,{\Delta _{ij,pq}}}^Q\left( {x,y} \right)} \right)}^2}dxdy} 
\\ \nonumber
&& = \iint_{{\Delta _{ij,pq}}} \left( f_{r,{\Delta _{ij,pq}}}^{}\left( {x,y} \right) - \hat f_{r,{\Delta _{ij,pq}}}^{}\left( {x,y} \right) + \hat f_{r,{\Delta _{ij,pq}}}^{}\left( {x,y} \right) - \hat f_{r,{\Delta _{ij,pq}}}^Q\left( {x,y} \right) \right)^2dxdy
\\ \nonumber
&& = \iint\limits_{{\Delta _{ij,pq}}} {{{\left( {f_r^{}\left( {x,y} \right) - \hat f_r^{}\left( {x,y} \right)} \right)}^2}dxdy} + \iint\limits_{{\Delta _{ij,pq}}} {{{\left( {\hat f_r^{}\left( {x,y} \right) - \hat f_r^Q\left( {x,y} \right)} \right)}^2}dxdy}
\\ \nonumber
&& \qquad {} + 2 \iint\limits_{{\Delta _{ij,pq}}} {\left( {f_r^{}\left( {x,y} \right) - \hat f_r^{}\left( {x,y} \right)} \right) \left( {\hat f_r^{}\left( {x,y} \right) - \hat f_r^Q\left( {x,y} \right)} \right)dxdy}
\\ \nonumber
&& = A\left( {{\Delta _{ij,pq}}} \right) \cdot MSE_{{f_r}}\left( {{\Delta _{ij,pq}}} \right) + \iint_{{\Delta _{ij,pq}}} {{{\left( {\sum {\sum\limits_{\left( {k,l} \right) \in \Omega } {\left( {F_{kl} - F_{kl}^Q} \right) \cdot {\Phi _{kl}}\left( {x,y} \right)} } } \right)}^2}dxdy} 
\\ \nonumber
&& \qquad {} + 2 \cdot \iint_{{\Delta _{ij,pq}}} \left( {\sum {\sum\limits_{\left( {k,l} \right) \in {\Omega ^C}} {F_{kl} \cdot {\Phi _{kl}}\left( {x,y} \right)} } } \right) \cdot \left( {\sum {\sum\limits_{\left( {k,l} \right) \in \Omega } {\left( {F_{kl} - F_{kl}^Q} \right) \cdot {\Phi _{kl}}\left( {x,y} \right)} } } \right)dxdy
\\ \nonumber
&& = A\left( {{\Delta _{ij,pq}}} \right) \cdot MSE_{{f_r}}\left( {{\Delta _{ij,pq}}} \right) + \sum {\sum\limits_{\left( {k,l} \right) \in \Omega } {\Gamma _{kl}^2} }
\\
\end{IEEEeqnarray}
Using the last result we calculate the MSE of the residual block with the quantized coefficients:
\begin{IEEEeqnarray}{rCl}
\label{eq:MSE of the quantized representation over a sub-slice}
MSE_{{f_r}}^Q\left( {{\Delta _{11}}} \right) & = & \frac{1}{{{\beta ^2}}} \sum\limits_{p = 1}^{\beta} {\sum\limits_{q = 1}^{\beta} {{\beta ^2}MN} \cdot \iint\limits_{\Delta _{ij,pq}^{}} {{{\left( {{f_r}\left( {x,y} \right) - \hat f_r^Q\left( {x,y} \right)} \right)}^2}dxdy}}
\\ \nonumber
& = & \frac{1}{{{\beta ^2}}} \sum\limits_{p = 1}^{\beta} \sum\limits_{q = 1}^{\beta} \left( MS{E_{{f_r}}}\left( {{\Delta _{ij,pq}}} \right) + {\beta ^2}MN \sum\limits_{}^{} {\sum\limits_{\left( {k,l} \right) \in \Omega }^{} {\Gamma _{kl}^2} }  \right)
\end{IEEEeqnarray}
We take expectation of the MSE and continue to develop the expression:
\begin{IEEEeqnarray}{rCl}
\label{eq:expected MSE of the quantized representation over a sub-slice - appendix}
E\left[ {MSE_{{f_r}}^Q\left( {{\Delta _{11}}} \right)} \right] & = & \frac{1}{{{\beta ^2}}} \sum\limits_{p = 1}^{\beta} \sum\limits_{q = 1}^{\beta} E\left[ MS{E_{{f_r}}}\left( {{\Delta _{ij,pq}}} \right) + {\beta ^2}MN \sum\limits_{}^{} {\sum\limits_{\left( {k,l} \right) \in \Omega }^{} {\Gamma _{kl}^2} } \right]
\\ \nonumber
& = & E\left[ {MSE_{{f_r}}\left( {{\Delta _{ij,pq}}} \right)} \right] + {\beta ^2}MN  \sum {\sum\limits_{\left( {k,l} \right) \in \Omega } {E\left[ {\Gamma _{kl}^2} \right]} }
\\ \nonumber
& = & {R_{{\Delta _{ij}}}}\left( {0,0} \right) - {\beta ^2}MN \sum {\sum\limits_{\left( {k,l} \right) \in \Omega } {\left( {E\left[ {F_{kl}^2} \right] - E\left[ {{{\left( {F_{kl} - F_{kl}^Q} \right)}^2}} \right]} \right)} }
\end{IEEEeqnarray}


\ifCLASSOPTIONcaptionsoff
  \newpage
\fi



\bibliographystyle{IEEEtran}
\bibliography{IEEEabrv,Refs_for_Improving_Low_Bit_Rate_Video_Coding}
%

%
%





\end{document}